\documentclass[twocolumn,floatfix,showpacs,prd,aps,tightenlines,superscriptaddress]{revtex4}
\usepackage{amsmath}
\usepackage{graphicx}
\usepackage{psfrag}
\usepackage{color}
\usepackage{dcolumn}% Align table columns on decimal point
\usepackage{bm}% bold math
\usepackage{longtable}
\usepackage[mathscr]{eucal}
\usepackage{mathrsfs}
\usepackage{xspace}
\newcommand{\bea}{\begin{eqnarray}}
\newcommand{\eea}{\end{eqnarray}}
\newcommand{\beq}{\begin{equation}}
\newcommand{\eeq}{\end{equation}}

\newcommand{\lazev}{{\sc LazEv}\xspace}
\newcommand{\twopunctures}{{\sc TwoPunctures}\xspace}
\newcommand{\carpet}{{\sc Carpet}\xspace}
\newcommand{\cactus}{{\sc Cactus}\xspace}
\newcommand{\ET}{{\sc EinsteinToolkit}\xspace}

\begin{document}

\def\fun#1#2{\lower3.6pt\vbox{\baselineskip0pt\lineskip.9pt
  \ialign{$\mathsurround=0pt#1\hfil##\hfil$\crcr#2\crcr\sim\crcr}}}
\def\lap{\mathrel{\mathpalette\fun <}}
\def\gap{\mathrel{\mathpalette\fun >}}
\def\kms{{\rm km\ s}^{-1}}
\def\vk{V_{\rm recoil}}

\def\figwidth{0.95\columnwidth}

\title{Accuracy Issues for Numerical Waveforms}

\author{Yosef Zlochower}
\affiliation{Center for Computational Relativity and Gravitation,\\
and School of Mathematical Sciences, Rochester Institute of
Technology, 85 Lomb Memorial Drive, Rochester, New York 14623}

\author {Marcelo Ponce}
\affiliation{Center for Computational Relativity and Gravitation,\\
Technology, 85 Lomb Memorial Drive, Rochester, New York 14623}

\affiliation{Department of Physics, University of Guelph, Guelph,
Ontario N1G 2W1, Canada}

\author {Carlos O. Lousto}
\affiliation{Center for Computational Relativity and Gravitation,\\
and School of Mathematical Sciences, Rochester Institute of
Technology, 85 Lomb Memorial Drive, Rochester, New York 14623}

\begin{abstract}
We study the convergence properties of our implementation of the {\it
moving punctures} approach at very high resolutions for an equal-mass,
nonspinning, black-hole binary. We find
convergence of the Hamiltonian constraint on the
horizons and the $L_2$ norm of the Hamiltonian constraint in the bulk
for sixth- and eighth-order finite difference implementations.
 The momentum constraint is
more sensitive, and its $L_2$ norm shows clear convergence for a
system with consistent sixth-order finite differencing,
while the momentum and BSSN constraints on the horizons show
convergence for both sixth- and eighth-order systems.
We analyze the gravitational waveform error from the late inspiral,
merger, and ringdown.
We find that using several lower-order techniques for increasing the
speed of numerical relativity simulations actually lead to apparently
nonconvergent errors. Even when using standard high-accuracy
techniques, rather than seeing clean convergence, where the
waveform phase is a monotonic function of grid resolution, we find
that the phase tends to oscillate with resolution, possibly due to
stochastic errors induced by grid refinement boundaries. 
Our results seem to indicate that one can obtain gravitational 
waveform phases to within $0.05$ rad. (and possibly as small as
0.015 rad.), 
while the amplitude error can be reduced to $0.1\%$. We then compare
with the waveforms obtained using the CCZ4 formalism. We find that the
CCZ4 waveforms have larger truncation errors for a given resolution,
but the Richardson extrapolation phase of the CCZ4 and BSSN waveforms
agrees to within $0.01$ rad., even during the ringdown.
\end{abstract}

 \maketitle

\section{Introduction}\label{sec:Introduction}

Numerical relativity (NR) has progressed rapidly since the
breakthroughs of 2005~\cite{Pretorius:2005gq, Campanelli:2005dd,
Baker:2005vv} that  allowed for the long-term evolution of black-hole
binaries (BHBs). Among NR's significant achievements are its
contributions towards the modeling of astrophysical gravitational wave
sources that will be relevant for the first direct detection and
parameter estimation by gravitational wave
observatories~\cite{Aylott:2009ya, Aylott:2009tn}.  NR has also made
contributions to the modeling of astrophysical sources, notably, the
modeling of the recoil kick imparted to the remnant BH from a BHB
merger due to unequal masses~\cite{Herrmann:2006ks, Baker:2006vn,
Gonzalez:2006md}, the remarkable discovery of unexpectedly large
recoil velocities from the merger of certain spinning
BHBs~\cite{Herrmann:2007ac, Campanelli:2007ew, Campanelli:2007cga,
Lousto:2008dn, Pollney:2007ss, Gonzalez:2007hi, Brugmann:2007zj,
Choi:2007eu, Baker:2007gi, Schnittman:2007ij, Baker:2008md,
Healy:2008js, Herrmann:2007zz, Herrmann:2007ex, Tichy:2007hk,
Koppitz:2007ev, Miller:2008en, Lousto:2011kp, Zlochower:2010sn,
Lousto:2010xk, Lousto:2011kp, Lousto:2012su}, and the application of
the numerical techniques to combined systems of BHs and neutron
stars~\cite{Sekiguchi:2010ja, Etienne:2011ea, Etienne:2008re,
Etienne:2007jg, Rezzolla:2011da, Hotokezaka:2011dh, Sekiguchi:2011zd,
Foucart:2011mz, Duez:2008rb}. More mathematical aspects of relativity
have also recently been investigated, including the evolution of
N-black holes~\cite{Lousto:2007rj, Campanelli:2007ea, Galaviz:2010te},
the exploration of the no-hair theorem~\cite{Campanelli:2008dv,
Owen:2010vw}, and cosmic~\cite{Campanelli:2006uy} and topological
censorship~\cite{Ponce:2010fq}, as well as BHBs in dimensions higher
than four~\cite{Shibata:2010wz, Zilhao:2010sr, Witek:2010xi}.
 The current state of
the art simulations can simulate  BHBs with mass ratios as small as
$q=1/100$~\cite{Lousto:2010ut, Sperhake:2011ik} and highly spinning
BHBs with intrinsic spins $\alpha=S_H/M_H^2$ up to (at least)
$0.97$~\cite{Lovelace:2010ne, Lovelace:2011nu}.  Currently these runs
are very costly and it is hard to foresee the possibility of
completely covering the parameter space densely enough for match
filtering the data coming from advanced laser interferometric
detectors by the time they become operational.

To reduce the computational costs, several low-accuracy approximations
are sometimes used. Among them are the techniques introduced in
Ref.~\cite{Brugmann:2008zz} where the number of buffer zones at AMR
boundaries is reduced by lowering the order of finite differencing by
successive orders near the AMR boundaries, the use of simple
interpolations of spectral initial data rather than using the complete
spectral expansion~\cite{Ansorg:2004ds}, and copying the initial
data to the two past time levels for use in prolongation at the
initial timestep. All of these approximations proved to be useful for
numerical simulations, but each one also has the side effect of
introducing a (hopefully) small ${\cal O}(h)$ error.  In this paper,
we examine the effects of these approximations by performing
high-resolution simulations of equal-mass, nonspinning BHBs,
 a problem generally considered well under control. We
show that a nonconvergent error, that cannot be detected by simple
means, is present when these techniques are used together. However,
even when low-accuracy approximations are eliminated, an apparently
stochastic error in the waveform phase is still present that prevents
us from seeing clean convergence of the waveform even at very high
resolutions. We estimate
this stochastic phase error is controllable  to within NINJA and NRAR
accuracy requirements, but does make it very difficult to get an
unambiguous measurement of the waveform phase and phase error.

Here we examine in detail the case of a nonspinning equal-mass binary.
The idea is that, any issues of accuracy found for these simple
systems will only be compounded by the introduction of different mass
ratios (which may require more AMR refinement levels and have a lower
effective resolution) and spins (which reduces the smoothness of the
data and leads to more complicated motion of the binary).

\section{Numerical Relativity Techniques}\label{Sec:Numerical}
To compute the numerical initial data, we use the puncture
approach~\cite{Brandt97b} along with the {\sc
TwoPunctures}~\cite{Ansorg:2004ds} thorn (see Table~\ref{tab:ID} for
the initial data parameter).  In this approach the
3-metric on the initial slice has the form $\gamma_{a b} = (\psi_{BL}
+ u)^4 \delta_{a b}$, where $\psi_{BL}$ is the Brill-Lindquist
conformal factor, $\delta_{ab}$ is the Euclidean metric, and $u$ is
(at least) $C^2$ on the punctures. The extrinsic curvature is given by
$K_{ij}=(\psi + u)^{-2} \hat K_{ij}$, where $\hat K_{ij}$ is a
superposition of Bowen-York solution~\cite{Bowen80} for BHs with spin
$\vec S$ (here zero) and momentum $\vec p$. The Brill-Lindquist
conformal factor is given by $ \psi_{BL} = 1 + \sum_{i=1}^n m_{i}^p /
(2 |\vec r - \vec r_i|), $ where $n$ is the total number of
``punctures'', $m_{i}^p$ is a parameter (not the horizon mass), and $\vec r_i$ is the coordinate location of puncture $i$.  We
evolve these BHB data sets using the \lazev~\cite{Zlochower:2005bj} implementation of the moving puncture
approach~\cite{Campanelli:2005dd, Baker:2005vv} with the conformal
function $W=\sqrt{\chi}=\exp(-2\phi)$ suggested by
Ref.~\cite{Marronetti:2007wz} (where $\chi$ is the evolution variable
introduced in~\cite{Campanelli:2005dd}).  The moving puncture approach
is based on the Baumgarte-Shapiro-Shibata-Nakamura (BSSN, BSSN-NOK)
formalism~\cite{Nakamura87, Shibata95, Baumgarte99}, where
the
gauge and evolution variables are adapted such that the system is
finite at the punctures.
 For the runs presented
here, we use centered, eighth-order finite differencing in
space~\cite{Lousto:2007rj} and a fourth-order Runge Kutta time
integrator. (Note that we do not upwind the advection terms.)
For points near (but not on) the computational domain boundary,
we reduce the order of finite differencing such that the stencils
fit in the computational domain (i.e.\ reduce from eighth to sixth
to fourth to second).
\begin{table}
  \caption{Initial data parameters. All runs used the same
   parameters. The punctures are locates at $\vec r=\pm(x,0,0)$,
   with momenta $\pm(p_r,p_t,0)$, puncture mass parameters $m_p$,
   and zero spin. The ADM mass is $1.00000003$. We use
 $N_{\rm colo.}\times N_{\rm colo.}\times N_{\rm colo.}$ collocation
points for the \twopunctures spectral solver.}
  \label{tab:ID}
  \begin{ruledtabular}
  \begin{tabular}{ll|ll|ll}
   $x$ &  4.250 & $m_p$ & 0.4887922277 & $p_r$ & -0.00114088 \\
    $p_t$ & 0.11081837 & $N_{\rm colo.}$ & 40 & $M_H$ & 0.50580 \\
  \end{tabular}
  \end{ruledtabular}
\end{table}

On the boundary points themselves, we use radiative boundary conditions 
for all variables, which on the
$x=x_{\rm min}$ and $x=x_{\rm max}$ faces, have the form
\begin{equation}
  \partial_t f|_x = -v \left(\frac{r}{x} \partial_x f
+\frac{f-f\infty}{r}\right),
\end{equation}
where the wavespeed $v$ and asymptotic value of the function
$f_\infty$
are parameters. We set the wavespeed to 1 for $\tilde  \gamma_{ij},
\tilde A_{ij}$, and $\beta^i$, while we set the wavespeed to
$\sqrt{2}$ for $\alpha, K,$ and $W$. We calculate $\partial_x f$
using a
one-sided, second-order stencil. We set $f_\infty=0$  for all variables
except  $\tilde \gamma_{ii}$, $\alpha$ and $W$, where we set $f_\infty=1$.

In the {\it method of lines} approach, for every evolved function
$f$, there is a corresponding function $f_{\rm RHS}$, where 
$\partial_t f = f_{\rm RHS}$. The evolution code fills in 
all the RHS gridfunctions (including on the boundaries).
No additional boundary conditions (except symmetry boundary conditions)
are applied to the evolution variables themselves.

After each {\it substep} of the RK4 integration, we enforce the
algebraic constraints, $\tilde \gamma = \det(\tilde \gamma_{ij}) = 1$ 
and $\tilde A= \tilde
\gamma^{ij}\tilde A_{ij}=0$ by
renormalizing $\tilde \gamma_{i j}$ and subtracting the trace from
$\tilde A_{ij}$, i.e.\ $$
\tilde \gamma_{ij} \to \frac{1}{\tilde \gamma^{3}} \tilde
\gamma_{ij},$$ and
$$
\tilde A_{ij} \to \tilde A_{ij} - \frac{1}{3} \tilde \gamma_{ij}
\tilde A.
$$

Our code uses the \cactus/\ET~\cite{cactus_web,
einsteintoolkit, Loffler:2011ay} infrastructure.  We use the \carpet~\cite{Schnetter-etal-03b} mesh refinement driver to provide a
``moving boxes'' style of mesh refinement. In this approach refined
grids of fixed size are arranged about the coordinate centers of both
holes.  The \carpet code then moves these fine grids about the
computational domain by following the trajectories of the two BHs.

We use {\sc AHFinderDirect}~\cite{Thornburg2003:AH-finding} to locate
apparent horizons.  We measure the magnitude of the horizon spin using
the Isolated Horizon algorithm detailed in Ref.~\cite{Dreyer02a}.
Note that once we have the horizon spin, we can calculate the horizon
mass via the Christodoulou formula
\begin{equation} {m^H} =
\sqrt{m_{\rm irr}^2 + S_H^2/(4 m_{\rm irr}^2)} \,,
\end{equation}
where $m_{\rm irr} = \sqrt{A/(16 \pi)}$ and $A$ is the surface area of
the horizon (we neglect the effects of momentum). 

For the gravitational waveform, we calculate the Newman-Penrose
$\psi_4$ Weyl scalar using fourth-order centered finite differencing
and then decompose $\psi_4$ at fixed radii using spin-weighted
spherical harmonics, $\psi_r = \sum_2^N c_{\ell,m} {}_{-2}Y_{\ell m}$.
 We use a fourth-order accurate integration to
calculate the $(\ell,m)$ modes of $\psi_4$.
We also  calculate the Hamiltonian (${\cal H}$), momentum (${\cal C}^i$),
and BSSN constraint (${\cal G}^i$) violations using fourth-order
centered finite differencing. We use fourth-order finite differencing
to reduce the computational cost of calculating analysis
quantities.
 
For our gauge conditions, we use a  modified 1+log lapse and a
modified Gamma-driver shift condition~\cite{Alcubierre02a,
Campanelli:2005dd, vanMeter:2006vi}, and an initial lapse $\alpha(t=0)
= 2/(1+\psi_{BL}^{4})$.  The lapse and shift are evolved with
\begin{subequations}
\label{eq:gauge}
  \begin{eqnarray}
(\partial_t - \beta^i \partial_i) \alpha &=& - 2 \alpha K,\\
 \partial_t \beta^a &=& (3/4) \tilde \Gamma^a - \eta \beta^a \,,
 \label{eq:Bdot}
 \end{eqnarray}
 \end{subequations}
where we use $\eta=2$ for all simulations presented below.

For our tests of the fast, but low-accuracy techniques, we performed a
set of 11  simulations.  In all cases, the physical boundaries of our
computational domains were located at $400M$ in all coordinate
directions, although we did use $\pi$-symmetry and z-reflection
symmetry to reduce the computational domain by a factor of 4. We chose
configurations with a base outer resolution of $h_{0}=400M/70$ with 9
levels of refinement. The finest resolution was  $M/44.8$. We then
used global refinements of this base configuration with outer
resolutions of $h_1=400M/84$, $h_2=400M/100$, and $h_3=400M/120$. The
ratio between the timestep and spatial resolution on the coarsest grid
was set to $0.06125$, the next two levels had 1/2 this timestep, the
following three had 1/4, and then each additional level had time
refinement of a factor of 2 with respect to the next coarsest grid.
The ratio $\kappa$ of the timestep and spatial resolution [Courant
factor (CFL)] on the finest timelevel was $\kappa=dt/h=0.49$.  In
addition, we performed simulations with all CFL factors reduced by
factors of $2$, $2\sqrt{2}$, and $4$.  For the low accuracy study we
used the reduction of order technique proposed
in~\cite{Brugmann:2008zz} to reduce the number of buffer zones at
the refinement boundaries to six. This is accomplished using the
following algorithm. During the four Runge-Kutta substeps, we used the
largest centered finite differences stencil located within a given AMR
level (including buffer zones), i.e.,for interior points at least 4 
points from the
AMR boundaries (in all direction), we used standard eighth-order
finite differencing, while for points  closer to the AMR boundaries,
we used progressively lower orders. On the AMR boundaries themselves,
we used a simple copy procedure, where data from the previous time
level was copied onto the AMR boundary. Then, after the subcycling
is complete, we overwrite the AMR boundary points, and an additional
5 points inside the AMR boundary via prolongation. Note that this
means that during the four substeps of the RK4 integration,
information from the AMR boundaries can propagate outside the buffer
region (and hence reduce the order of convergence of the bulk).

To initialize the past timelevels of the initial slice, we used a
simple copy. This introduces on ${\cal O}(h)$ error in the
prolongation of the first time level. Since all subsequent field
points depend on the data on this time level, a small ${\cal O}(h)$
error should persist throughout the evolution.  We also  used the {\tt
interpolation} option of {\sc TwoPunctures} to quickly interpolate the
spectral solution on the grid. This approximation has the effect of
introducing high-frequency noise (which does dissipate away) and
constraint violation (which does not).

\begin{figure}
  \includegraphics[width=.95\columnwidth]{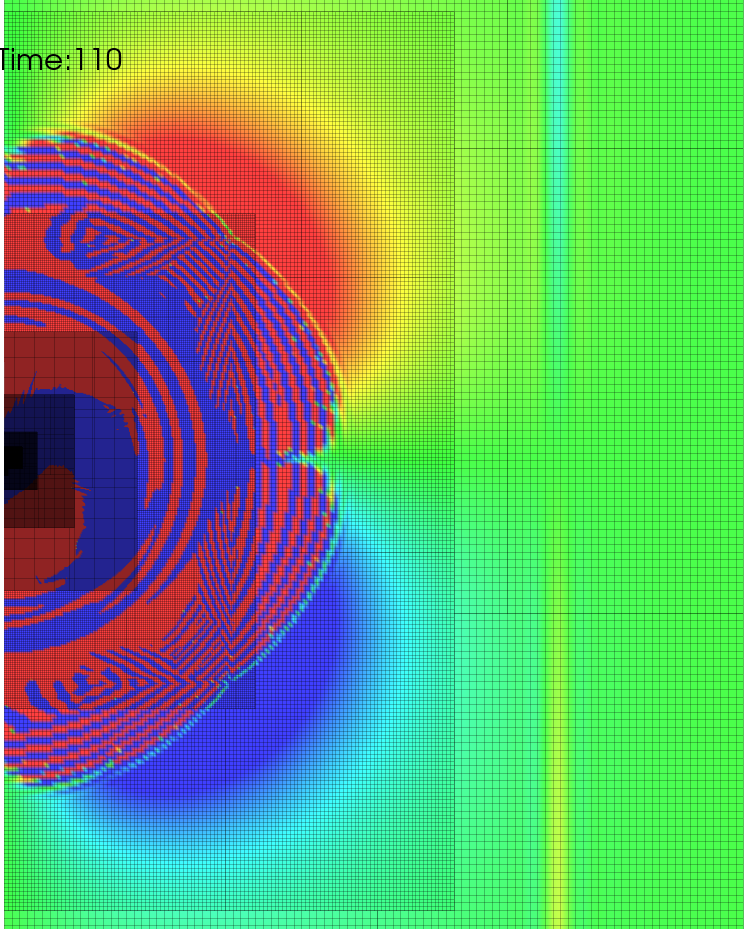}
  \caption{A plot of the $\Re\{\psi_4\}$ on the $xy$ plane at $t=110M$.
Note the interference pattern that develops as part of the wave
reflects off the AMR boundary. Also note that the reflection
originates in the buffer region and not on the AMR boundary itself.
The reflected waves will soon interfere producing still higher spatial
frequencies.}
\label{fig:amr_ref}
\end{figure}

We performed 25 high-accuracy simulations. For these simulations, we used nearly identical spatial
grid structure, but had identical CFL factors on each gridlevel. In
addition, we used resolution $h_1\cdots h_3$ above and a higher
resolution $h_4=400M/144$ (we also performed a single run with
resolution $h_5=400M/174$). We used CFL factors of $\kappa=1/4$,
$\kappa=1/(4\sqrt{2})$, and $\kappa=1/8$. A CFL factor of $\kappa=1/2$ was not stable (due
to low-resolution in the outer grid~\cite{Schnetter:2010cz}).  These
runs also differed from the low-accuracy runs in that a full
complement of 16 buffer zones were also used and no reduction of order
near AMR boundaries was necessary. In addition we used the {\tt
init\_3\_timelevels} algorithm to initialize the grid, which properly
initializes the past two timelevels of the initial data slice, and we
used the computationally expensive {\tt evaluation} option for
{\sc TwoPunctures}. 
The final two parameters that determine the accuracy of our
simulations are the dissipation order and prolongation order. We found
that fifth-order Kreiss-Oliger dissipation was most effective at
suppressing the high-frequency noise that originates as reflections from
AMR boundaries (see Fig.~\ref{fig:amr_ref}). We similarly  used fifth-order prolongation operators.
As an additional check, we compare all runs with two other systems,
an eighth-order system with seventh-order dissipation and prolongation
(we use seventh rather than ninth to reduce the number of required
buffer points) and a fully consistent sixth-order system with
seventh-order dissipation and prolongation operators.
\begin{figure}
  \includegraphics[width=\figwidth]{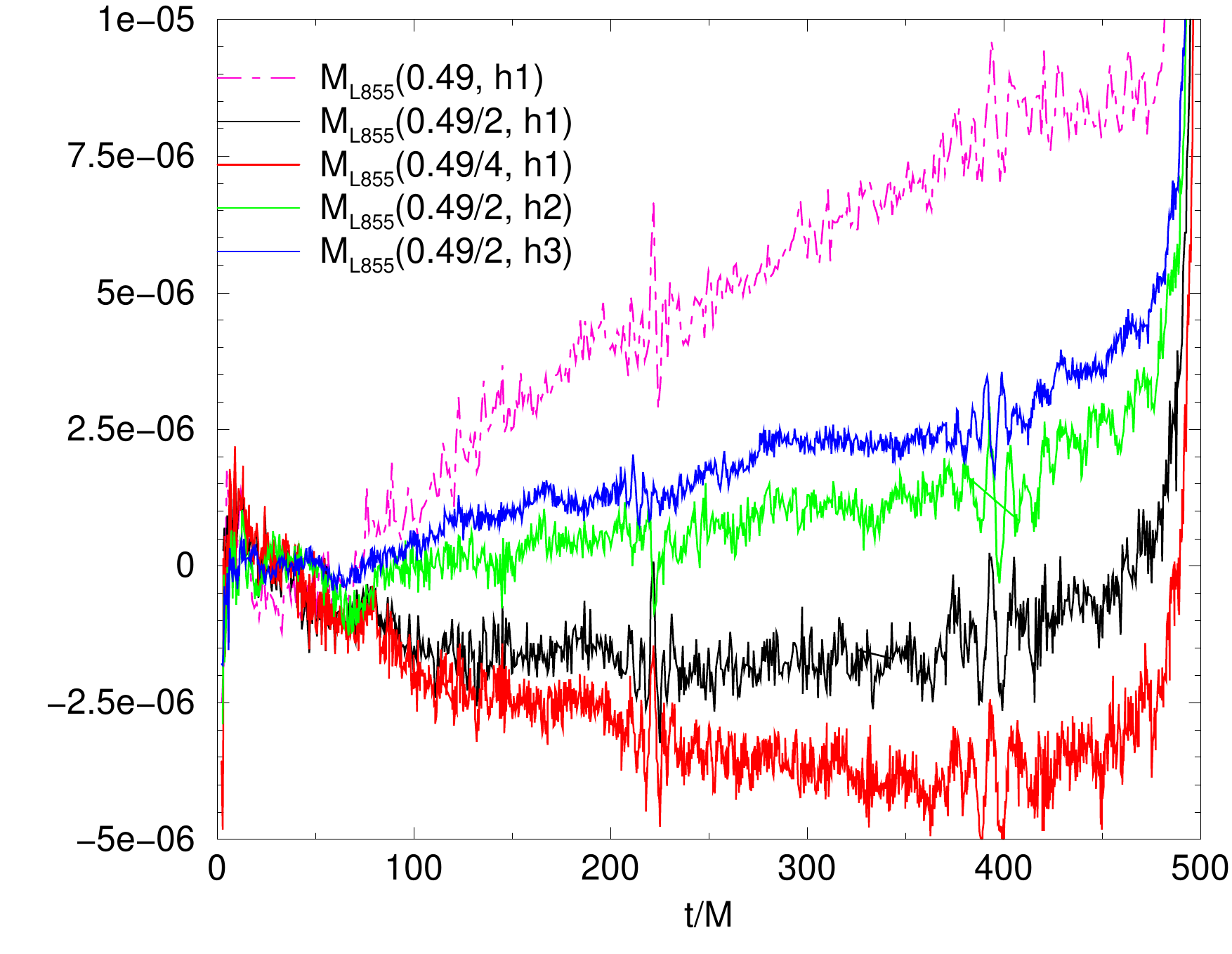}
  \includegraphics[width=\figwidth]{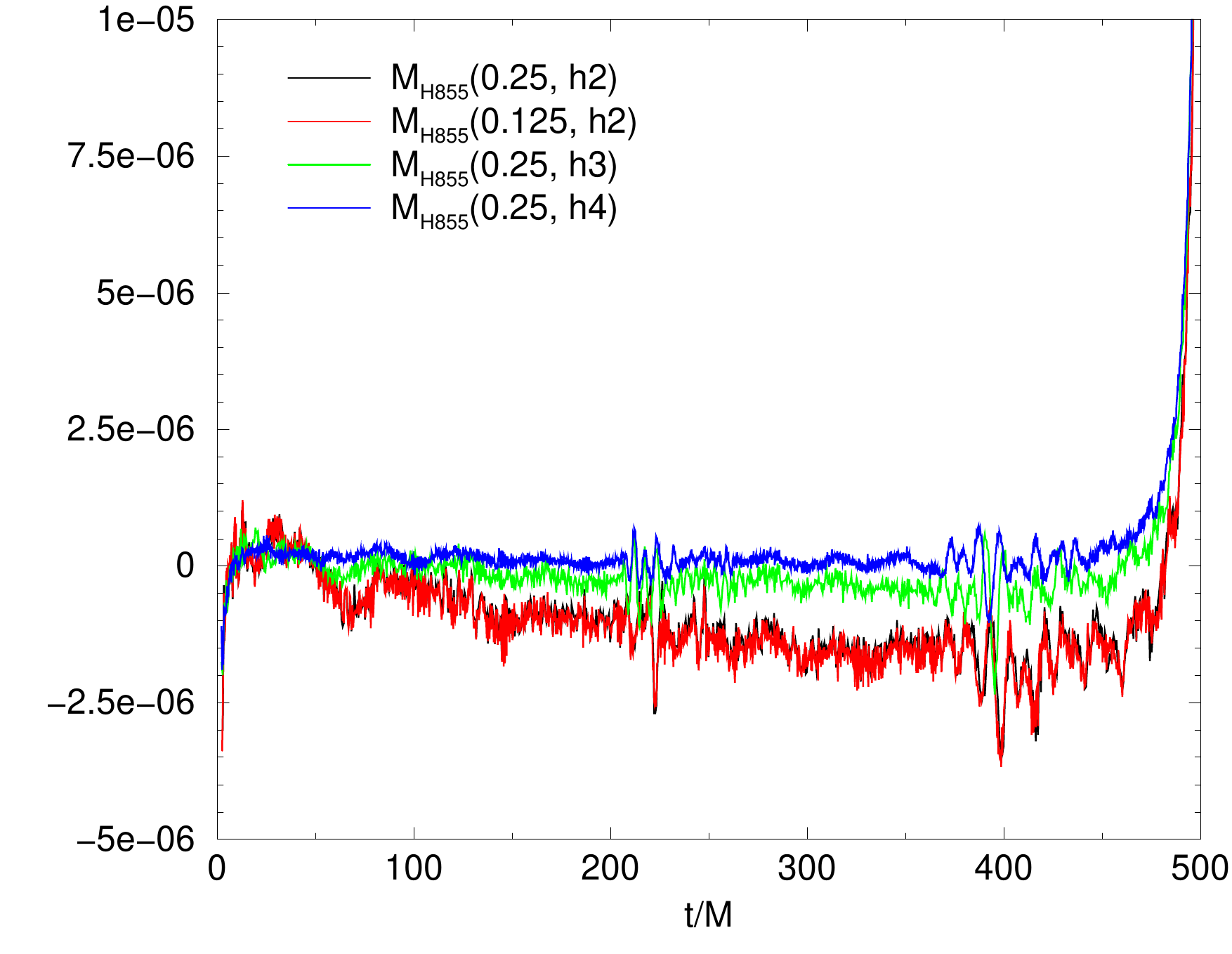}
  \caption{The horizon mass as a function of time for different
resolutions and CFL factors for the L855 
algorithm (top) and H855 algorithm (bottom). Note the poor
mass conservation for the low accuracy simulations. Also note that the
mass for the H855 simulations is unaffected by CFL (i.e.\
the curves corresponding to
different CFLs lie on top of each other), while
the
mass is strongly affected by CFL for the low accuracy simulations.
}
  \label{fig:hmass}
\end{figure}

In the current work, we are interested in computing the phase of the
dominant $(\ell=2,m=2)$ mode of $\psi_4$ at finite radius. While there
can be significant phase errors associated with extrapolation to
$r=\infty$, these errors are controllable by using larger
computational domains, multipatch methods~\cite{Pollney:2009yz}, 
pseudospectral methods~\cite{Szilagyi:2009qz}, and Cauchy-Characteristic
extraction~\cite{Babiuc:2010ze, Reisswig:2009us}. Ideally, with
high, but still practical, resolution, the phase error in the waveform
would be at or below the phase differences inherent in choosing
different background metric ansatze for the same binary
configuration, which was found to be $< 10^{-2}$ rad. using spectral
codes~\cite{Garcia:2012dc}.

In the figures and tables below, we denote the low accuracy
simulations by ``L855'' and higher-accuracy simulations by
``H855'', ``H877'', and ``H677''. Here the first digit indicates the
spatial finite-difference order, the second indicates the dissipation
order, and the third indicates the spatial prolongation order.
 
\begin{table}[t]
  \caption{Differences between the low-accuracy, and high-accuracy
    configurations. 
 ``N Buffer'' is the number of buffer zones at refinement boundaries,
 ``Initial Data'' refers to the method used for computing the conformal
factor from the spectral coefficients,
 ``AMR Initialization'' refers to the order of approximation used to
initialize the past time levels of the initial slice (used for
prolongation only). Here, ``radius'' refers to the ``half-width'' in all
directions of each level and CFL refers to the ratio between
$dt$ and $h$ on that level.
All high-accuracy simulations (H855, H877, H677)
use the same grid structures and initialization parameters.}
   \label{tab:acc_diff}
\begin{ruledtabular}
\begin{tabular}{l|l|l}
& low & high\\
\hline
  N Buffer & 6 & 16 \\
  Initial Data & interpolation  & evaluation\\
  AMR Initialization & 1st order & 4th order \\
\end{tabular}
\begin{tabular}{l|ll|rr}
& low &  & high & \\
\hline
level & radius & CFL & radius & CFL \\
\hline
 0 & 400 & $\kappa/8$ & 400 & $\kappa$\\
 1 & 208 & $\kappa/8$ & 208 & $\kappa$\\
 2 & 115 & $\kappa/8$ & 115 & $\kappa$\\
 3 & 60 & $\kappa/4$ & 60 & $\kappa$\\
 4 & 30 & $\kappa/4$ & 30 & $\kappa$\\
 5 & 12 & $\kappa/2$ & 12 & $\kappa$\\
 6 & 5 & $\kappa$ & 5 & $\kappa$\\
 7 & 1.5 & $\kappa$ & 1.5 & $\kappa$\\
 8 & 0.75 & $\kappa$ & 0.75 & $\kappa$\\
 
\end{tabular}
\end{ruledtabular}

\end{table}

\section{Results}
\label{sec:results}
Our studies were motivated by phase errors in intermediate-mass-ratio
simulations~\cite{Nakano:2011pb}. In that study, we found that mass conservation
was an important criterion for an accurate evolution. Here we study
the conservation of the horizon mass as a function of resolution and CFL
factor for both the L855 and H855 configurations.

\subsection{Horizon Mass}
The horizon mass for nonspinning BHs is given by the irreducible mass
$M_{\rm irr} = \sqrt{A_H}/16$, where
\begin{equation}
  A_H = \oint d^2s,
\end{equation} 
the integral is performed over the surface of the horizon and
$d^2s$ is the proper {\it volume} element on the horizon 
 associated with the induced metric
on the horizon.
We can also define the horizon average of a function $H$ as
\begin{equation}
  \langle H\rangle = \frac{\oint H d^2s} { \oint d^2 s}.
\end{equation}
Similarly, we can describe horizon fluxes of a vector field ${\cal
C}^i$
with the flux integral
\begin{equation}
  \langle {\cal C}^i\rangle = \frac{\oint {\cal C}^i R_i d^2s} { \oint d^2 s},
\end{equation}
where $R_i$ is the unit norm on the horizon.

As shown in Fig.~\ref{fig:hmass}, the conservation of the horizon mass
for the L855 simulations is strongly affected by the CFL
factor, with the largest deviations being between the $\kappa=0.49$
and $\kappa=0.49/4$ CFL factors. On the other hand, for the H855
simulations, the conservation of mass is unaffected by CFL, but is
rather affected only by an increase in spatial resolution. Also note
that the mass profiles are much flatter in the H855
simulations.

\begin{figure}
  \includegraphics[width=\figwidth]{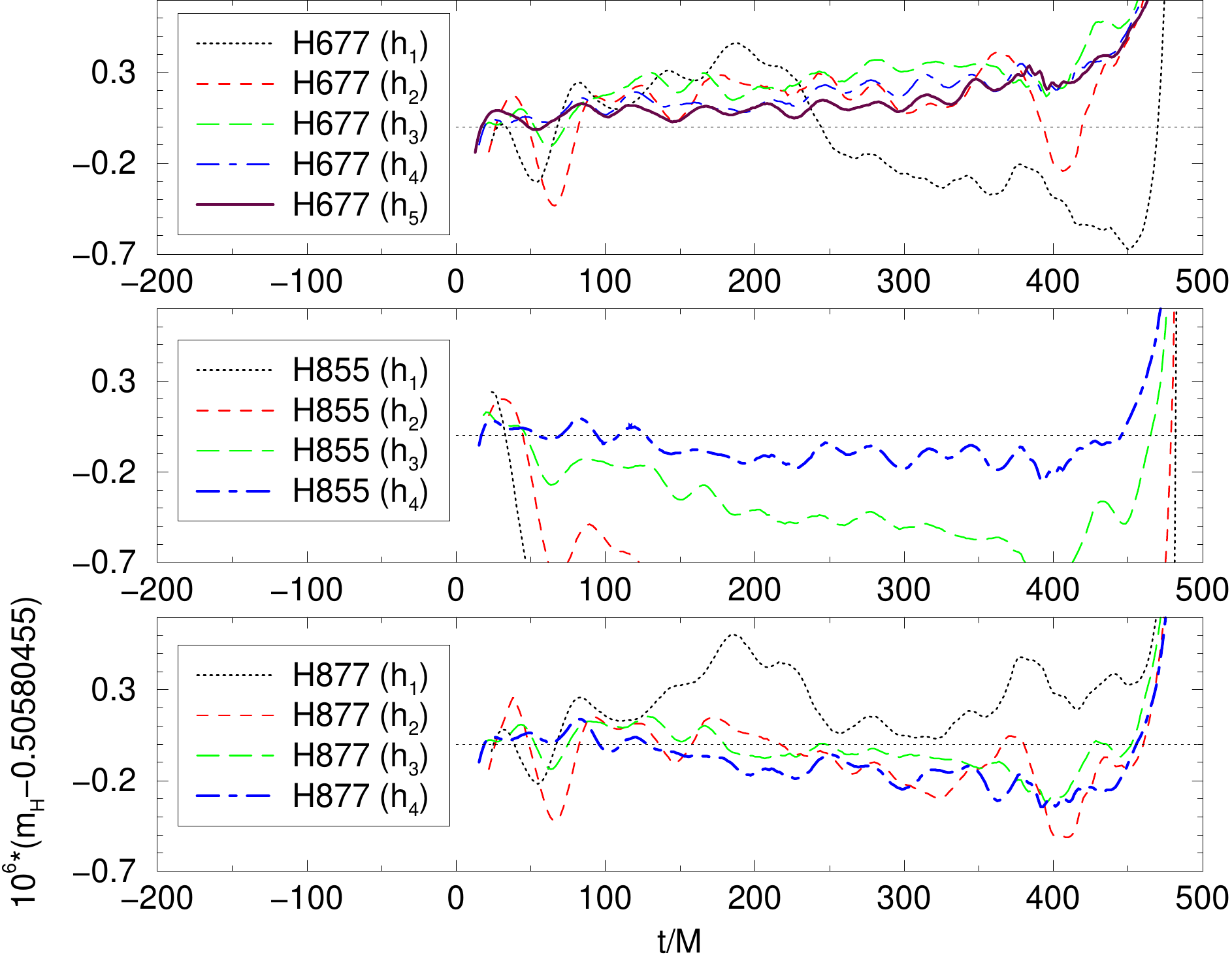}
  \caption{The horizon mass versus resolution for the H855, H877,
and H677 systems (all with CFL=0.25).
 Here the horizon mass has been smoothed by using
a running average. At high resolutions, H855 and H677 both show very
flat masses.  H855 and H877 both show a small drop in mass, which is
unphysical, while H677 shows an increase in mass.}
\label{fig:hmass_cmp_Hs}
\end{figure}

We compare the horizon masses obtained with the H855, H877 and H677
systems (at a fixed CFL of 0.25) in Fig.~\ref{fig:hmass_cmp_Hs}. The H855 shows
 the most variation with resolution,
but appears to be comparable to H877 at high resolution.  H855 and H877
both show a small drop in mass, which is
unphysical, while H677 shows an increase in mass. At intermediate
resolutions, it appears that H877 or H677 are better than H855 in
terms of mass conservation. At high resolutions, H855 and H877 appear
to be slightly more accurate than H677 since they show flatter
profiles of the mass versus time.

\subsection{Constraints}
The size of the constraint violations is a measure of
how closely the simulations obey the Einstein equations, and,
in addition,
violations of the constraint can lead to mass loss.
We examined the Hamiltonian constraint violation both in the bulk and
the average of the Hamiltonian constraint violation on each horizon.
The idea of using the horizon average constraints is
to see how much {\it strange}
matter is falling into the BHs (and thus a source of mass fluctuation). 

\begin{figure}
  \includegraphics[width=\figwidth]{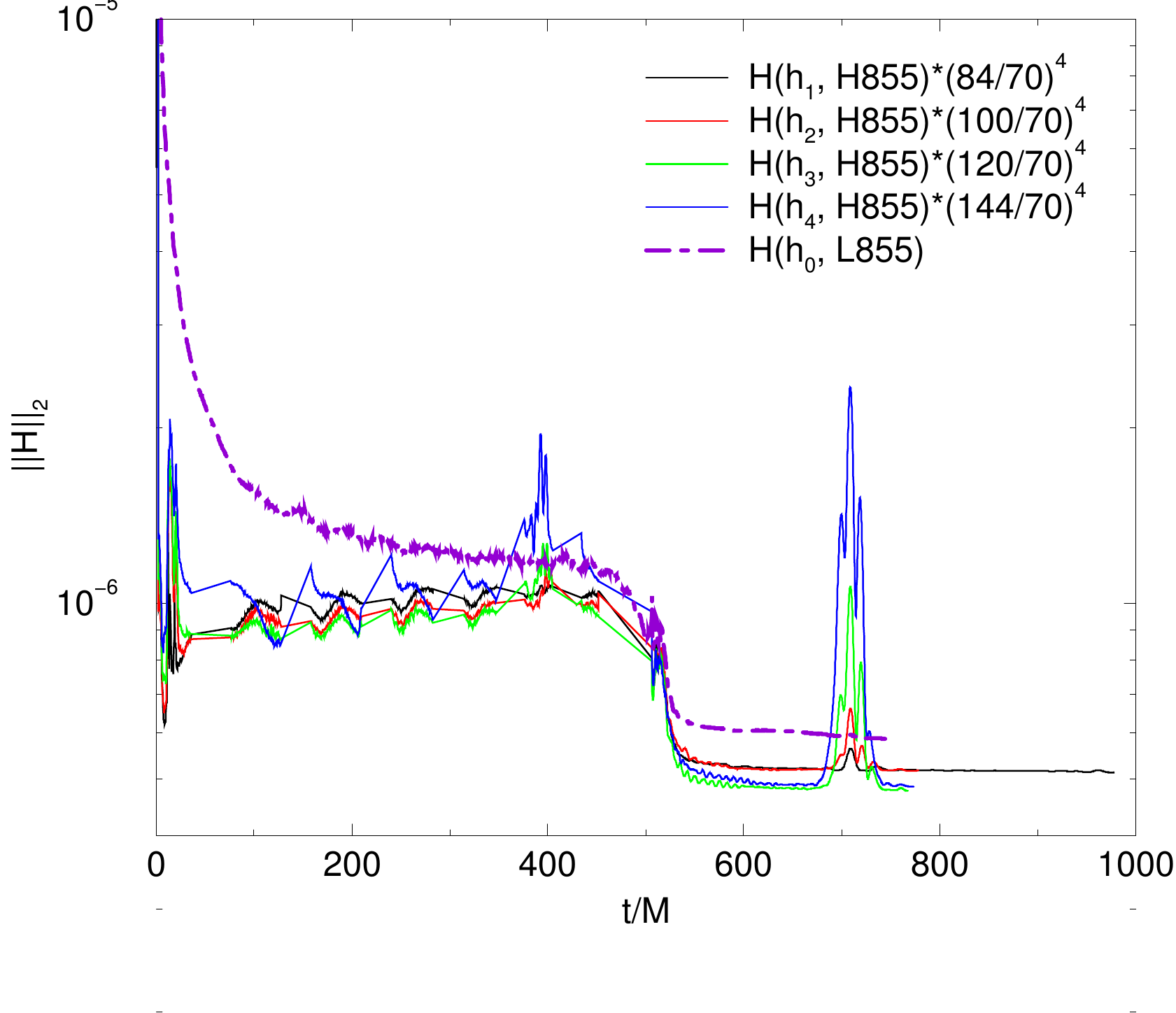}
  \caption{Convergence of the $L_2$ norm of the Hamiltonian
constraint in the volume between the horizons and a coordinate sphere
at $r=20M$. Fourth-order convergence, consistent with the constraint
calculation algorithm (and also the time integrator) is apparent for
the H855 simulations. The low accuracy simulation shows a
larger error during the inspiral, which may be the source of the
error in the simulation. Spikes associated with the BHs
crossing the symmetry boundary at $x=0$, which are an  artifact 
associated with $\pi$ symmetry, have been removed (see
Fig.~\ref{fig:H_pi_nopi}).}
  \label{fig:HC}
\end{figure}

\begin{figure}
  \includegraphics[width=\figwidth]{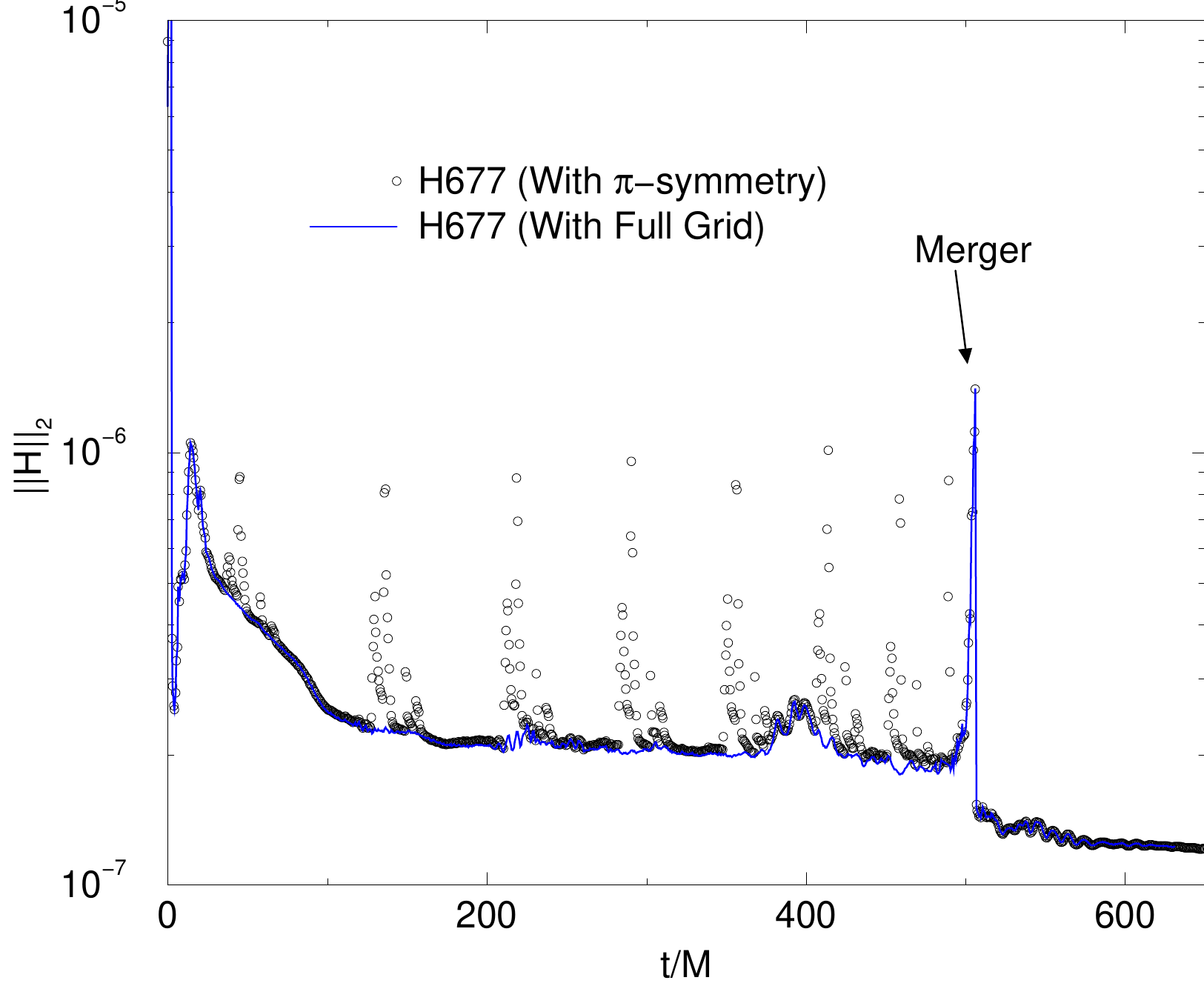}
  \includegraphics[width=\figwidth]{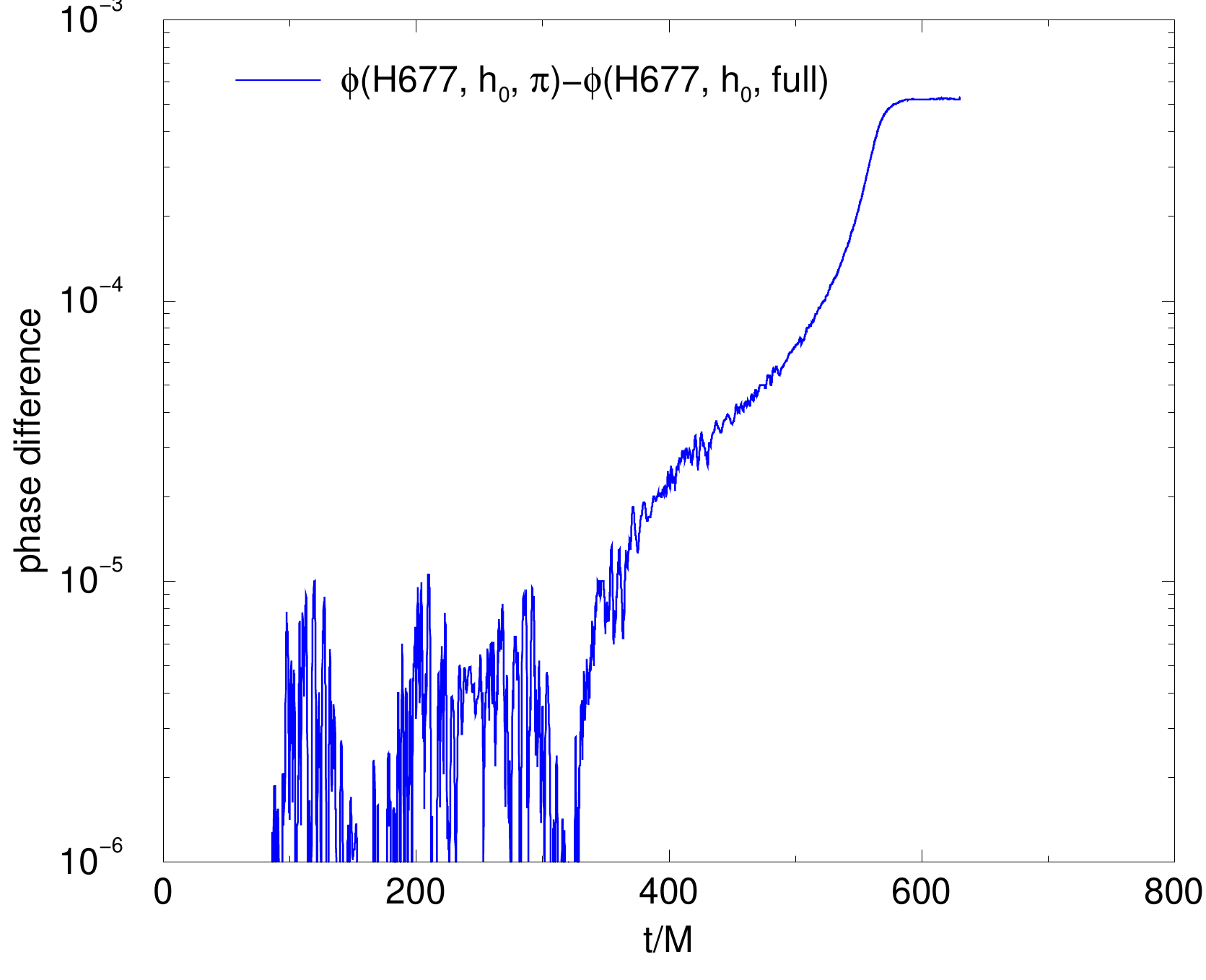}
  \caption{A plot showing that the spikes in the $L_2$ norm of the
Hamiltonian are an artifact of $\pi$-symmetry when the BHs cross the
symmetry boundary. The plot shows the $L_2$ norm
for identical runs, with the exception that one does not use
$\pi$ symmetry. Outside the spikes themselves, the two runs agree.
The bottom plot shows the phase differences between (otherwise
identical) runs with and without $\pi$-symmetry.}
\label{fig:H_pi_nopi}

\end{figure}

\begin{figure}
  \includegraphics[width=\figwidth]{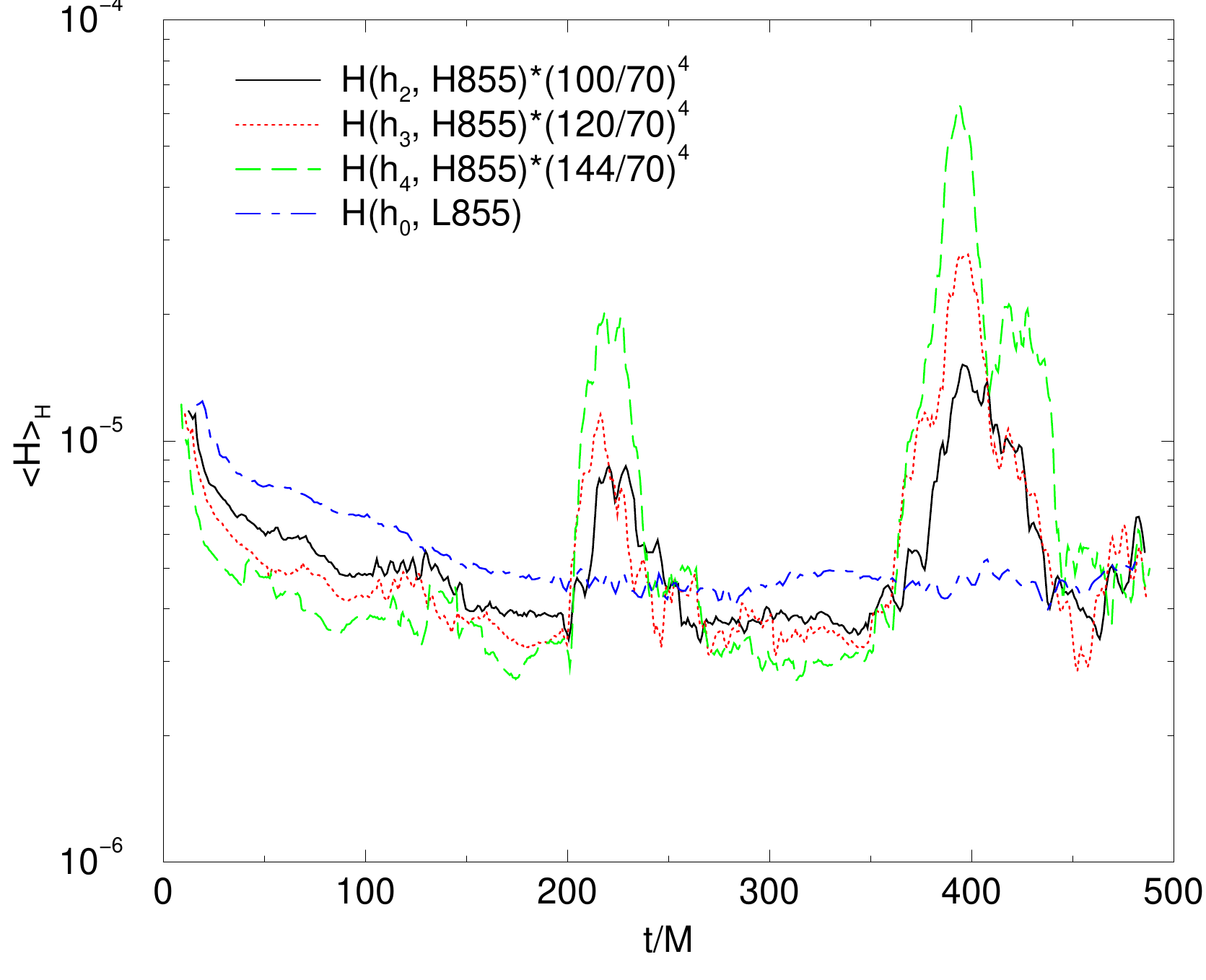}
  \caption{Convergence of the horizon averaged value of the
Hamiltonian constraint
violation for the H855 system. The horizon average constraint
violation of H855 is consistent with that of L855 (after accounting
for differences in resolution).}
  \label{fig:HC_havg}
\end{figure}

 As seen in
Figs.~\ref{fig:HC} and \ref{fig:HC_havg}, the Hamiltonian constraint violation in the bulk is
significantly larger for the L855 simulations than the higher accuracy
H855 simulations, with the
constraint violation appearing immediately at the start of the
simulation and decreasing towards the values seen in the H855
simulations. The H855 simulations, on the other hand, show
clear fourth-order convergence. The constraints on the horizons
converge to fourth-order for the H855 simulations, with
similar values to the L855. We can therefore
conclude that absorption of constraint violation by the horizon is not
the ultimate source of either the lack of conservation in the mass or
the phase error in the waveform. Spikes in the
$L_2$ norm of the Hamiltonian, which are an artifact of the
of $\pi$-symmetry (possibly in the way the grids around each BH 
need to be enlarged near the symmetry boundary, or even an
analysis artifact),
as shown
in Fig.~\ref{fig:H_pi_nopi}, have been removed from Fig.~\ref{fig:HC}.
In Fig.~\ref{fig:H_pi_nopi}, we show the the phase difference
between a run that uses $\pi$-symmetry and one that does not. The
differences in the phase are under $0.001$ rad. over the entire
waveform.

\begin{figure}
  \includegraphics[width=\figwidth]{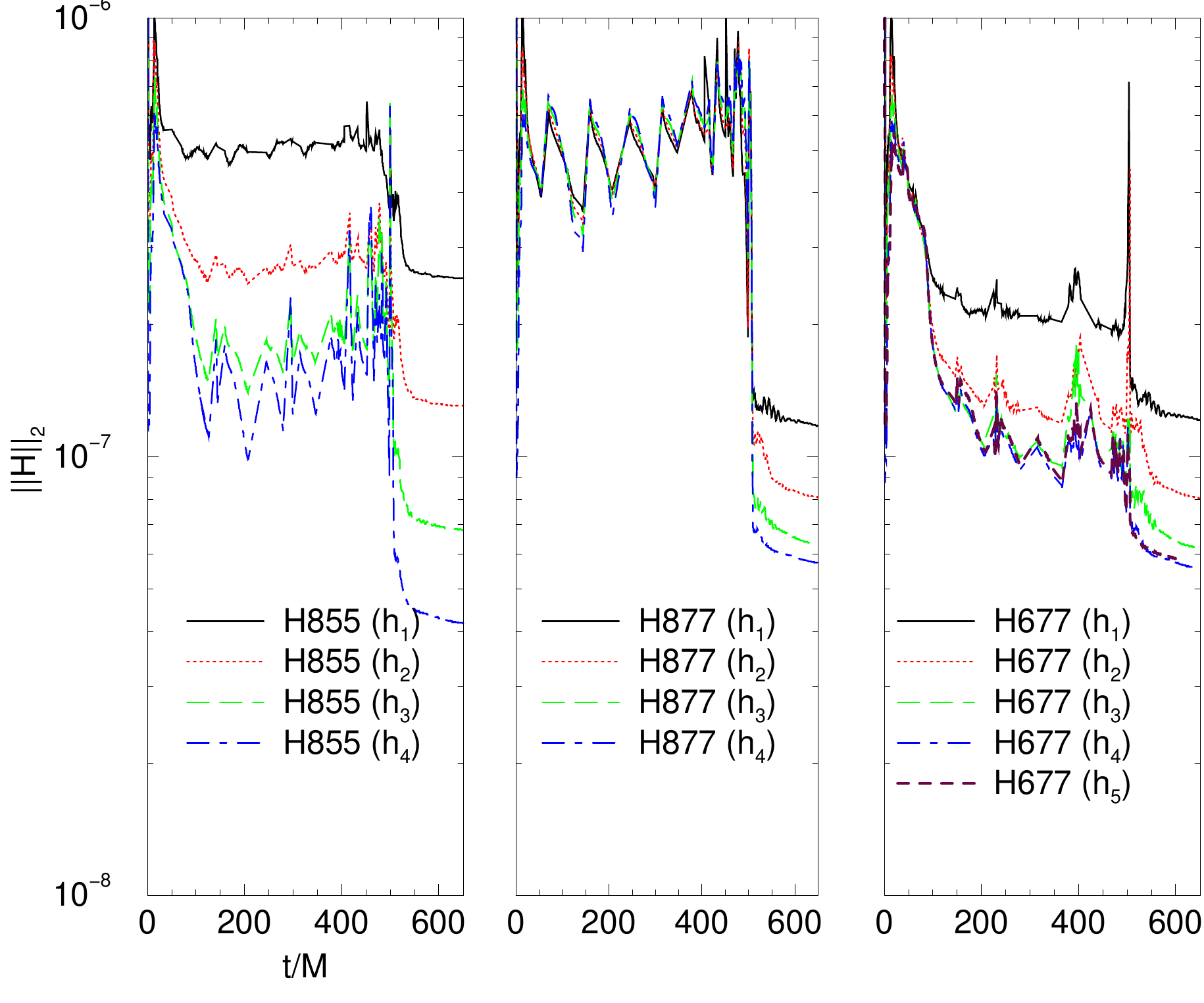}
  \caption{The $L_2$ norm of the Hamiltonian constraint for the H855,
H877, and H677 systems. H877 shows clear convergence to a finite
value at all resolutions. For H855, the constraint decreases with
resolution but appears to converge to a smaller finite value. H677
shows the best behavior, but also converges to (a still smaller)
finite value. Post merger, H855 gives the smallest violation (but only
at the highest resolution). At lower resolution, H855 gives the
largest violation. Note that the curves have not been rescaled.}
  \label{fig:HC_hcmp}
\end{figure}
\begin{figure}
  \includegraphics[width=\figwidth]{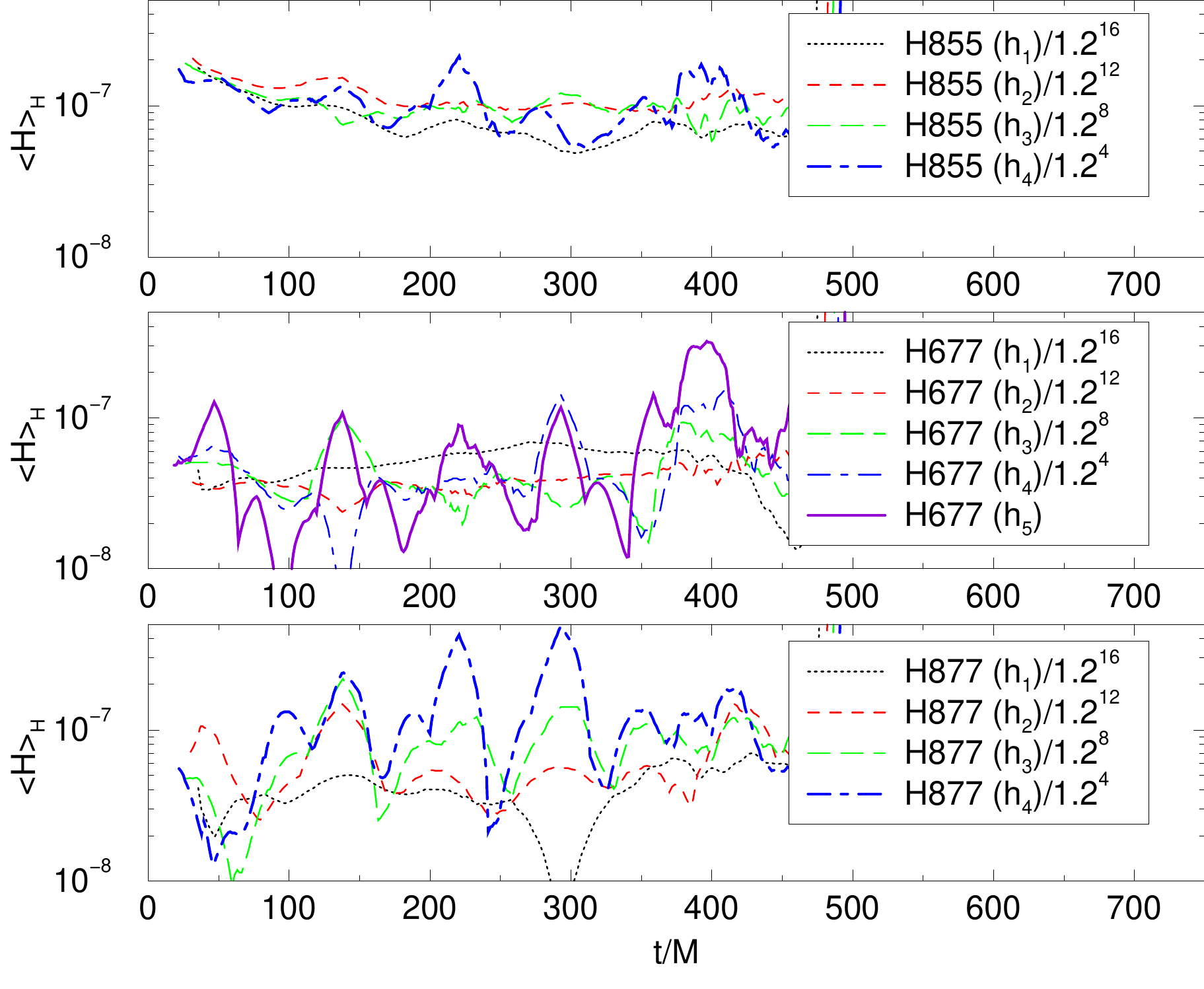}
  \caption{The horizon average value of the Hamiltonian constraint for
the H855, H877, and H677 systems. In all cases, the CFL was set to
0.25. The constraint violation have been rescaled assuming
fourth-order convergence. In all cases, convergence to zero is poorest
at the highest resolutions. H677 shows the most convergent behavior,
while H877 shows the poorest.}
  \label{fig:HC_avg_cmp}
\end{figure}
\begin{figure}
  \includegraphics[width=\figwidth]{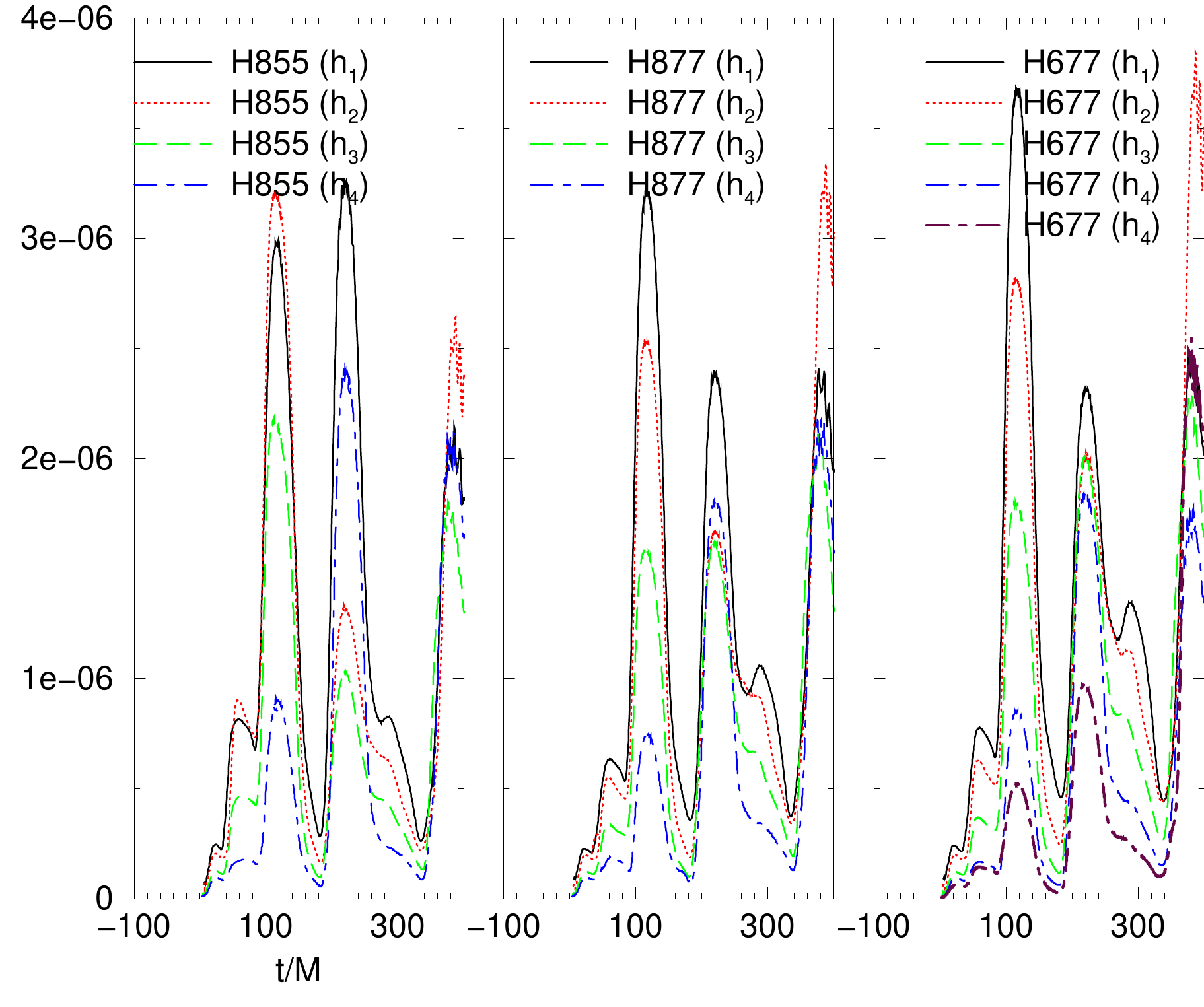}
  \caption{The $L_2$ norm of the $x$-component of the
Momentum constraint for the H855,
H877, and H677 systems. Only the H677 system shows consistent
convergence to zero with resolution. Note that the constraints have
not been rescaled.}
  \label{fig:mC_nm2}
\end{figure}

In Figs.~\ref{fig:HC_hcmp}~and~\ref{fig:HC_avg_cmp}, we compare the
$L_2$ norm and horizon-averaged values of the Hamiltonian constraint
violations for the H855, H877, and H677
systems (all with CFL 0.25). Unlike the horizon mass, here H855 shows
better behavior than H877. That is H855 shows convergence of
the $L_2$ norm to zero, while H877 shows convergence to a finite
value.
 In Fig.~\ref{fig:mC_nm2}, we show the $L_2$
norm of the $x$ component of the momentum constraint (other components
are qualitatively similar). Here, only the H677 simulations show
consistent convergence with resolution.

To summarize, at medium resolutions, both H877 and H677 preserve the
horizon mass better than H855. At high resolution, H855 appears to be
a little better than H877 or H677 (slightly flatter profile). The
horizon-averaged constraint violations appear to be smallest for H677,
followed by H855. The $L_2$ norm of the constraint violation favors
H677 prior to merger (and after $t=150M$), and H855 (high resolution) 
post merger. Prior to $t=150M$, high resolution H855 is better.

\begin{figure}
   \includegraphics[width=\figwidth]{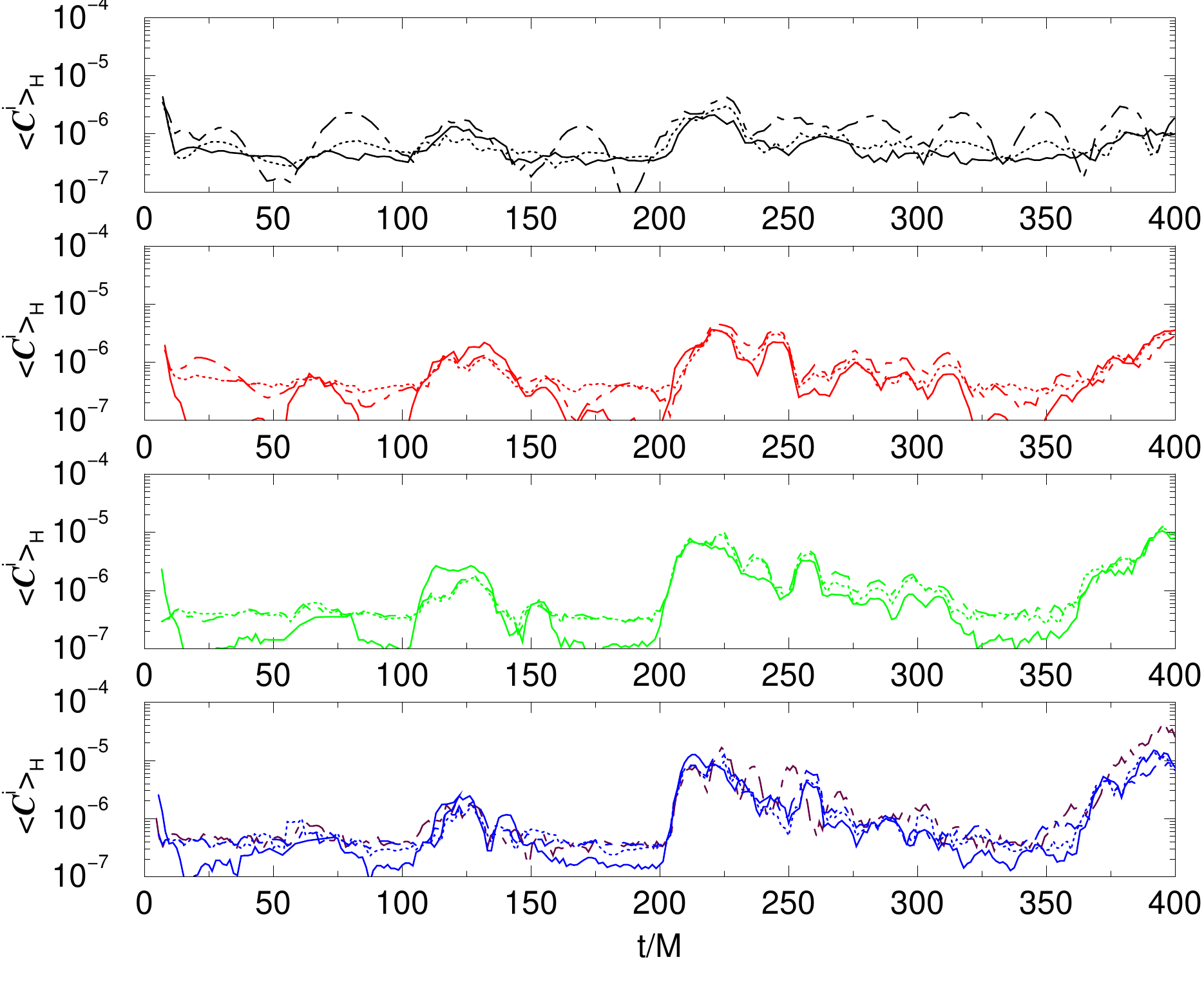}
   \includegraphics[width=\figwidth]{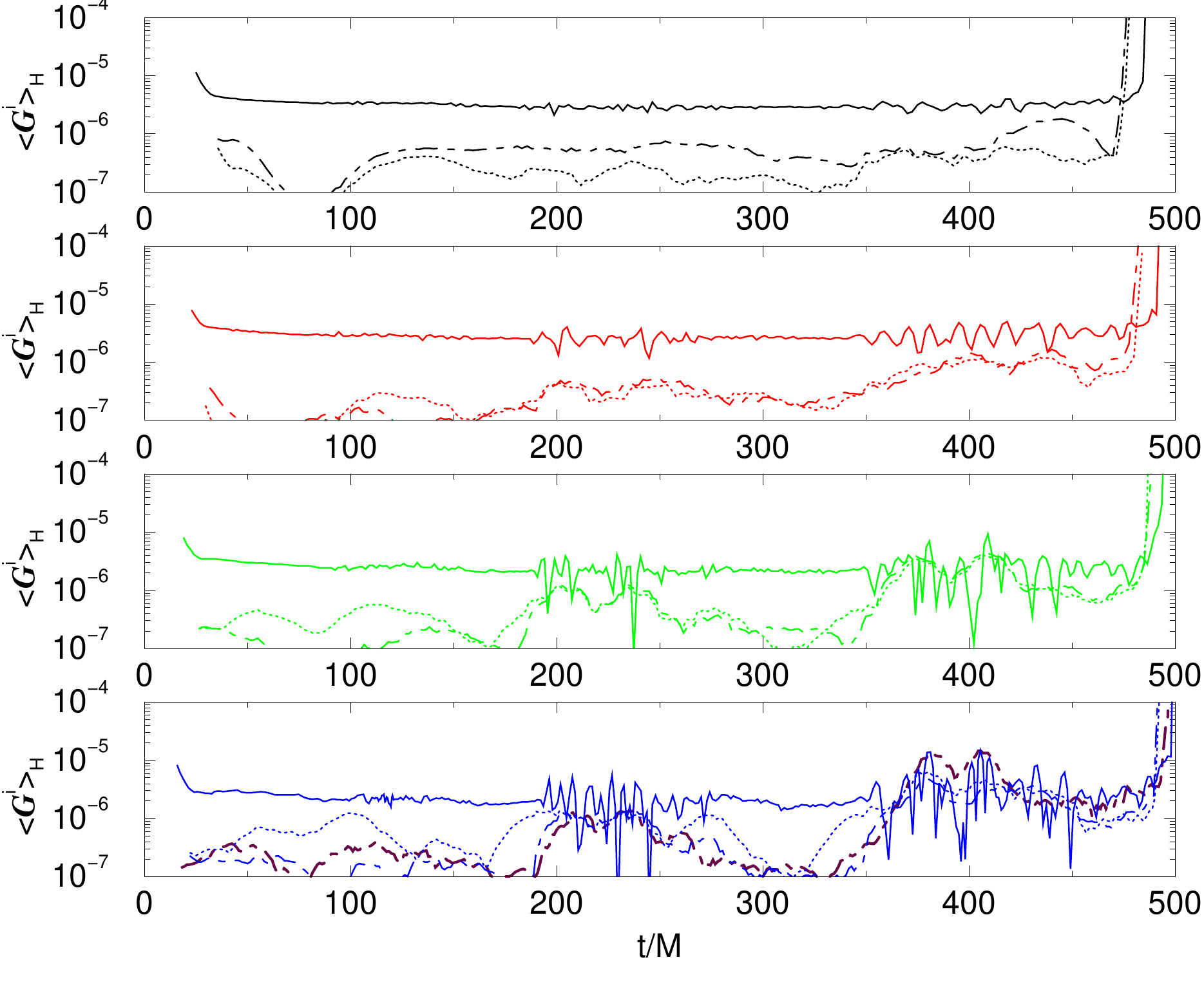}
  \caption{The horizon momentum constraint flux $\langle {\cal
C}^i\rangle_H$ and BSSN constraint flux $\langle {\cal
G}^i\rangle_H$. Here H677 results are displayed using dot-dashed
curves, H877 using dashed curves, and H855 using solid curves.
 The subplots are stacked by resolution, with $h_1$ on
top, followed by $h_2$, $h_3$. The bottom subplot shows results
for $h_4$, as well as  $h_5$ (thick solid line) for H677.
Note that both the momentum and BSSN the constraints 
have been rescaled by
a factor if $(h_i/h_5)^4$ (i.e.\ assuming fourth-order convergence).}
\label{fig:horizon_flux}
\end{figure}

In Fig.~\ref{fig:horizon_flux} we plot the horizon average of the
momentum constraint $\langle {\cal C}^i\rangle_H$ and BSSN constraint 
$\langle {\cal G}^i\rangle_H$, where ${\cal G}^i = \tilde \Gamma^i + \partial_j
\tilde \gamma ^{i,j}$, which is identically zero if the underlying ADM
equations are solved exactly. These two horizon quantities
 can be thought of as the
flux of constraint violation entering the horizons. For the momentum
constraints, the three systems seem to have the same level
of constraint violation.
On the other hand, for the BSSN constraint, the H855 system shows
larger constraint violations at all resolutions than the other two.

Based on these results, one may expect that the high-resolution H677
is 
the most accurate. The horizon-averaged Hamiltonian and momentum
constraint (but not BSSN constraint) seem to indicate that the next
most accurate simulation, at high resolution, is H855. 
While it is surprising that H877 does not perform as
well as H677, this may be due to effects of reduced dissipation combined with
the eighth-order algorithm.

\subsection{Phase errors}

\begin{figure}
  \includegraphics[width=\figwidth]{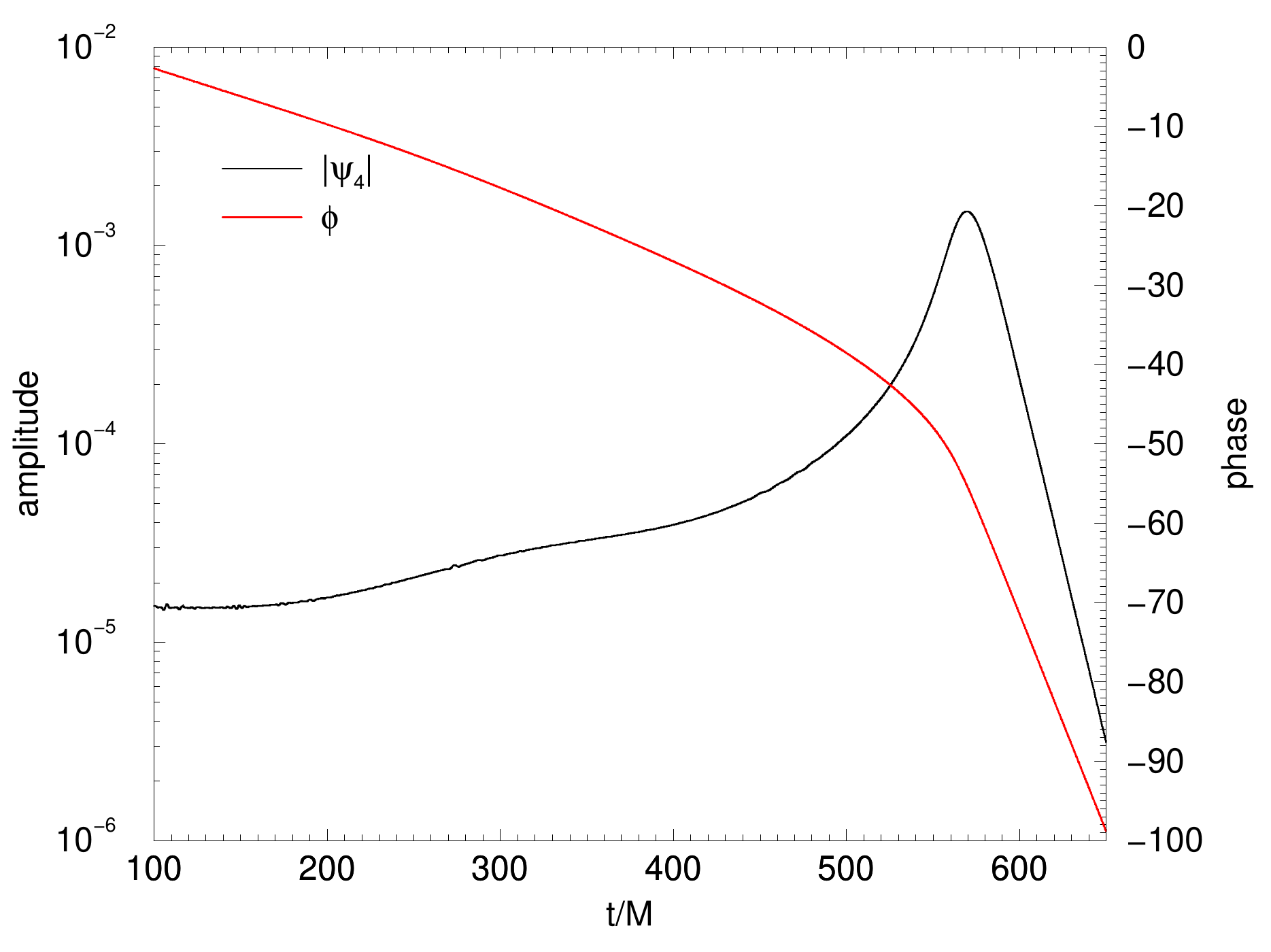}
  \caption{The magnitude and phase of the $(\ell=2,m=2)$ mode of
$\psi_4$ extracted at $r=50M$. The magnitude is shown on a log scale,
while the phase is shown on a linear scale. The exponential decay of
the quasinormal ringdown is evident after $t\gtrsim 585M$.}
  \label{fig:wave}
\end{figure}

\begin{widetext}

\begin{figure}
  \includegraphics[width=0.4\columnwidth]{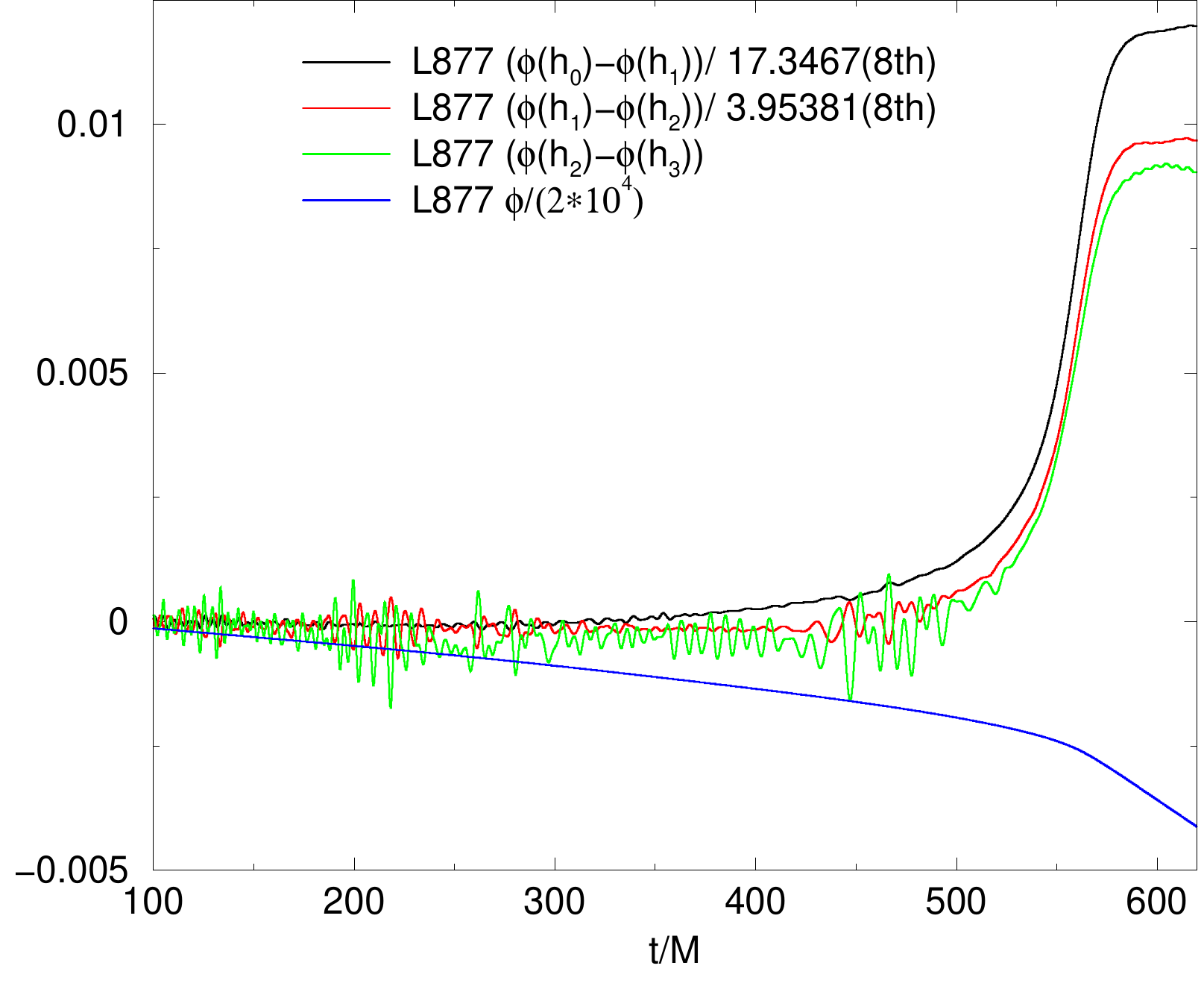}
  \includegraphics[width=0.4\columnwidth]{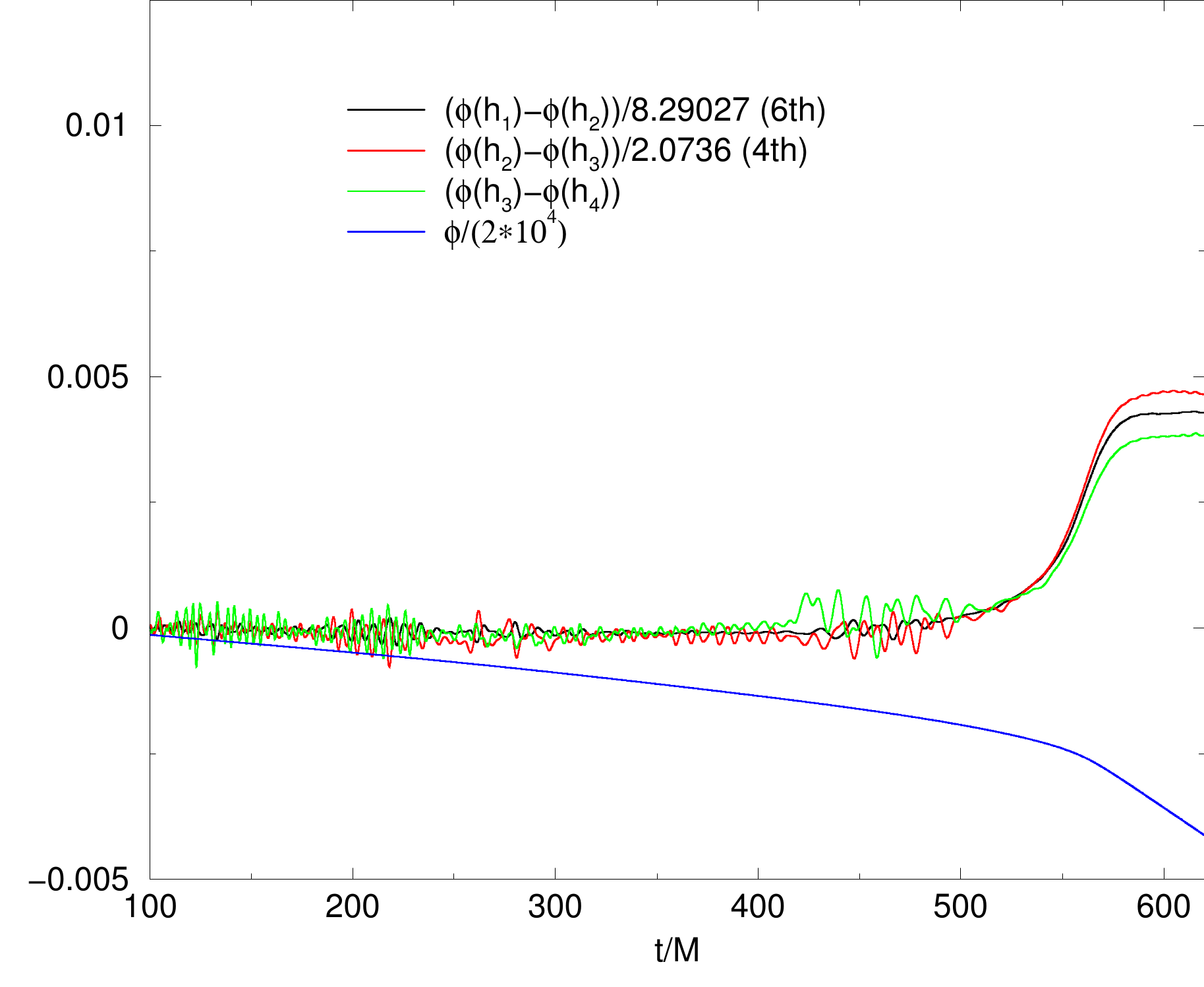}
  \caption{Convergence of the phase of the $(\ell=2,m=2)$ mode of
$\psi_4$ for the L855 (left) and H855 (right)
simulations. In all cases the CFL factor was 0.125 on the finest grid.
Higher-order convergence is apparent for the L855 system.
However, the solution itself has a nonconvergent error. }
  \label{fig:phase}
  \includegraphics[width=.40\columnwidth]{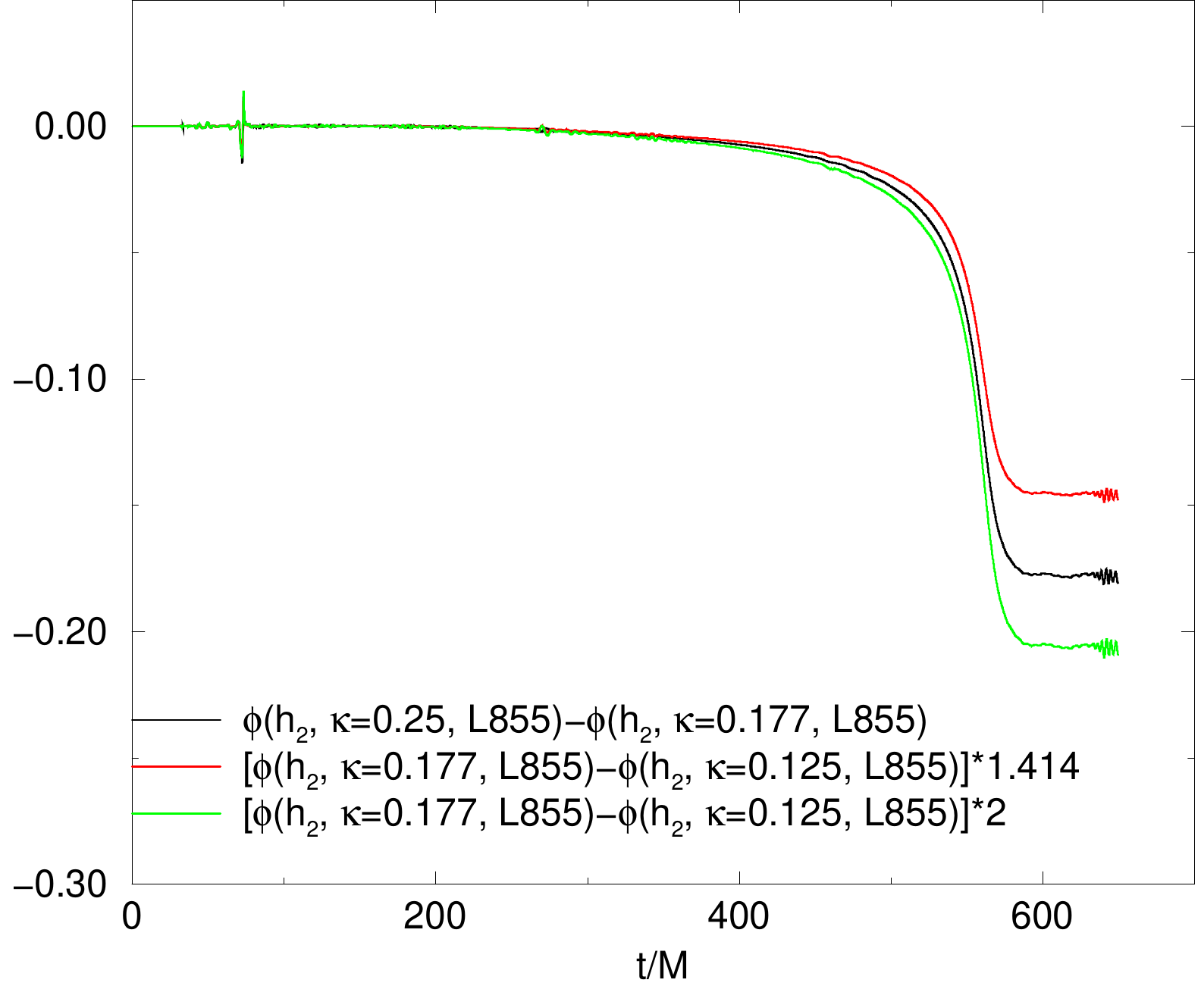}
  \includegraphics[width=.40\columnwidth]{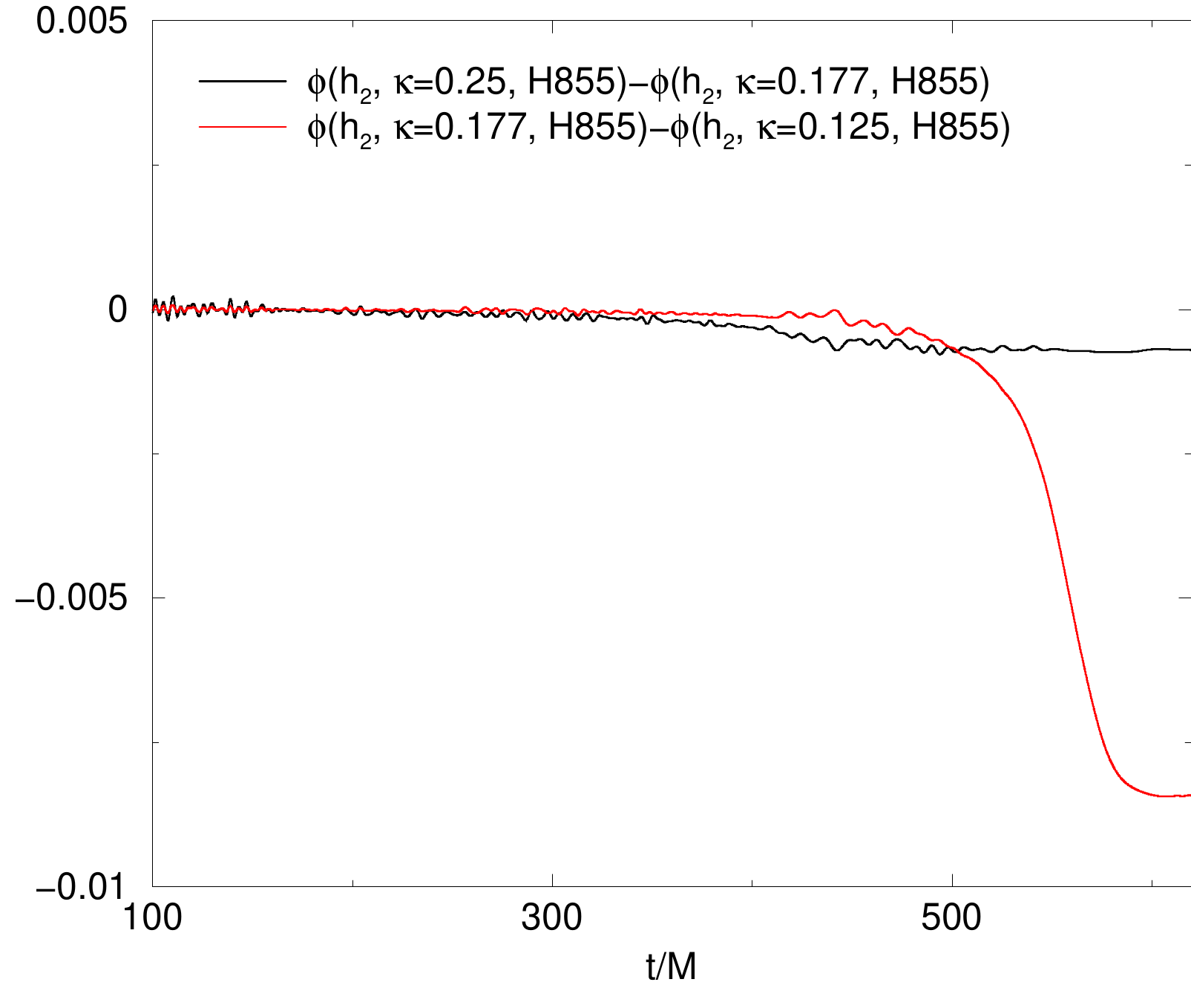}
  \includegraphics[width=.40\columnwidth]{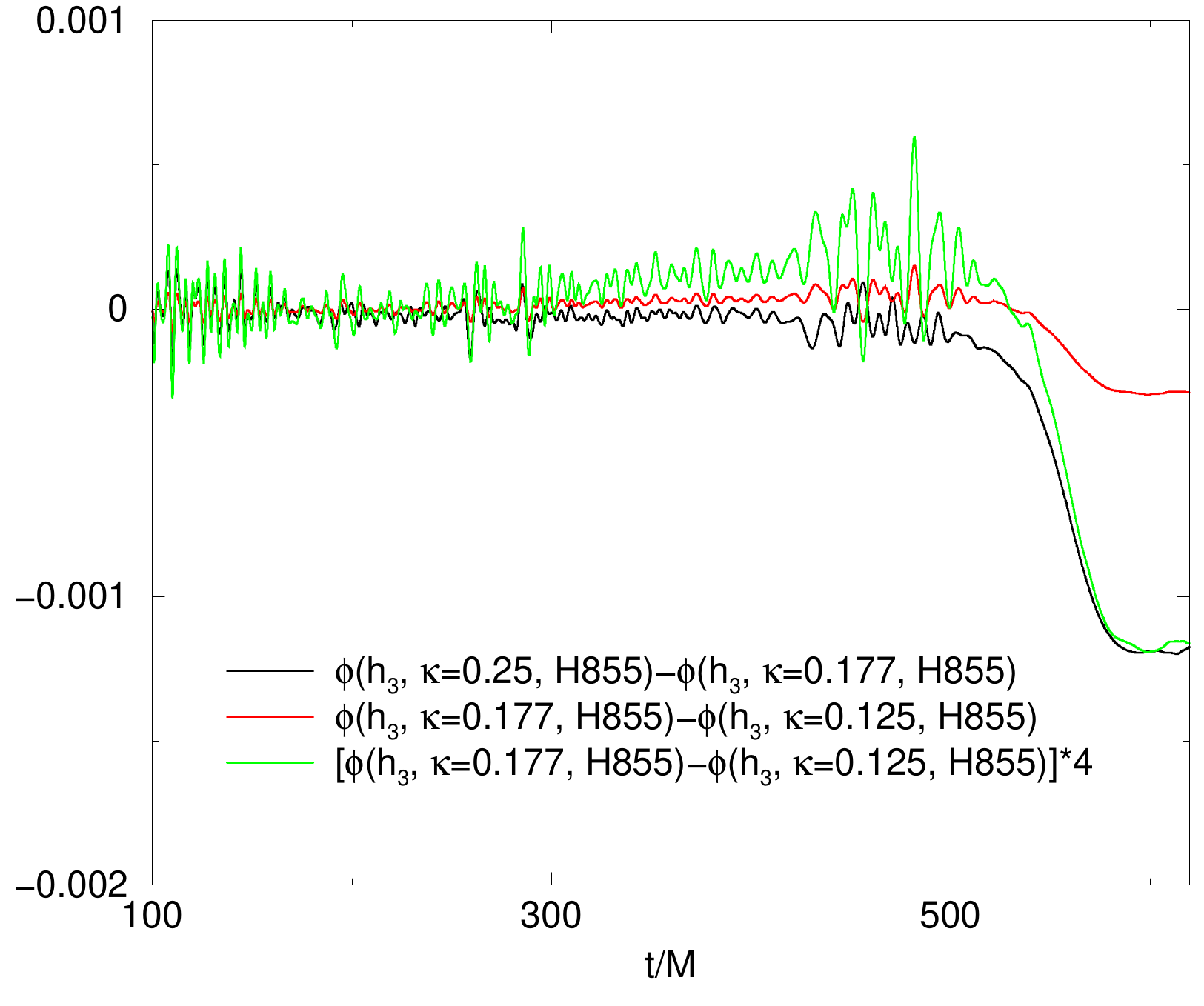}
  \includegraphics[width=.40\columnwidth]{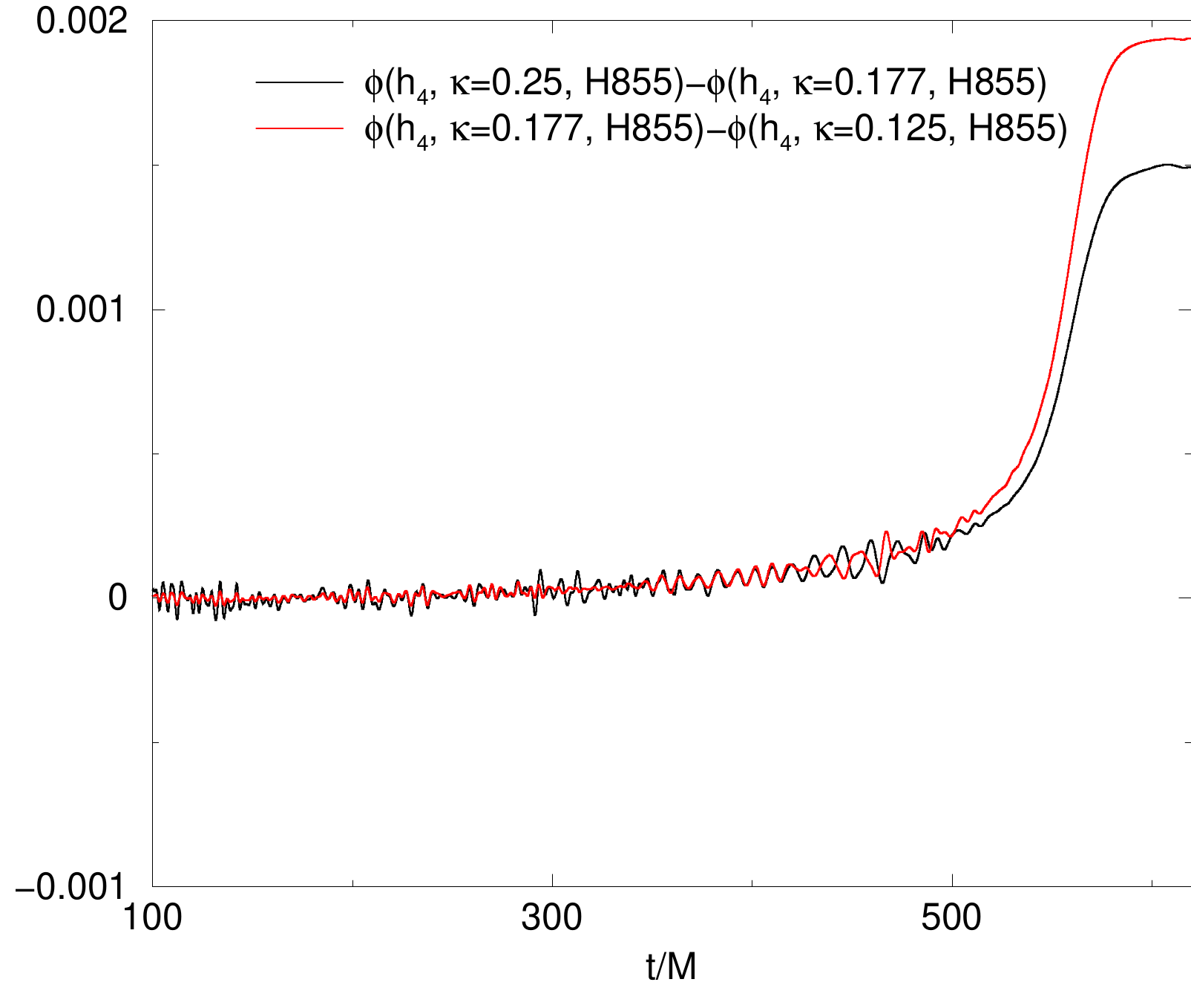}
  \caption{Convergence of the phase with changes to the CFL factor.
   The top left panel shows the convergence of the phase for the
L855 system. Slightly better than first-order convergence is
seen. The top right panel shows the convergence rate, at the same
spatial resolution, for the H855 simulations. Note the phase
differences are a factor of 10-100 smaller, but not convergent.
The bottom left shows the same convergence plot, but for the next
higher spatial resolutions. Clean 4th-order convergence is seen at
late times, at early times the differences are consistent with zero.
Finally, the bottom right panel shows the same plot, but for the
highest resolutions runs. Here again, convergence is not observed with
CFL. Also the phase differences with CFL are larger than for the
medium resolution.}
  \label{fig:phase_v_cfl}
\end{figure}

\end{widetext}

For the L855 configurations we saw clean convergence of the
phase of the $(\ell=2,m=2)$ mode of $\psi_4$ (see Fig.~\ref{fig:wave}) at the smallest CFL
factor (0.125). While eight-order convergence in the phase is apparent, the
actual phase error is relatively large, when effects of the CFL factor
are included. In Fig.~\ref{fig:phase} we show the
convergence of the phase for both the L855 and H855
simulations. Fourth-order convergence is apparent in the H855
simulations.

\begin{figure}
  \includegraphics[width=\figwidth]{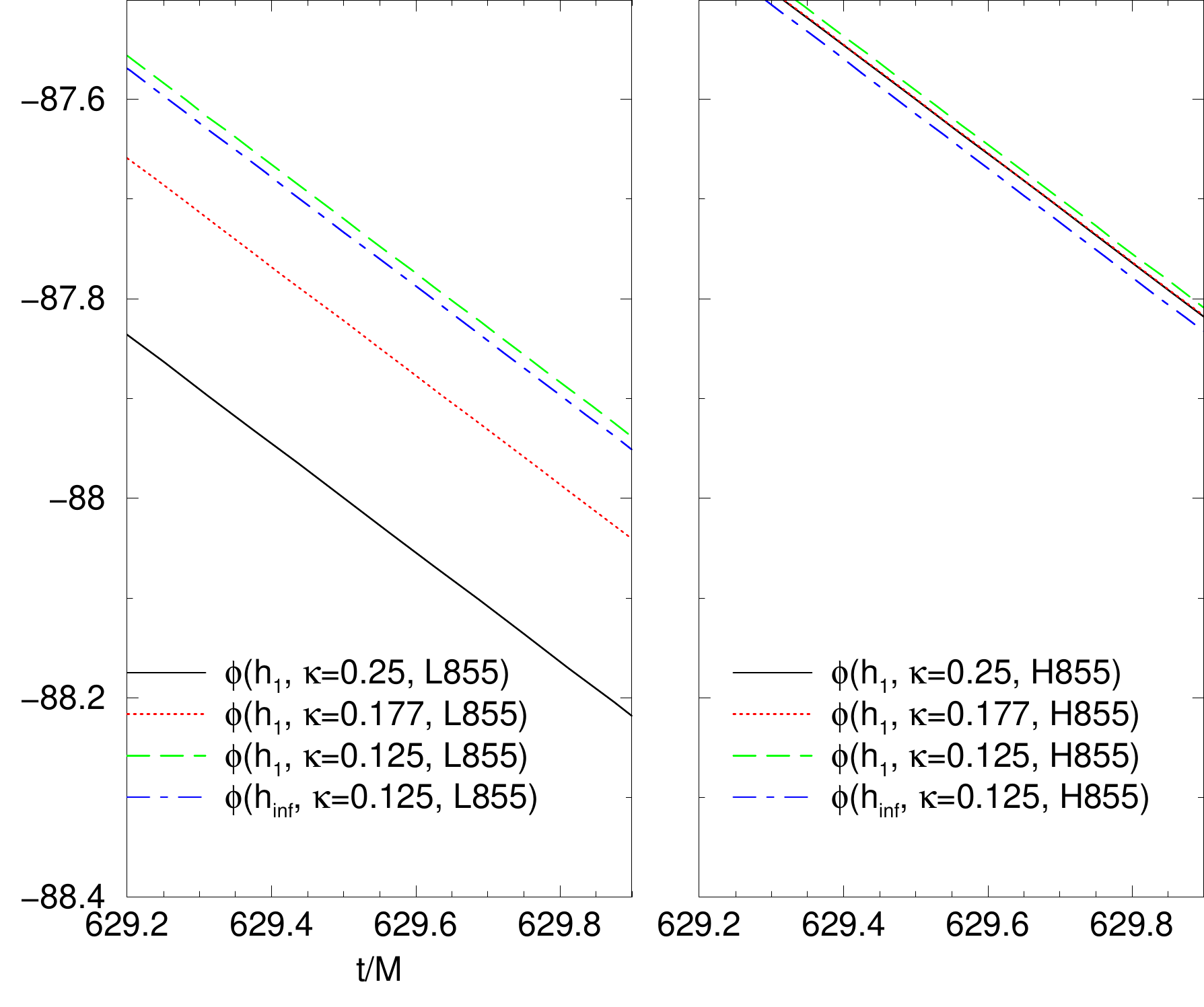}
  \caption{The phase of the $(\ell=2,m=2)$ mode of $\psi_4$ during the
ringdown phase. Shown are phases for the H855 and 
L855 at three different CFL factors, as well as the 
Richardson extrapolation phase based on the highest resolution runs
with the smallest CFL. Note how the phase converges to different
values depending on CFL for the L855 system, but
shows consistency between CFL values for the H855 algorithm.
}
  \label{fig:phase_nonconv}
\end{figure}

\begin{figure}
  \includegraphics[width=\figwidth]{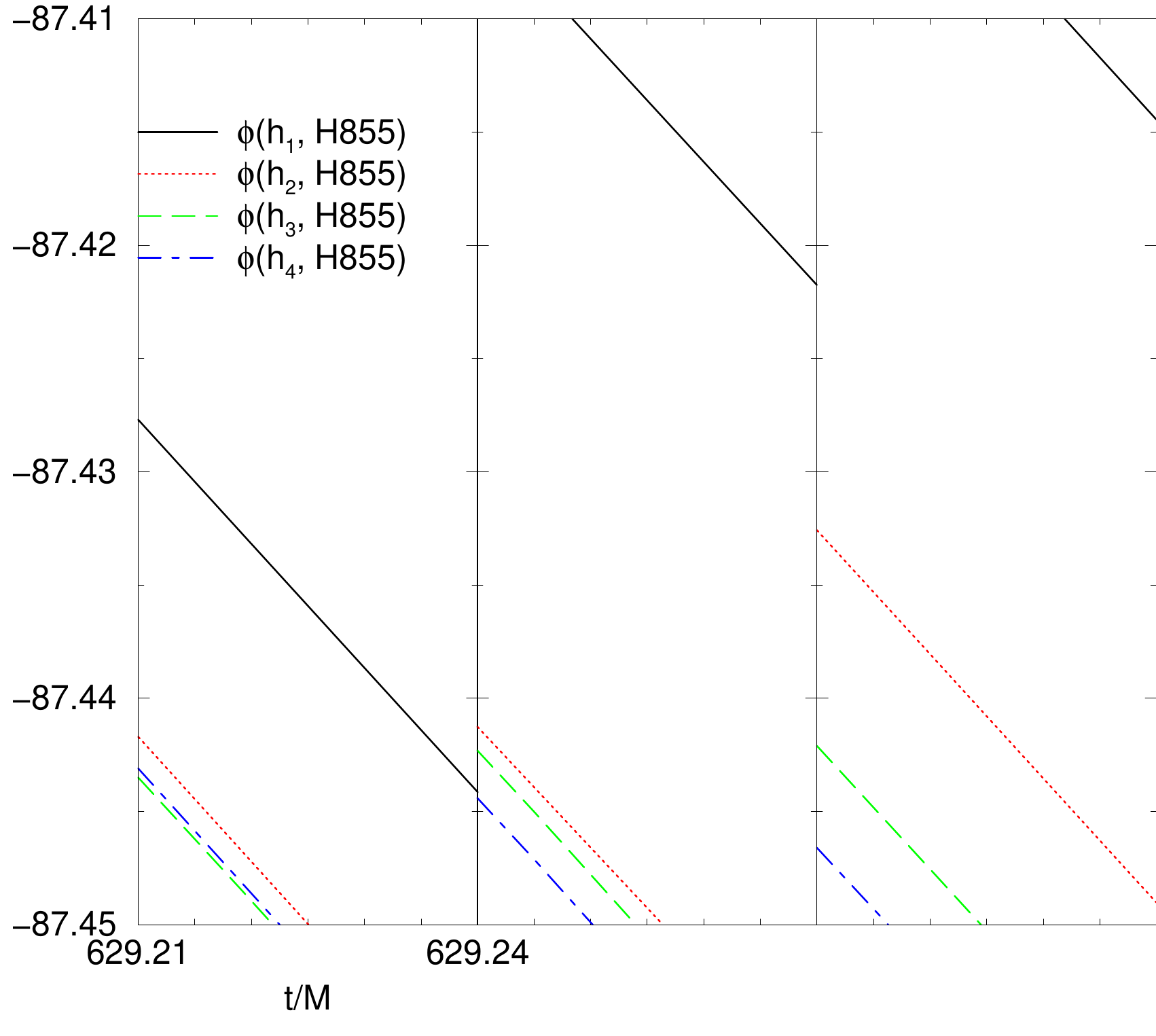}
  \includegraphics[width=\figwidth]{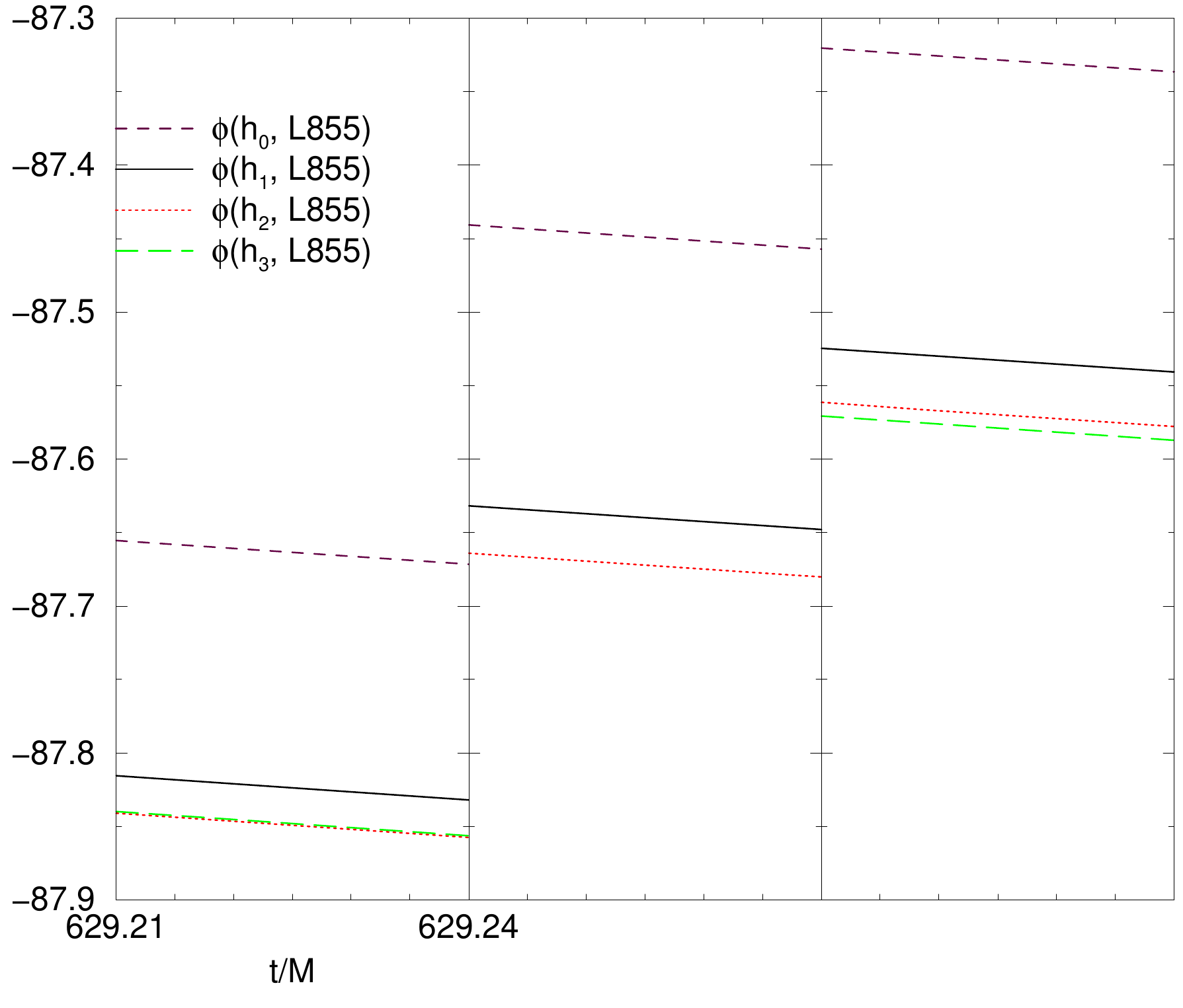}
  \caption{ Plot showing the phase for a short duration during
ringdown for simulations using the H855 (top) and
L855 (bottom) algorithms. The plots are arranged according to
CFL factor $\kappa$ with $\kappa=1/4$ on the left,
$\kappa=1/(4\sqrt{2})$ in the middle, and $\kappa=1/8$ on the right.
In each case, the $x$ and $y$ axes are the same for each panel for a
given algorithm.
For the H855 runs,  note that 
$\kappa=1/4$ curves are not in convergence order, while
$\kappa=1/(4\sqrt{2})$ are in convergence order but the curve
separations are not consistent with convergence. $\kappa=1/8$ shows
convergence, but the extrapolation is about 0.13 rad from the high
resolution run which is much larger than expected.
The L855 show apparently better convergence
properties but the phases are not convergent to each other.}
  \label{fig:phase_v_res_v_cfl}
\end{figure}

In Figs.~\ref{fig:phase_v_cfl} and \ref{fig:phase_nonconv}, we show the phase of the $(\ell=2,m=2)$ mode of
$\psi_4$ as a function of CFL factor $\kappa$. Note that this means
that only the timestep is refined, not the spatial resolution.
Once might expect that the error would decrease, or at least not
increase, as $dt$ is made smaller. Generally, this is the case,
however, we note that the error can increase as $dt$ is made smaller due to
the effects of dissipation. For example, for a simple first-order
discretization for the advection equation, at fixed spatial resolution,
the error in the solution for a marginally resolved
waveform increases as $\kappa\to 0$. In
Fig.~\ref{fig:phase_v_res_v_cfl}, we show the
convergence of the phase as a function of the CFL factor (at fixed
spatial resolution). The L855 simulations show clear
first-order convergence at all spatial resolutions, while the
H855 simulations actually show an increase in error with CFL
at higher resolutions. Combining the information from
Figs.~\ref{fig:phase_v_res_v_cfl}
and \ref{fig:phase_v_cfl}, we conclude that the L855 system shows a clear
trend where the waveform at infinite resolution is a function of CFL
factor. This means that at fixed CFL, which is how numerical
convergence studies are performed, there is a nonconvergent,
error in the phase. Fortunately, the phase for the H855 simulations
appears to be independent of CFL.

\begin{figure}
  \includegraphics[width=.95\columnwidth]{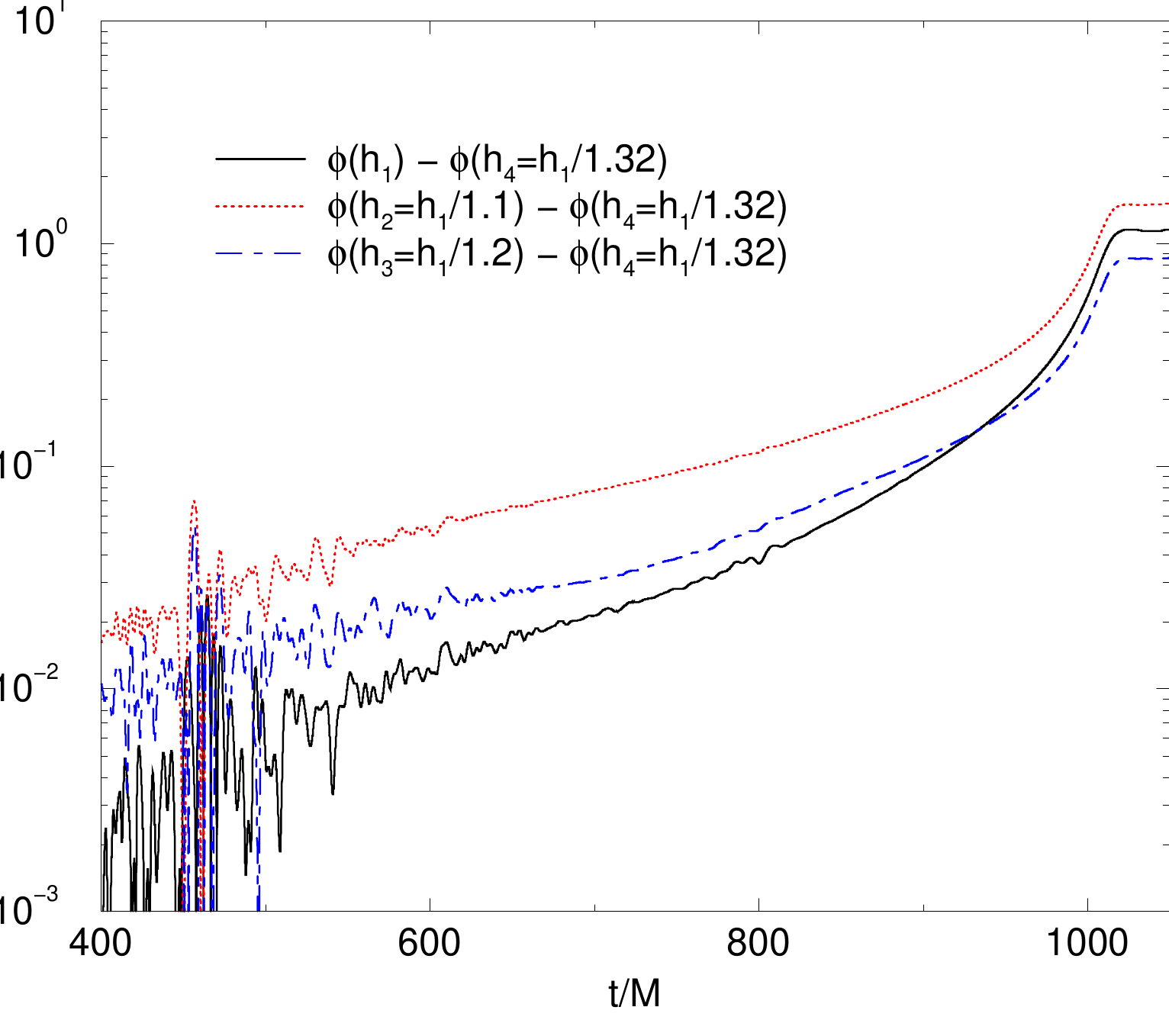}
  \caption{A plot showing the phase differences between three
resolutions of a $q=1/10$ nonspinning BHB and the corresponding
phase for a still higher resolution using the L855 system (with
$\kappa=1/4$). Note how the lowest resolution run has phase closest to
the highest resolution for the inspiral phase.
}
\label{fig:10to1}
\end{figure}

The above observed phase oscillations occur, but at a magnified level
for non-equal mass systems. In Fig.~\ref{fig:10to1}, we show the
waveform phase for a $q=1/10$ BHB evolved using the L855 system with
$\kappa=1/4$~\cite{Nakano:2011pb}.

\subsection{Comparing waveform phases using different dissipation and prolongation
operators}
\begin{figure}
  \includegraphics[width=\figwidth]{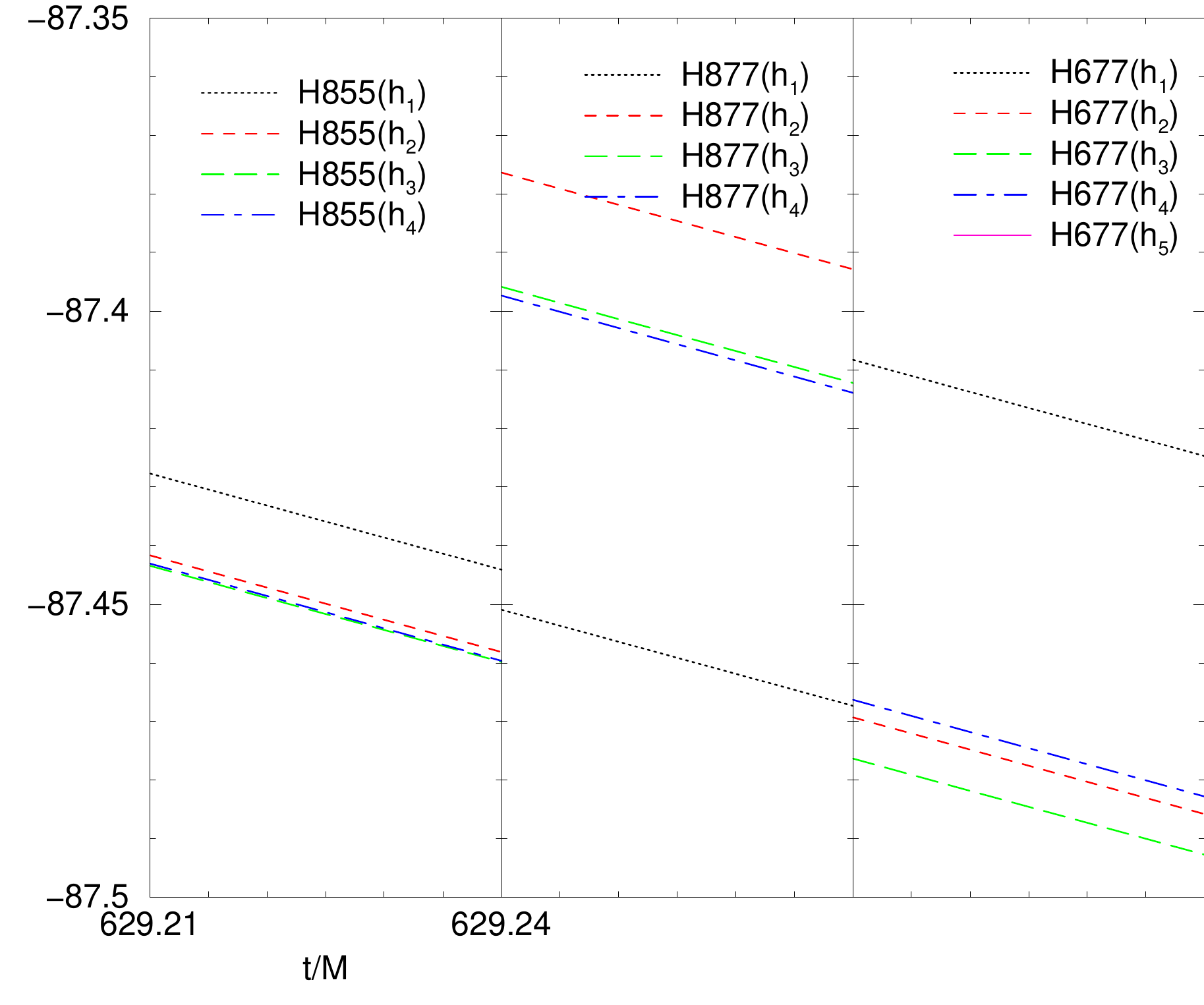}
  \caption{The phase of the $(\ell=2,m=2)$ mode of $\psi_4$ near
merger for the H855, H877, and H677 systems. Note how the more {\it
accurate} H877 and H677 runs actually show significantly more scatter at
lower resolutions than H855.
For both H877 and H855, the phases for the
$h_3$ and $h_4$ resolution lie nearly on top of each other. Given the
phase differences with the other resolutions, this is likely due to
an oscillation (i.e., nonmonotonic dependence) in the phase with resolution. 
	}
  \label{fig:cmp_diss_and_pro}
\end{figure}

In Fig~\ref{fig:cmp_diss_and_pro}, we show the phase of the
$(\ell=2,m=2)$ mode of $\psi_4$ near
merger for the H855, H877, and H677 systems. {\it A priori,} we might expect
that the H877 systems is the most accurate, and would therefore show
the smallest amount of scatter and spurious oscillations of the phase
with resolution. In fact,
this system, and H677, shows significantly larger scatter than
H855. A possible explanation is that the H877 and H677 systems are the
least dissipative, and therefore most sensitive to high-frequency
grid noise errors (these high frequency signals originate as
interference signals from the reflection of the initial data pulse at
the AMR boundaries).

\begin{figure}
  \includegraphics[width=\figwidth]{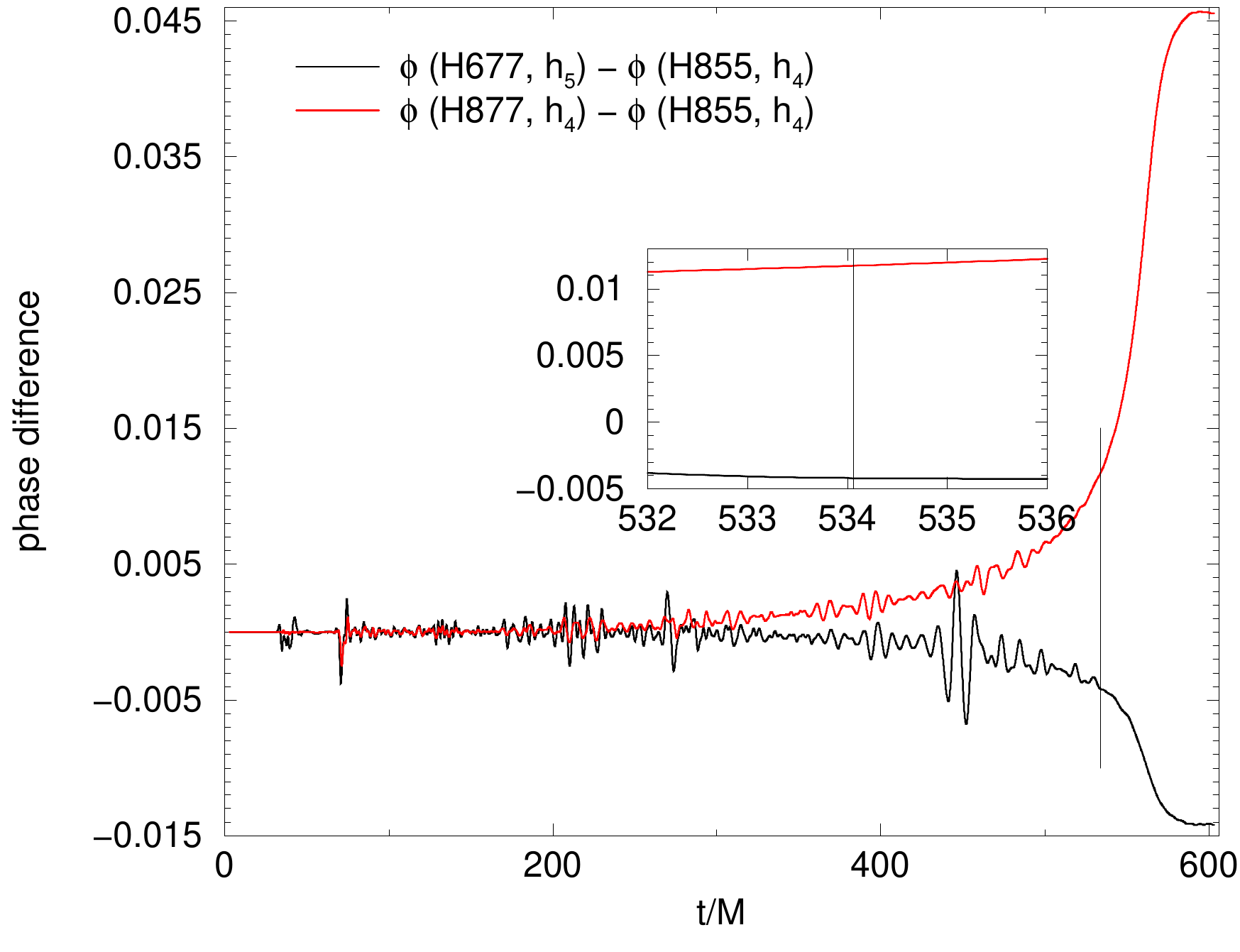}
  \caption{The difference between the phase of the $(\ell=2,m=2)$ mode 
  of $\psi_4$ as calculated using the H855 and H877, as well as the
H855 and H677 systems. The vertical line corresponds to the time when
the waveform frequency is $M \omega=0.2$. At this time, the phase
error appears to be as small as  $0.005$ ($0.012$, if we use the
differences between H855 and H877) rad. Given the results apparent in
Fig.~\ref{fig:cmp_diss_and_pro}, we
estimate this to be a reasonable error of the phase error.
}
  \label{fig:phase_error}
\end{figure}

The H877 system does not appear to be consistent with either H855 
or H677 if we assume that the error in the phases is given
by the differences between the two highest resolutions. 
However, we note that in all the cases, the phase is not a monotonic
function of the resolution, i.e.\ there are oscillations of the phase.
The differences between the phases for H855 and H877 and H855 and H677 
are better
measures of the error. In this case, we find that the phase error is
closer to 0.05. We plot the difference of the highest resolution H855
simulation with the highest resolution H877 and H677 simulations
in Fig.~\ref{fig:phase_error}. Note that, based on the differences
between the H855 and H677 simulations, we predict a phase error,
during ringdown, of  0.015 rad. (0.045 if we use H877). Interestingly,
this lower number is very close to the difference between the H855 Z4
simulations (described below) and the H855 simulations (see
Fig.~\ref{fig:phase_v_res_z4_bssn}).
We note that in~\cite{Garcia:2012dc},
where the SPEC code was used to compare waveforms using the same
evolution system, but different
initial data techniques, the waveform phases agreed to within
$\sim 0.002$ .rad, which was lower than the estimated overall waveform
error of $\sim 0.01$ .rad over a 4000M inspiral.

\begin{figure}
  \includegraphics[width=\figwidth]{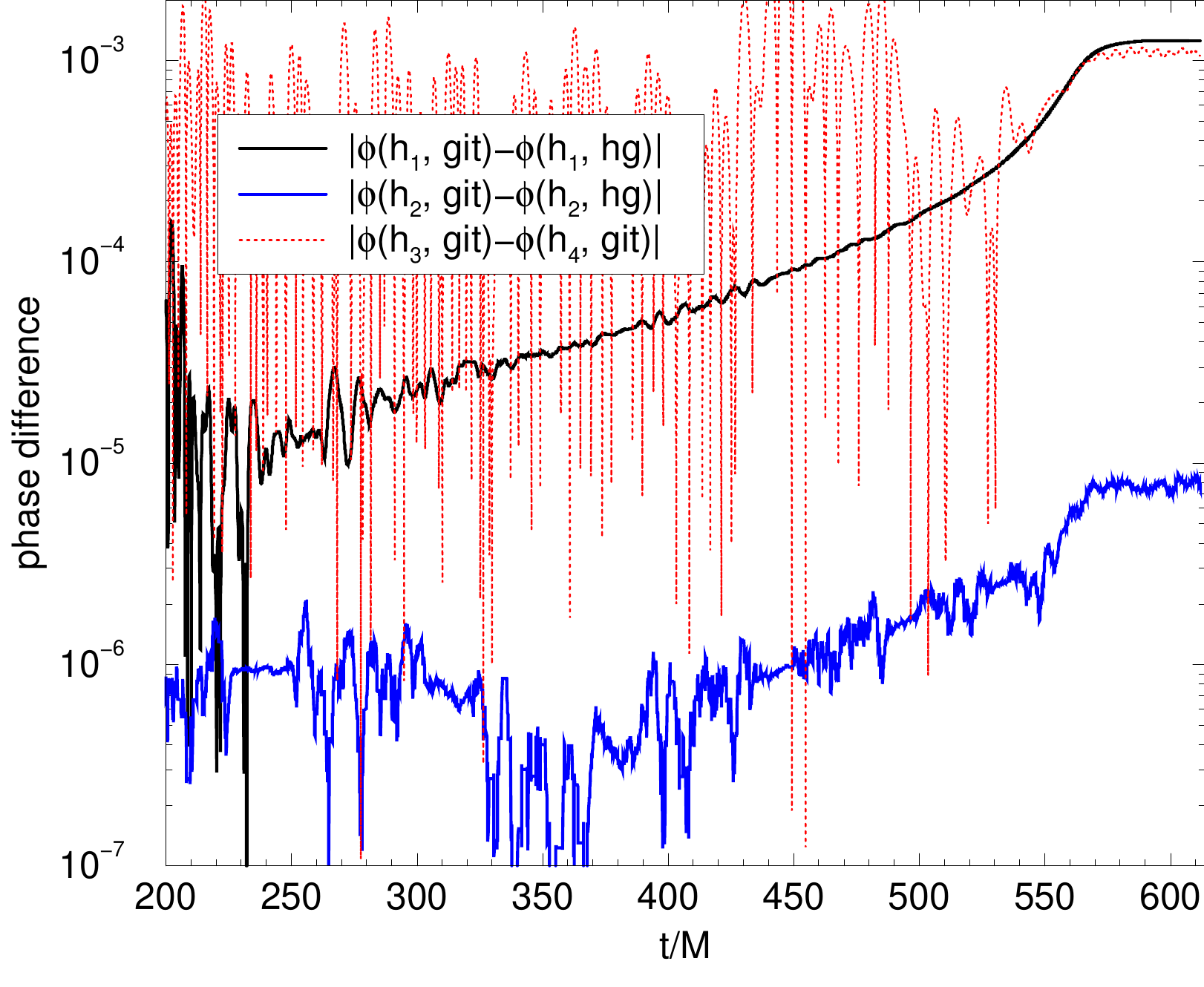}
  \caption{The phase of the $(\ell=2,m=2)$ mode of $\psi_4$ near
merger for the H855 systems with a CFL of $\kappa=0.25$ using both
the git and hg versions of the \carpet AMR driver.
The phase deviation between the git and hg versions increases
exponentially with time, but decreases rapidly with increasing
resolution. For comparison, the phase difference between the medium
and high resolutions runs is also shown. 
 }
  \label{fig:cmp_carpets}
\end{figure}

We also noticed differences in the phase between the ET\_2011\_10 (and
later) version of \carpet (see~\cite{einsteintoolkit}) and the
ET\_2011\_05 (and previous) versions. In Fig.~\ref{fig:cmp_carpets},
we use ``hg'' and ``git'' to denote the newer and older versions,
respectively.
 As seen in
Fig.~\ref{fig:cmp_carpets}, the differences converge away rapidly with
resolution. The source of the deviation appears to be a difference in
the prolongation operators that affects non smooth data. The
exponentially growing error in the phase apparent during the late 
inspiral is one of the main reasons why obtaining highly-accurate
phases is so difficult.


\begin{figure}
  \includegraphics[width=\figwidth]{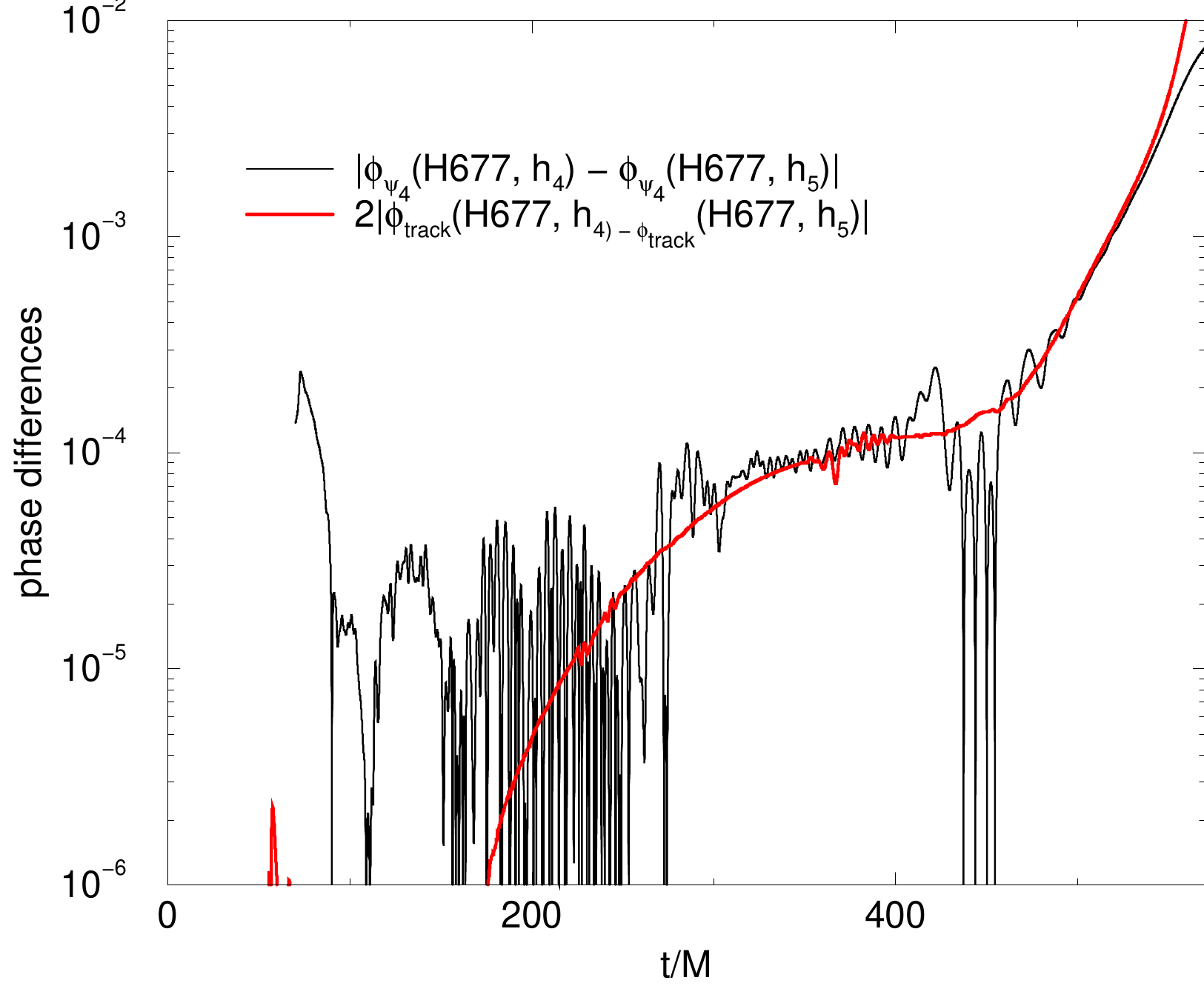}
  \includegraphics[width=\figwidth]{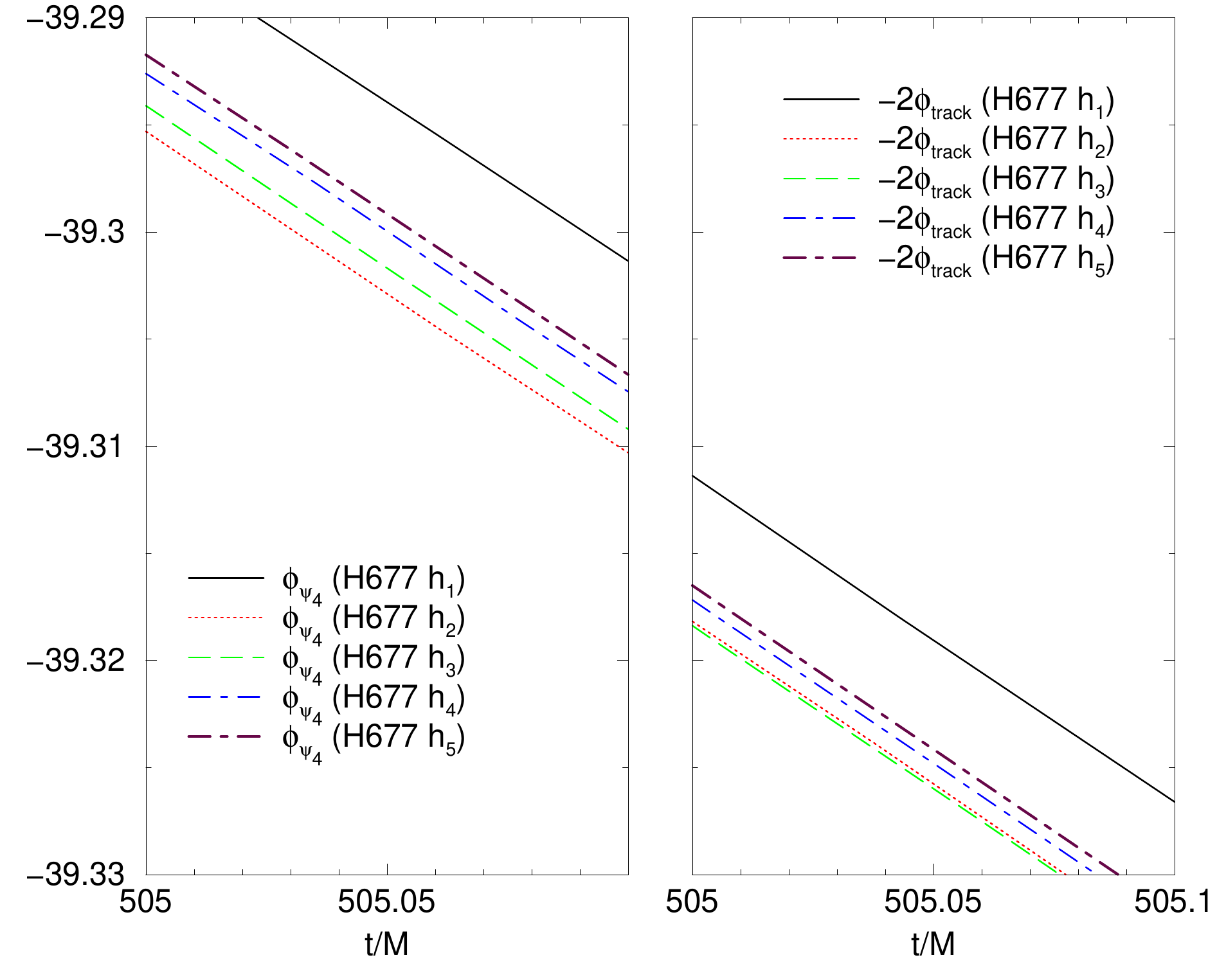}
  \caption{A comparison of the phase differences in both the
trajectory and waveform between the two
highest resolutions for the H677 system. The plot ends at the point
where the orbital separation is $0.05M$. The agreement between the two
curves indicates that the error in the phase of the waveform is
associated with the error in the phase of the trajectory. The bottom
plot shows the orbital and waveform phases in a small time interval for
the different resolutions. Note how the oscillation (with resolution)
in the phase of the waveform are matched by similar oscillations in
the orbital phase.}
\label{fig:track_wave_phase_cmp}
\end{figure}

To conclude this section, we mention that the phase error seen in the
waveform is not a result of the extraction technique, but is rather
associated with phase error in the orbital motion itself. This is
demonstrated in Fig.~\ref{fig:track_wave_phase_cmp}, where
oscillations in phase (with resolution) are apparent in
both the orbital phase and waveform phase at the same time.

\subsection{Analysis using the CCZ4 Formalism}
\begin{widetext}

\begin{figure}
  \includegraphics[width=0.4\columnwidth]{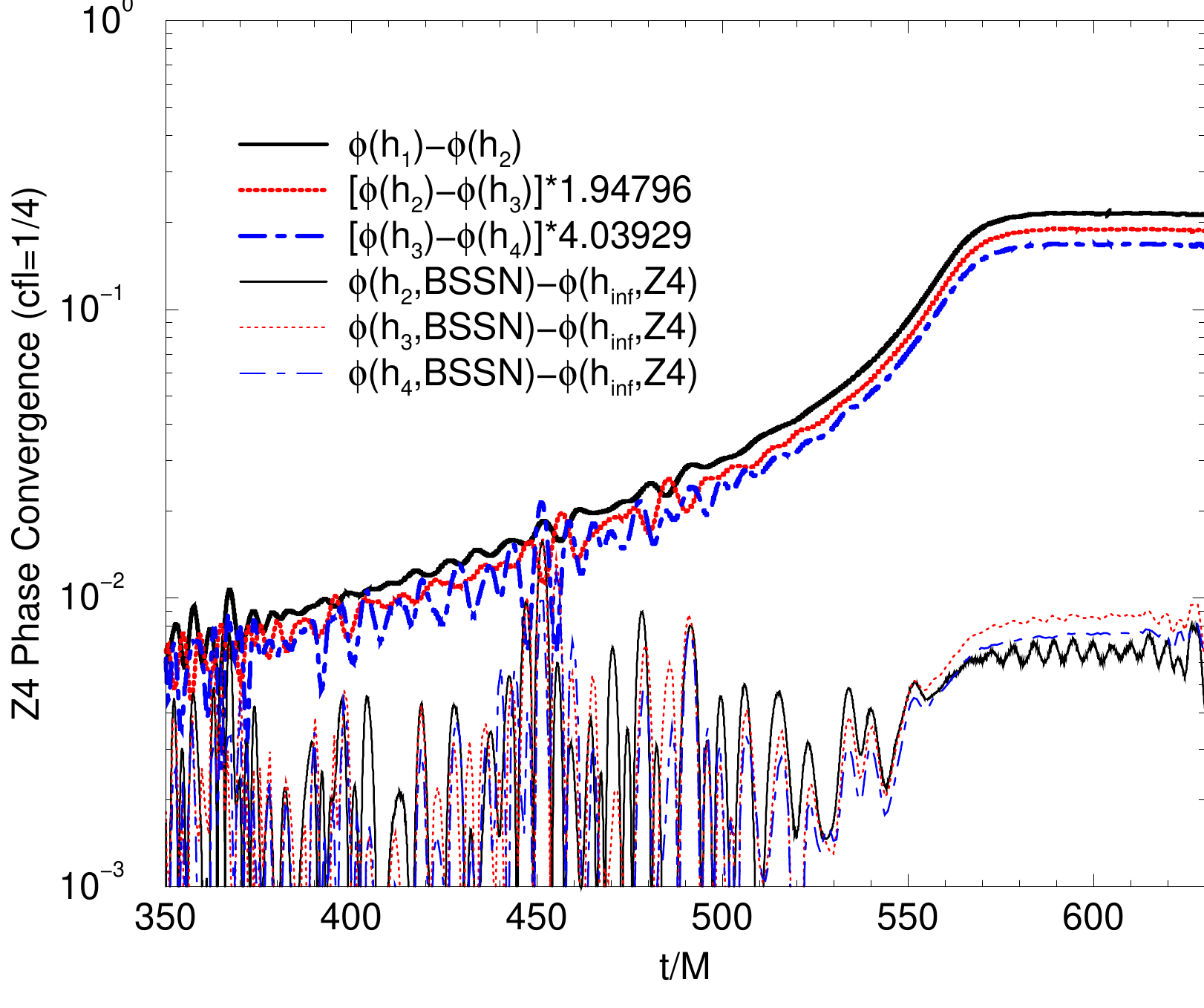}
  \includegraphics[width=0.4\columnwidth]{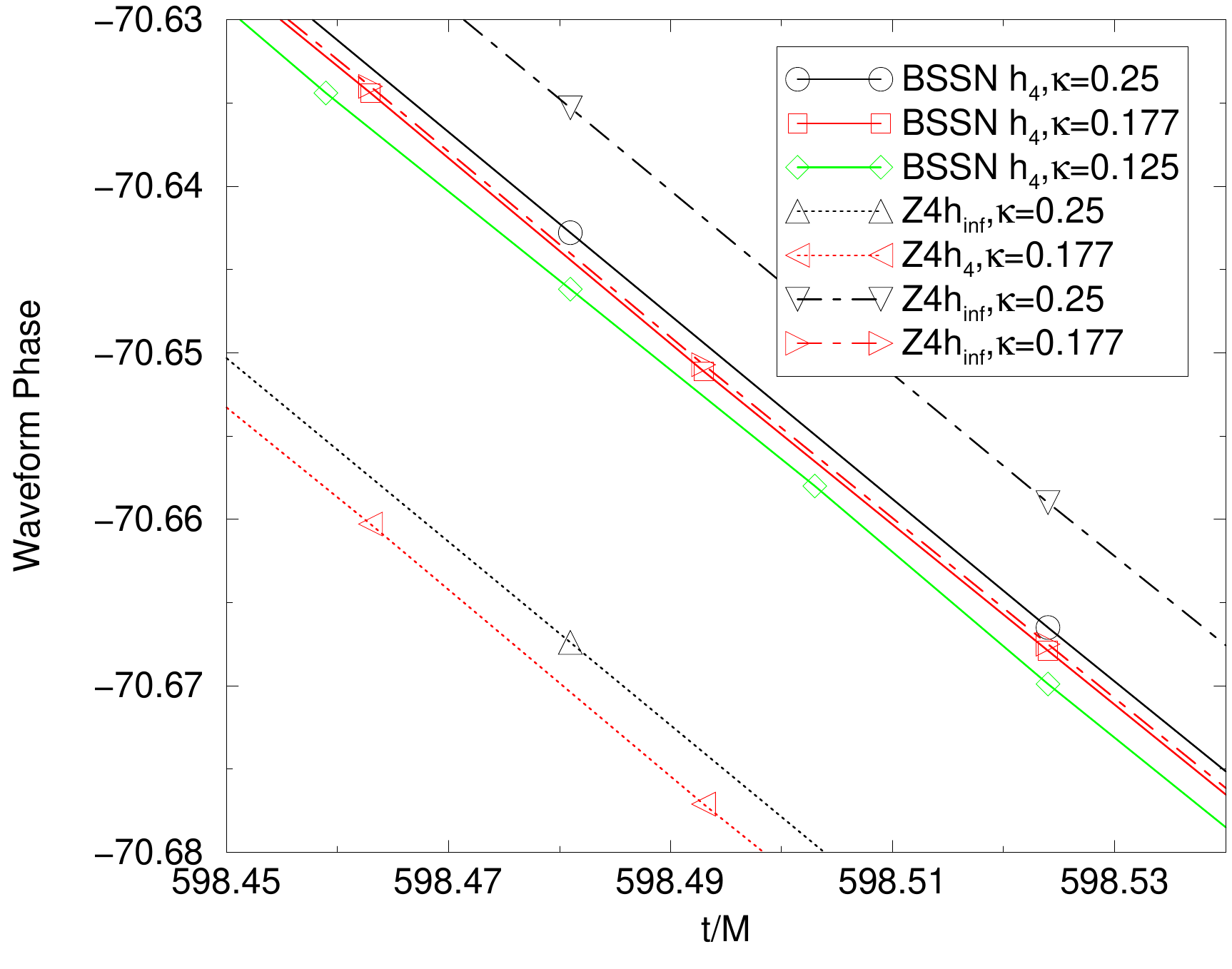}
  \includegraphics[width=0.4\columnwidth]{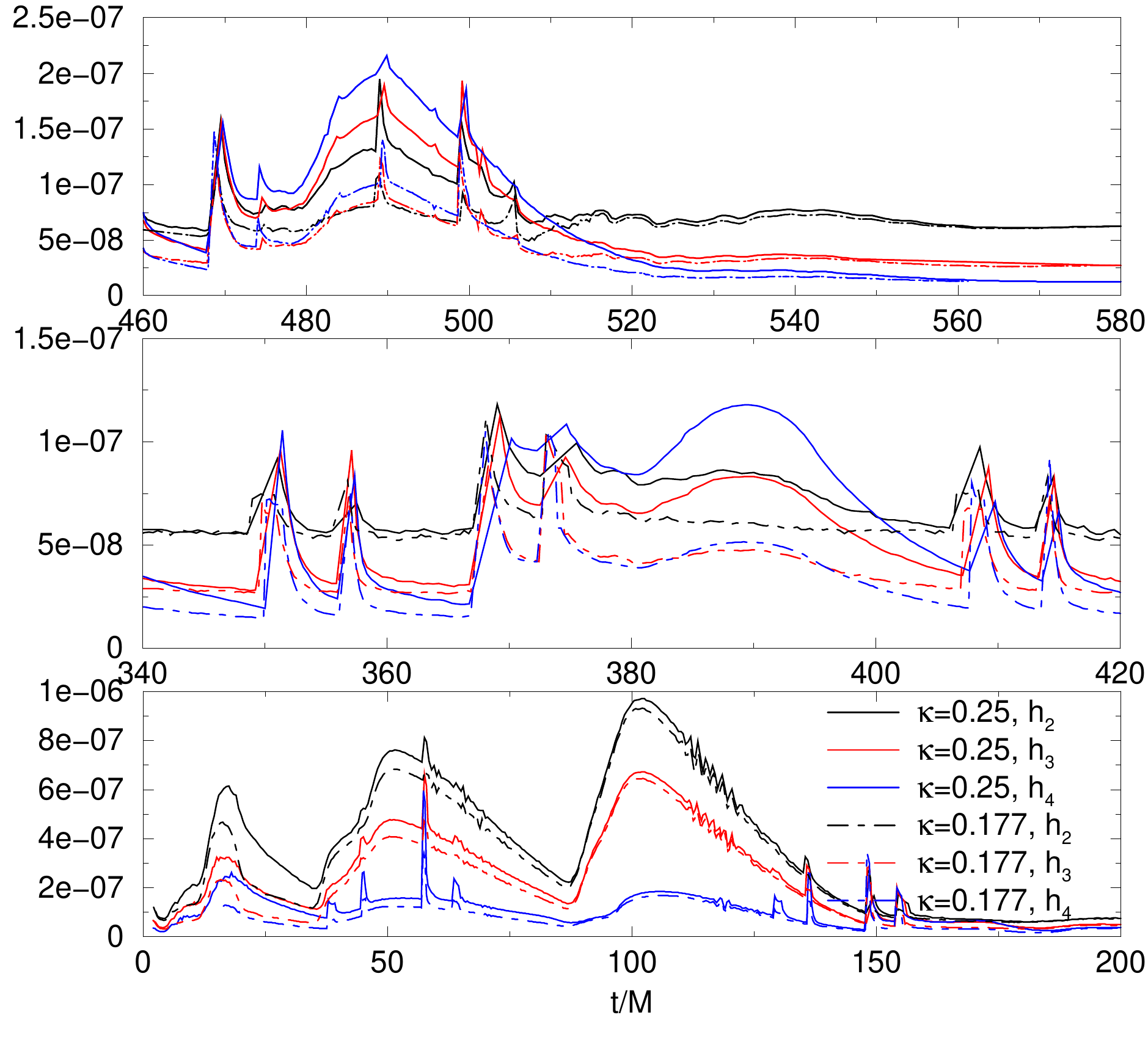}
  \includegraphics[width=0.4\columnwidth]{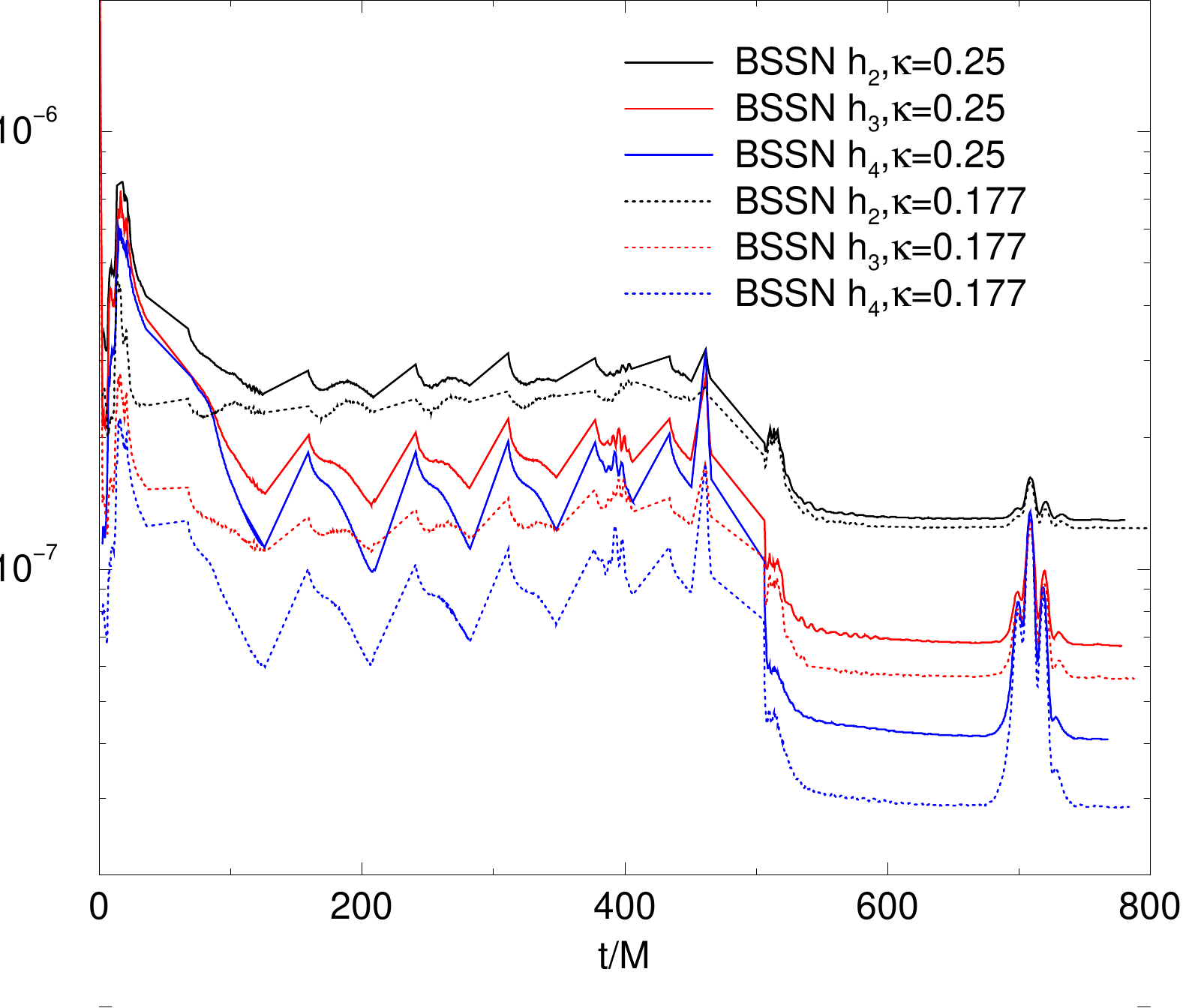}
  \caption{ (top left) The phase convergence for the Z4 evolutions
and a comparison with the Richardson extrapolated phase from the Z4
evolutions with the corresponding BSSN phases. (top right)
The waveform phase during the late ringdown. The BSSN phases all lie between
the highest resolution Z4  phases and the extrapolated phases.
However, there also seems to be a trend towards moving to more
negative phases as the CFL is reduced. This is observable in both BSSN
and Z4.
(bottom left) The $L_2$ norm of the Hamiltonian constraint
for the Z4 runs with 2 different CFLs. Convergence with resolution and
consistency between CFLs is apparent at early and late times. However,
inconsistencies between CFLs and nonconvergent behavior is apparent
in the later inspiral and merger phases. The oscillations in $||{\cal
H}||_2$ during the inspiral are consistent with the findings
of~\cite{Alic:2011gg}.
(bottom right) The $L_2$ norm of the Hamiltonian constraint for
the comparable BSSN runs. Here improvements in the CFL are apparent
when increasing resolution and decreasing the CFL factor. (Note that
the spikes apparent in Fig.~\ref{fig:H_pi_nopi} have been removed for clarity.)
}
  \label{fig:phase_v_res_z4_bssn}
  \label{fig:Z4_hc_norm2}

 \includegraphics[width=.4\columnwidth]{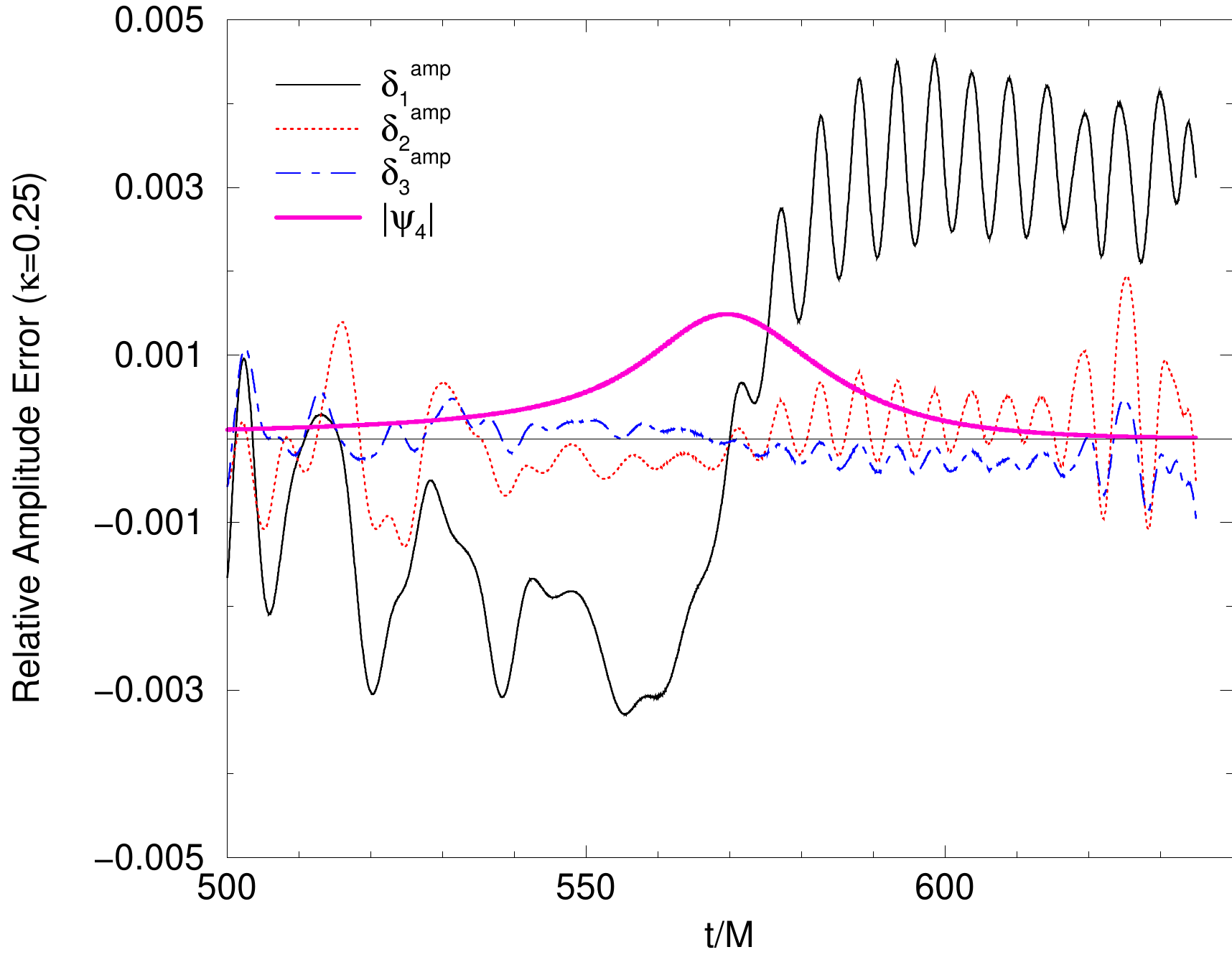}
 \includegraphics[width=.4\columnwidth]{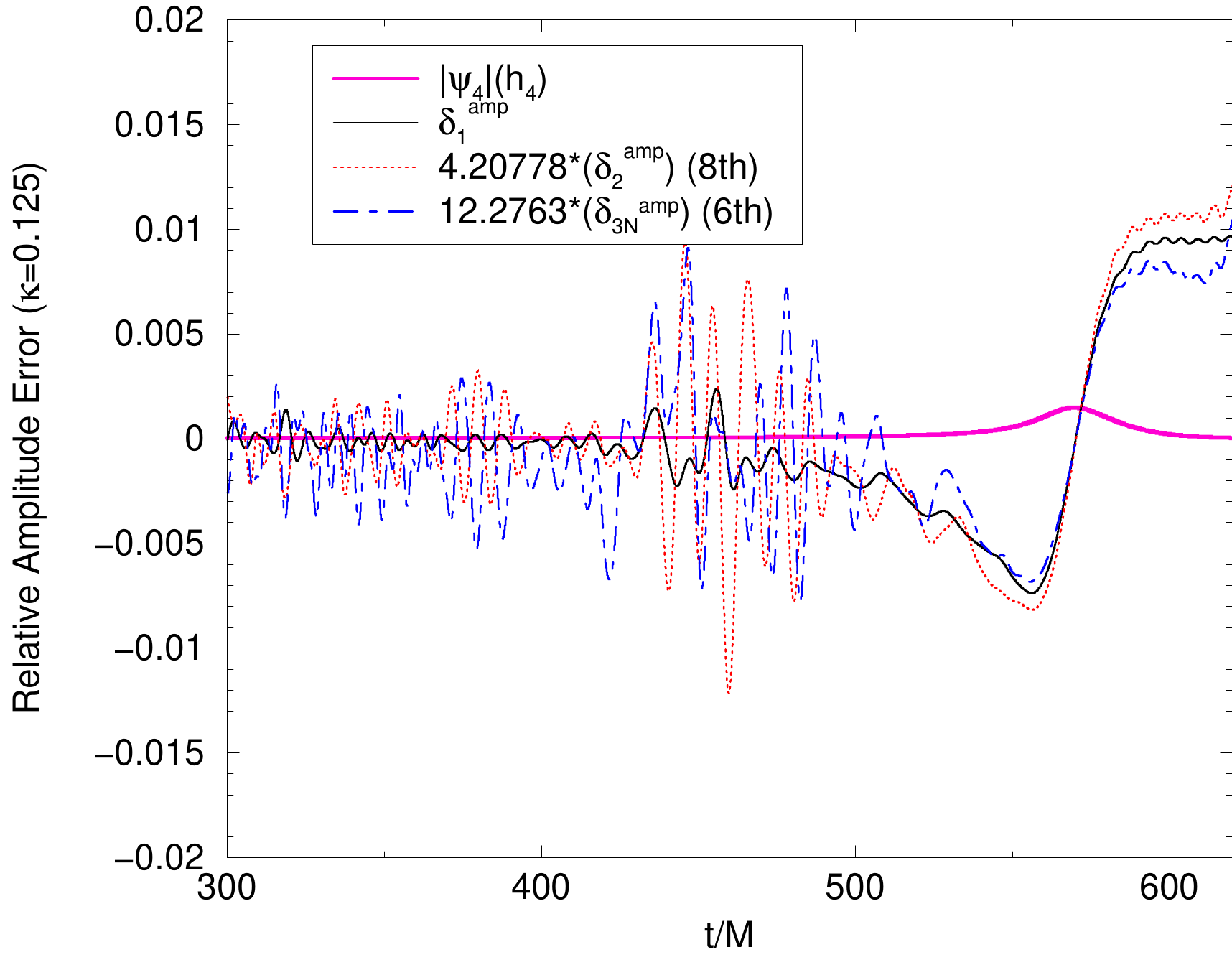}
 \caption{The error in the amplitude $\delta_{i}^{\rm amp}$ of the
$(\ell=2, m=2)$ mode of $\psi_4$ [$\delta_{i}^{\rm amp}=
(|\psi_4|(h_i) -
|\psi_4|(h_4))/|\psi_4|(h_4))$]. Clean convergence is seen for the
$\kappa=0.125$ runs, while a change in sign in $\delta_3$ is seen in the
$\kappa=0.25$ runs. The magnitude of $\delta_3$ is roughly a factor of
3  smaller for the $\kappa=0.25$ runs.}
\label{fig:amp_conv}
\end{figure}

\end{widetext}

The waveforms do not appear to converge in a consistent manner at high
resolutions. In order to try to elucidate the origin of this problem,
we repeated several of the simulations using the Conformal Z4
algorithm developed in Alic {\it et al}~\cite{Alic:2011gg}. 
For the $Z_4$ parameters,
we used $\kappa_1 = 0.1$, $\kappa_2=0$, and $\kappa_3=0$ (Alic  {\it
et al}
suggest setting $\kappa_3=1/2$, but our simulations were based on an
earlier version of their paper).  We used the \lazev
coding infrastructure to generate the Z4 evolution code. In addition,
we modified the gauge conditions to use Eq.~(\ref{eq:gauge}) above (with
$\tilde \Gamma^i$ replaced by the CCZ4 variable $\hat \Gamma^i$), rather
than the gauge condition used in~\cite{Alic:2011gg} (for the damping
parameters, we used the values given in~\cite{Alic:2011gg}). For these
Z4 runs, we used the H855 system, that is, eighth-order centered
finite differencing, fifth-order dissipation, fifth-order spatial
prolongation, a full complement of 16 buffer zones, and fourth-order
accurate initialization. As is evident in
Fig.~\ref{fig:phase_v_res_z4_bssn}, for a given spatial
resolution, the Z4 algorithm exhibits a larger truncation error
compared to BSSN. While the Z4 runs show better than 4th-order
convergence, with clean convergence at all resolutions attempted, the
size of the error makes comparison with BSSN difficult. That is, with
coarser resolutions, the BSSN algorithm also appears to be cleanly
convergent. Nevertheless, we use a Richardson extrapolation of the
waveform phase for Z4 and compare this extrapolated phase with the BSSN
phases, as shown in  Fig.~\ref{fig:phase_v_res_z4_bssn}. Here the BSSN
phases are all much closer to the extrapolated Z4 phase than any of
the individual Z4 phases. It would thus appear that the Z4 and BSSN
results are consistent.

\subsection{Convergence of the Amplitude}
\label{sec:amp}
  We measure the amplitude variation with resolution, relative to the
highest resolution, of the
$(\ell=2,m=2)$ mode of $\psi_4$ $$\delta_{i}^{\rm amp}= (|\psi_4|(h_i) -
|\psi_4|(h_4))/|\psi_4|(h_4))$$ for the $\kappa=0.25, 0.177, 0.125$
H855 configurations. We find that the deviations are smallest for
$\kappa=0.25$, larger for $\kappa=0.177$, and largest for
$\kappa=0.125$. Interestingly, the $\kappa=0.177$ results are more
noisy than either of the other two. As shown in Fig.~\ref{fig:amp_conv},
clear convergence is seen for the $\kappa=0.125$ configurations,
while an oscillation is seen in the $\kappa=0.25$ results, which is
consistent with our previous findings for the phase errors. That is,
while cleaner convergence is seen with $\kappa=0.125$, this appears to
be a consequence of the larger error associated with smaller CFLs.
Thus, for high accuracy runs, one should apparently not use
$\kappa=0.125$. The
overall relative amplitude error appears to be controllable to within
0.1\%.

\begin{figure}
  \includegraphics[width=\figwidth]{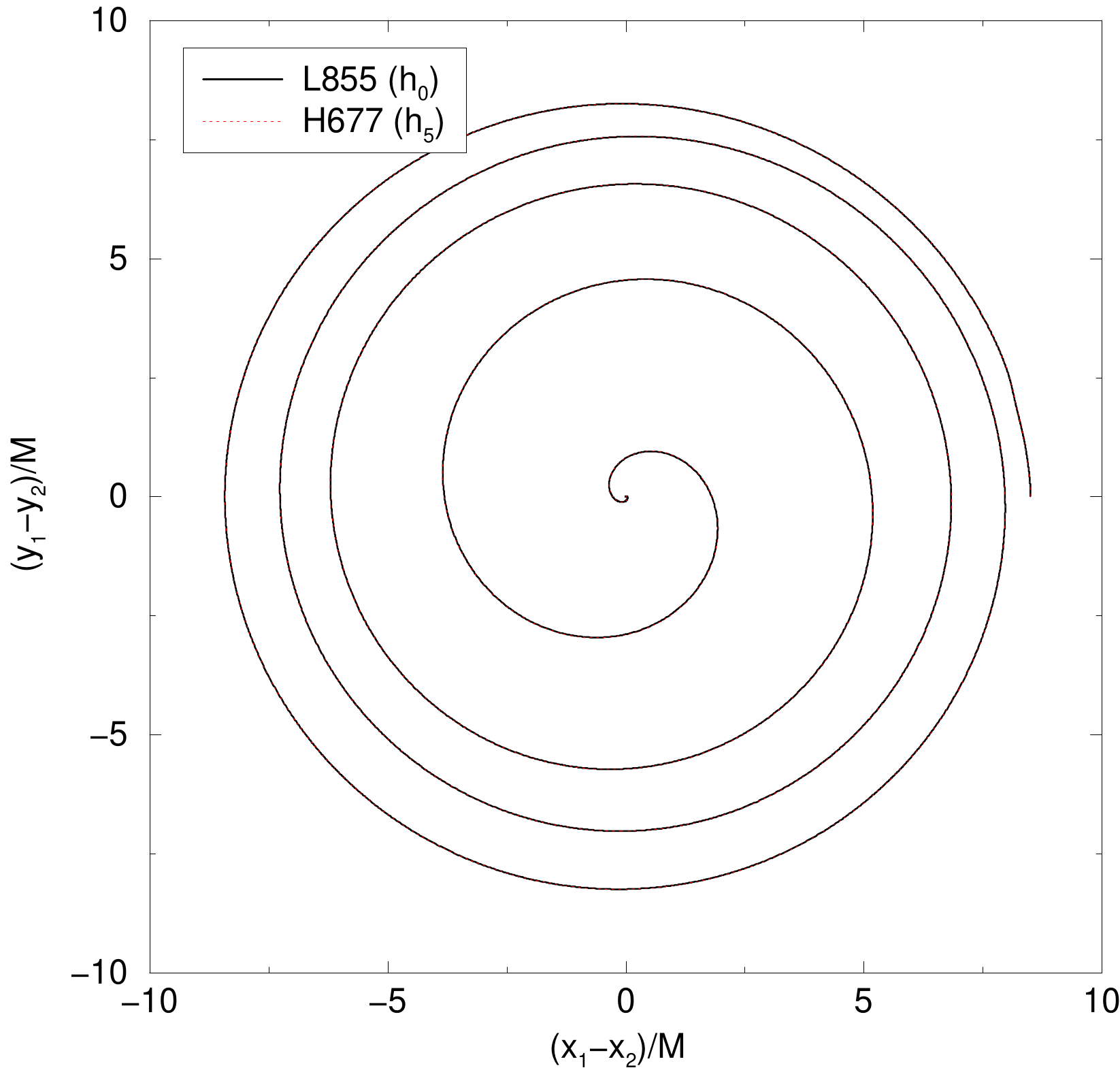}
  \includegraphics[width=\figwidth]{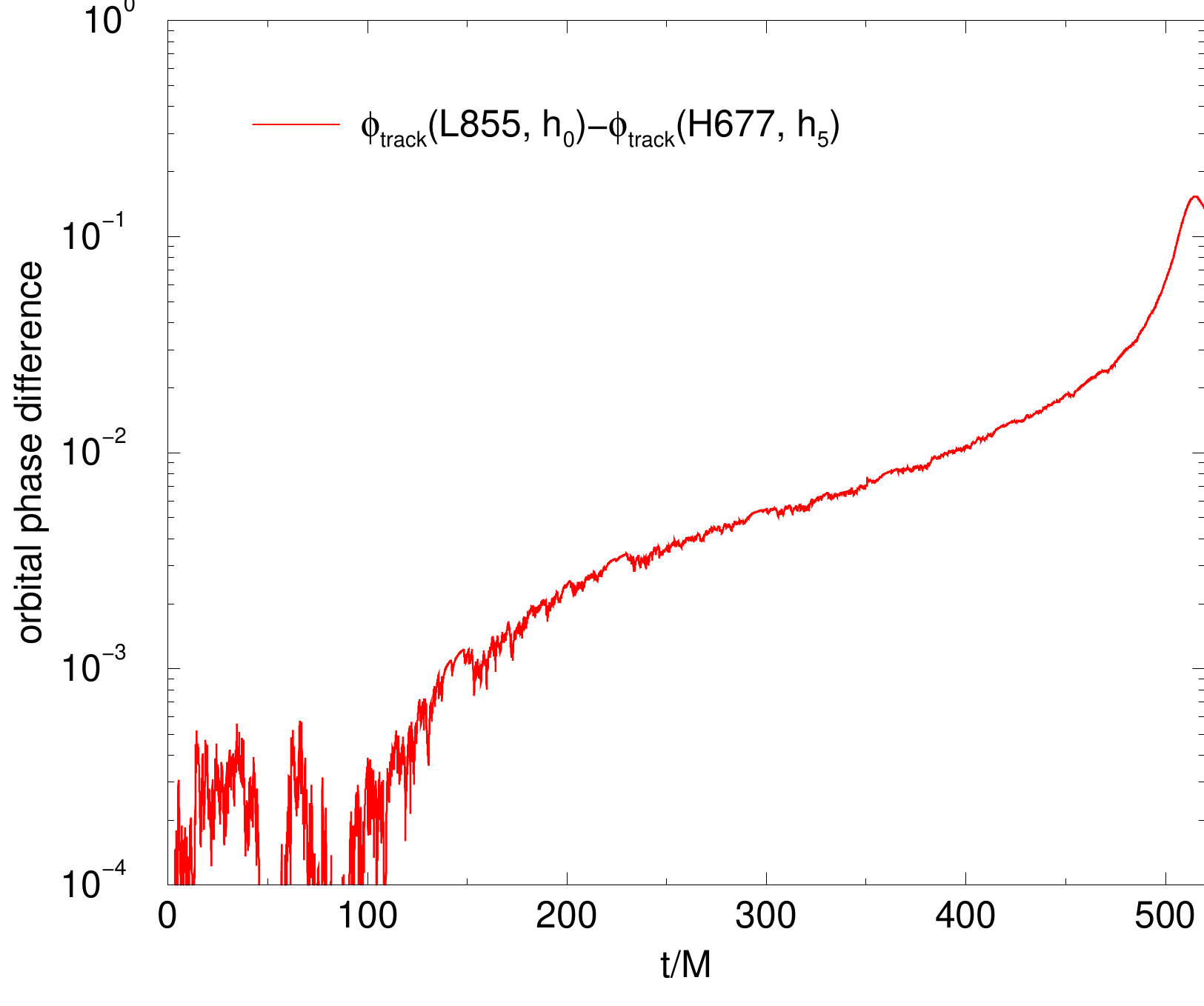}
\caption{A comparison of the orbital trajectory for the lowest
resolution L877 (low-accuracy) simulation and the highest resolution
H677 simulation. The phase error at merger is $0.15$ rad, which
corresponds to a waveform phase error of $0.30$ rad. The actual
post-merger phase difference is about $0.20$ rad.}
\label{fig:track}
\end{figure}

We conclude this section by noting that even the lowest accuracy L855
simulation is accurate enough for many applications where extreme
phase accuracy is not needed (e.g.\ recoil studies). In
Fig.~\ref{fig:track} we show the orbital trajectory for the lowest
resolution L855 and highest resolution H677 simulations. The
two orbital trajectories overlap throughout the entire simulation,
even at merger, where the phase error increases to $0.15$ rad.

\begin{figure}
  \includegraphics[width=\figwidth]{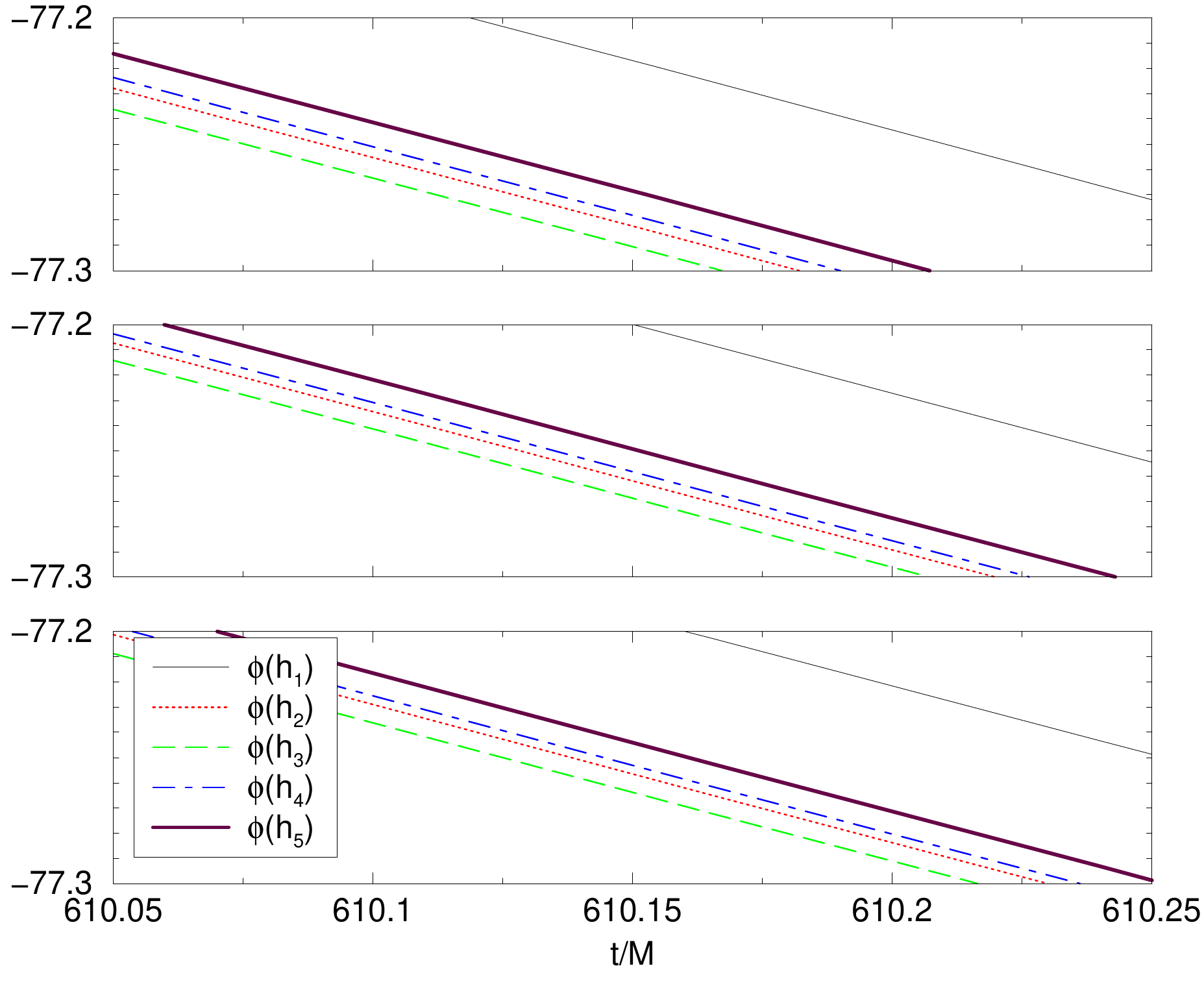}
\caption{The phase of the $(\ell=2,m-2)$ mode of $\psi_4$ as
calculated at $r=50M$ (bottom) and $r=60M$ (middle)  and $r=100M$
(top) 
for the H677
system. The $r=50M$ and $r=100M$ spheres intersect two AMR levels, while the
$r=60M$ lies on a single level (but does intersect a buffer region). No
difference in the oscillations in the phases of the waveform are
apparent. }
\label{fig:cmp_radii}
\end{figure}

Finally, we note that our AMR grids changed with resolution because
our chosen radii fell in between actual gridpoints. This means that
the AMR boundary sizes changed from resolution to resolution. In
fact, 
 the location of the AMR levels was determined during each
evolution by tracking the location of the ``puncture''. This means that
the AMR locations also changed from resolution to resolution. 
In addition, our $r=50$ waveform was generated by integrating over a
sphere that crossed AMR boundaries. 
While
it is difficult to determine the effect of the former on our results,
we can show that the latter effect (due to integrating over AMR
boundaries) was not a significant source of error. In
Fig.~\ref{fig:cmp_radii}, we plot the phase from the five H677
simulations at both $r=50$ and $r=100$ (the plots have been
translated). Note how the oscillations in the phase at $r=50$ are
mirrored in the phase at $r=100$.

\clearpage
\section{Discussion}

After removing various low-accuracy approximations, we were able to
reduce variation in the horizon mass of the two BHs in a BHB to the
level of $10^{-6}M$. We find the Hamiltonian constraint violations
(both the $L_2$ norm in the bulk, and horizon-averaged values) are
convergent. The momentum and BSSN constraints also converge to
fourth-order on the horizons.  Convergence of the momentum constraint
in the bulk is less clear, with only H677 showing clean convergence.

We find a stochastic error in the phase of the waveform that prevents
us from seeing convergence at high resolution. We speculate that this
stochastic error is associated with the high-frequency, unresolved,
wave from the initial-data burst of radiation that reflects off the AMR
boundaries, producing a complex, highly-variable interference pattern
on the grid (see Fig.~\ref{fig:amr_ref}).
Without a clean convergence of the high-resolution runs, it is
difficult to predict the actual error in the phase of the waveform at
merger and ringdown. We note that the oscillations in the phase (with
resolution) appear to be controllable to about the level of $0.015$
rad., which is accurate enough for data analysis purposes (i.e.\
exceeds NRAR and NINJA accuracy goals). Whether the phase error really
is this small depends on the source of the oscillations. If these are
due to essentially stochastic errors introduced by grid reflections,
then it seems reasonable to assume the phase error is controllable to 
$0.015$ rad. However, if the source of the oscillation is actually
lower-order errors becoming more dominant at higher resolutions, then
we need far higher-resolution runs. Since the lowest-order error is
the second-order prolongation (in time) error, one may need to perform
a convergence study (at very high resolutions) where the grids are
refined by factors of two between resolutions (with coarsest
resolution as fine, or finer than, our finest resolution).
 Such a study would be
computationally very expensive, but may ultimately be necessary.

One reason why grid refinement boundaries may be a significant issue
has to do with reflection of the (relatively) high-frequency initial
burst of radiation.
By construction, the AMR grids are well adapted to
evolving the relatively low-frequency waveform signal (with smallest
period of about 10M), but the initial burst has frequencies a factor
of four higher. To properly evolve this pulse, may require a factor of
four increase in resolution, which would require a factor of 256 in
computational resources. Fortunately, this signal is not physical,
and one only needs to confirm that the effects 
on the rest of the waveform of poorly 
resolving this pulse are within data analysis tolerances.
However, use of data with smaller spurious high-frequency wave
content~\cite{Hannam:2006zt,Lovelace:2008hd} and more correct
low-frequency wave content~\cite{Mundim:2010hu,Kelly:2009js,Kelly:2007uc} 
should help reduce AMR reflections
and the corresponding stochastic phase error. We expect to recover
full convergence of the moving puncture formalism in the unigrid limit
performing tests similar to those in Ref.~\cite{Babiuc:2007vr}.

In order to obtain an independent measure of the phase error in our
simulations, we  compare the phase of the waveform obtained using both the BSSN
and CCZ4 formalisms. We find agreement at the level of $0.01$ radian,
but only after a Richardson extrapolation.
 Because both codes used
very similar technologies and the same AMR implementation, 
 it would be very useful to compare the results from
these simulations with other codes, analogous to what was done with
the Samurai project~\cite{Hannam:2009hh}. Comparisons with other
AMR-based codes, such as BAM~\cite{Brugmann:2008zz}, and
pseudospectral codes, such as SPEC~\cite{Scheel:2006gg,Boyle:2006ne},
would be especially useful. 

Our simulations show that the L855 system is inappropriate for use
where high phase-accuracy is needed. However, for situations where lower
accuracy is acceptable, e.g.\ for recoil studies, it can be used. Our
tests also show that, even when using high-accuracy methods with
a state-of-the-art AMR code for numerical relativity, clean
convergence remains elusive, but we have seen improvements with the
prolongation order in systems such as H677 and H877, with signs of
consistency for H677.
We note that~\cite{Pollney:2009yz} use 
the numerically more expensive H899 algorithm, which also
includes upwinded finite differencing stencils. They found that
upwinding increases the accuracy of the simulation.

\section{conclusion}
  We analyze the accuracy of a BHB simulation by examining the
preservation of the individual horizon masses, convergence of the
constraints, and convergence of the magnitude and phase of the
$(\ell=2,m=2)$ mode $\psi_4$ extracted at $r=50M$. As seen in
Fig.~\ref{fig:hmass}, the fast, but low-accuracy approximations lead to
poor conservation of the mass compared to the slower, but more
accurate, techniques.  For the high-accuracy techniques, we find
convergence of the constraints to the expected order, as seen in
Fig.~\ref{fig:H_pi_nopi}. A residual, possibly stochastic, phase error is
seen in the waveform itself. By comparing the waveforms generated
using different techniques and different evolution systems, as shown
~in Figs.~\ref{fig:phase_error}~and~\ref{fig:phase_v_res_z4_bssn}, we find consistency in the phase at the
level
of $0.05$ rad. over the entire waveform.

\acknowledgments 

The authors thank D.Alic,  I.Hinder, F. L\"offler, C.Palenzuela,
and Z.Etienne for
helpful discussions.
We gratefully acknowledge the NSF for financial support from Grants
AST-1028087, PHY-0929114, PHY-0969855, PHY-0903782, OCI-0832606,
PHY-1229173, PHY-1212426, and
DRL-1136221,  and NASA for financial support from Grant No.
07-ATFP07-0158. Computational resources were provided by the Ranger
system at the Texas Advance Computing Center (Teragrid allocation
TG-PHY060027N), which is supported in part by the NSF, and by
NewHorizons at Rochester Institute of Technology, which was supported
by NSF grant No. PHY-0722703, PHY-1229173, DMS-0820923, and AST-1028087.

\bibliographystyle{apsrev}
\bibliography{../../Bibtex/references}

\begin{thebibliography}{86}
\expandafter\ifx\csname natexlab\endcsname\relax\def\natexlab#1{#1}\fi
\expandafter\ifx\csname bibnamefont\endcsname\relax
  \def\bibnamefont#1{#1}\fi
\expandafter\ifx\csname bibfnamefont\endcsname\relax
  \def\bibfnamefont#1{#1}\fi
\expandafter\ifx\csname citenamefont\endcsname\relax
  \def\citenamefont#1{#1}\fi
\expandafter\ifx\csname url\endcsname\relax
  \def\url#1{\texttt{#1}}\fi
\expandafter\ifx\csname urlprefix\endcsname\relax\def\urlprefix{URL }\fi
\providecommand{\bibinfo}[2]{#2}
\providecommand{\eprint}[2][]{\url{#2}}

\bibitem[{\citenamefont{Pretorius}(2005)}]{Pretorius:2005gq}
\bibinfo{author}{\bibfnamefont{F.}~\bibnamefont{Pretorius}},
  \bibinfo{journal}{Phys. Rev. Lett.} \textbf{\bibinfo{volume}{95}},
  \bibinfo{pages}{121101} (\bibinfo{year}{2005}), \eprint{gr-qc/0507014}.

\bibitem[{\citenamefont{Campanelli
  et~al.}(2006{\natexlab{a}})\citenamefont{Campanelli, Lousto, Marronetti, and
  Zlochower}}]{Campanelli:2005dd}
\bibinfo{author}{\bibfnamefont{M.}~\bibnamefont{Campanelli}},
  \bibinfo{author}{\bibfnamefont{C.~O.} \bibnamefont{Lousto}},
  \bibinfo{author}{\bibfnamefont{P.}~\bibnamefont{Marronetti}},
  \bibnamefont{and}
  \bibinfo{author}{\bibfnamefont{Y.}~\bibnamefont{Zlochower}},
  \bibinfo{journal}{Phys. Rev. Lett.} \textbf{\bibinfo{volume}{96}},
  \bibinfo{pages}{111101} (\bibinfo{year}{2006}{\natexlab{a}}),
  \eprint{gr-qc/0511048}.

\bibitem[{\citenamefont{Baker et~al.}(2006{\natexlab{a}})\citenamefont{Baker,
  Centrella, Choi, Koppitz, and van Meter}}]{Baker:2005vv}
\bibinfo{author}{\bibfnamefont{J.~G.} \bibnamefont{Baker}},
  \bibinfo{author}{\bibfnamefont{J.}~\bibnamefont{Centrella}},
  \bibinfo{author}{\bibfnamefont{D.-I.} \bibnamefont{Choi}},
  \bibinfo{author}{\bibfnamefont{M.}~\bibnamefont{Koppitz}}, \bibnamefont{and}
  \bibinfo{author}{\bibfnamefont{J.}~\bibnamefont{van Meter}},
  \bibinfo{journal}{Phys. Rev. Lett.} \textbf{\bibinfo{volume}{96}},
  \bibinfo{pages}{111102} (\bibinfo{year}{2006}{\natexlab{a}}),
  \eprint{gr-qc/0511103}.

\bibitem[{\citenamefont{Aylott et~al.}(2009{\natexlab{a}})}]{Aylott:2009ya}
\bibinfo{author}{\bibfnamefont{B.}~\bibnamefont{Aylott}} \bibnamefont{et~al.},
  \bibinfo{journal}{Class. Quant. Grav.} \textbf{\bibinfo{volume}{26}},
  \bibinfo{pages}{165008} (\bibinfo{year}{2009}{\natexlab{a}}),
  \eprint{0901.4399}.

\bibitem[{\citenamefont{Aylott et~al.}(2009{\natexlab{b}})}]{Aylott:2009tn}
\bibinfo{author}{\bibfnamefont{B.}~\bibnamefont{Aylott}} \bibnamefont{et~al.},
  \bibinfo{journal}{Class. Quant. Grav.} \textbf{\bibinfo{volume}{26}},
  \bibinfo{pages}{114008} (\bibinfo{year}{2009}{\natexlab{b}}),
  \eprint{0905.4227}.

\bibitem[{\citenamefont{Herrmann et~al.}(2006)\citenamefont{Herrmann,
  Shoemaker, and Laguna}}]{Herrmann:2006ks}
\bibinfo{author}{\bibfnamefont{F.}~\bibnamefont{Herrmann}},
  \bibinfo{author}{\bibfnamefont{D.}~\bibnamefont{Shoemaker}},
  \bibnamefont{and} \bibinfo{author}{\bibfnamefont{P.}~\bibnamefont{Laguna}},
  \bibinfo{journal}{AIP Conf.} \textbf{\bibinfo{volume}{873}},
  \bibinfo{pages}{89} (\bibinfo{year}{2006}), \eprint{gr-qc/0601026}.

\bibitem[{\citenamefont{Baker et~al.}(2006{\natexlab{b}})}]{Baker:2006vn}
\bibinfo{author}{\bibfnamefont{J.~G.} \bibnamefont{Baker}}
  \bibnamefont{et~al.}, \bibinfo{journal}{Astrophys. J.}
  \textbf{\bibinfo{volume}{653}}, \bibinfo{pages}{L93}
  (\bibinfo{year}{2006}{\natexlab{b}}), \eprint{astro-ph/0603204}.

\bibitem[{\citenamefont{Gonz\'alez
  et~al.}(2007{\natexlab{a}})\citenamefont{Gonz\'alez, Sperhake, Brugmann,
  Hannam, and Husa}}]{Gonzalez:2006md}
\bibinfo{author}{\bibfnamefont{J.~A.} \bibnamefont{Gonz\'alez}},
  \bibinfo{author}{\bibfnamefont{U.}~\bibnamefont{Sperhake}},
  \bibinfo{author}{\bibfnamefont{B.}~\bibnamefont{Brugmann}},
  \bibinfo{author}{\bibfnamefont{M.}~\bibnamefont{Hannam}}, \bibnamefont{and}
  \bibinfo{author}{\bibfnamefont{S.}~\bibnamefont{Husa}},
  \bibinfo{journal}{Phys. Rev. Lett.} \textbf{\bibinfo{volume}{98}},
  \bibinfo{pages}{091101} (\bibinfo{year}{2007}{\natexlab{a}}),
  \eprint{gr-qc/0610154}.

\bibitem[{\citenamefont{Herrmann
  et~al.}(2007{\natexlab{a}})\citenamefont{Herrmann, Hinder, Shoemaker, Laguna,
  and Matzner}}]{Herrmann:2007ac}
\bibinfo{author}{\bibfnamefont{F.}~\bibnamefont{Herrmann}},
  \bibinfo{author}{\bibfnamefont{I.}~\bibnamefont{Hinder}},
  \bibinfo{author}{\bibfnamefont{D.}~\bibnamefont{Shoemaker}},
  \bibinfo{author}{\bibfnamefont{P.}~\bibnamefont{Laguna}}, \bibnamefont{and}
  \bibinfo{author}{\bibfnamefont{R.~A.} \bibnamefont{Matzner}},
  \bibinfo{journal}{Astrophys. J.} \textbf{\bibinfo{volume}{661}},
  \bibinfo{pages}{430} (\bibinfo{year}{2007}{\natexlab{a}}),
  \eprint{gr-qc/0701143}.

\bibitem[{\citenamefont{Campanelli
  et~al.}(2007{\natexlab{a}})\citenamefont{Campanelli, Lousto, Zlochower, and
  Merritt}}]{Campanelli:2007ew}
\bibinfo{author}{\bibfnamefont{M.}~\bibnamefont{Campanelli}},
  \bibinfo{author}{\bibfnamefont{C.~O.} \bibnamefont{Lousto}},
  \bibinfo{author}{\bibfnamefont{Y.}~\bibnamefont{Zlochower}},
  \bibnamefont{and} \bibinfo{author}{\bibfnamefont{D.}~\bibnamefont{Merritt}},
  \bibinfo{journal}{Astrophys. J.} \textbf{\bibinfo{volume}{659}},
  \bibinfo{pages}{L5} (\bibinfo{year}{2007}{\natexlab{a}}),
  \eprint{gr-qc/0701164}.

\bibitem[{\citenamefont{Campanelli
  et~al.}(2007{\natexlab{b}})\citenamefont{Campanelli, Lousto, Zlochower, and
  Merritt}}]{Campanelli:2007cga}
\bibinfo{author}{\bibfnamefont{M.}~\bibnamefont{Campanelli}},
  \bibinfo{author}{\bibfnamefont{C.~O.} \bibnamefont{Lousto}},
  \bibinfo{author}{\bibfnamefont{Y.}~\bibnamefont{Zlochower}},
  \bibnamefont{and} \bibinfo{author}{\bibfnamefont{D.}~\bibnamefont{Merritt}},
  \bibinfo{journal}{Phys. Rev. Lett.} \textbf{\bibinfo{volume}{98}},
  \bibinfo{pages}{231102} (\bibinfo{year}{2007}{\natexlab{b}}),
  \eprint{gr-qc/0702133}.

\bibitem[{\citenamefont{Lousto and Zlochower}(2009)}]{Lousto:2008dn}
\bibinfo{author}{\bibfnamefont{C.~O.} \bibnamefont{Lousto}} \bibnamefont{and}
  \bibinfo{author}{\bibfnamefont{Y.}~\bibnamefont{Zlochower}},
  \bibinfo{journal}{Phys. Rev.} \textbf{\bibinfo{volume}{D79}},
  \bibinfo{pages}{064018} (\bibinfo{year}{2009}), \eprint{0805.0159}.

\bibitem[{\citenamefont{Pollney et~al.}(2007)}]{Pollney:2007ss}
\bibinfo{author}{\bibfnamefont{D.}~\bibnamefont{Pollney}} \bibnamefont{et~al.},
  \bibinfo{journal}{Phys. Rev.} \textbf{\bibinfo{volume}{D76}},
  \bibinfo{pages}{124002} (\bibinfo{year}{2007}), \eprint{0707.2559}.

\bibitem[{\citenamefont{Gonz\'alez
  et~al.}(2007{\natexlab{b}})\citenamefont{Gonz\'alez, Hannam, Sperhake,
  Brugmann, and Husa}}]{Gonzalez:2007hi}
\bibinfo{author}{\bibfnamefont{J.~A.} \bibnamefont{Gonz\'alez}},
  \bibinfo{author}{\bibfnamefont{M.~D.} \bibnamefont{Hannam}},
  \bibinfo{author}{\bibfnamefont{U.}~\bibnamefont{Sperhake}},
  \bibinfo{author}{\bibfnamefont{B.}~\bibnamefont{Brugmann}}, \bibnamefont{and}
  \bibinfo{author}{\bibfnamefont{S.}~\bibnamefont{Husa}},
  \bibinfo{journal}{Phys. Rev. Lett.} \textbf{\bibinfo{volume}{98}},
  \bibinfo{pages}{231101} (\bibinfo{year}{2007}{\natexlab{b}}),
  \eprint{gr-qc/0702052}.

\bibitem[{\citenamefont{Brugmann et~al.}(2008)\citenamefont{Brugmann, Gonzalez,
  Hannam, Husa, and Sperhake}}]{Brugmann:2007zj}
\bibinfo{author}{\bibfnamefont{B.}~\bibnamefont{Brugmann}},
  \bibinfo{author}{\bibfnamefont{J.~A.} \bibnamefont{Gonzalez}},
  \bibinfo{author}{\bibfnamefont{M.}~\bibnamefont{Hannam}},
  \bibinfo{author}{\bibfnamefont{S.}~\bibnamefont{Husa}}, \bibnamefont{and}
  \bibinfo{author}{\bibfnamefont{U.}~\bibnamefont{Sperhake}},
  \bibinfo{journal}{Phys. Rev.} \textbf{\bibinfo{volume}{D77}},
  \bibinfo{pages}{124047} (\bibinfo{year}{2008}), \eprint{0707.0135}.

\bibitem[{\citenamefont{Choi et~al.}(2007)}]{Choi:2007eu}
\bibinfo{author}{\bibfnamefont{D.-I.} \bibnamefont{Choi}} \bibnamefont{et~al.},
  \bibinfo{journal}{Phys. Rev.} \textbf{\bibinfo{volume}{D76}},
  \bibinfo{pages}{104026} (\bibinfo{year}{2007}), \eprint{gr-qc/0702016}.

\bibitem[{\citenamefont{Baker et~al.}(2007)}]{Baker:2007gi}
\bibinfo{author}{\bibfnamefont{J.~G.} \bibnamefont{Baker}}
  \bibnamefont{et~al.}, \bibinfo{journal}{Astrophys. J.}
  \textbf{\bibinfo{volume}{668}}, \bibinfo{pages}{1140} (\bibinfo{year}{2007}),
  \eprint{astro-ph/0702390}.

\bibitem[{\citenamefont{Schnittman et~al.}(2008)}]{Schnittman:2007ij}
\bibinfo{author}{\bibfnamefont{J.~D.} \bibnamefont{Schnittman}}
  \bibnamefont{et~al.}, \bibinfo{journal}{Phys. Rev.}
  \textbf{\bibinfo{volume}{D77}}, \bibinfo{pages}{044031}
  (\bibinfo{year}{2008}), \eprint{0707.0301}.

\bibitem[{\citenamefont{Baker et~al.}(2008)}]{Baker:2008md}
\bibinfo{author}{\bibfnamefont{J.~G.} \bibnamefont{Baker}}
  \bibnamefont{et~al.}, \bibinfo{journal}{Astrophys. J.}
  \textbf{\bibinfo{volume}{682}}, \bibinfo{pages}{L29} (\bibinfo{year}{2008}),
  \eprint{0802.0416}.

\bibitem[{\citenamefont{Healy et~al.}(2009)}]{Healy:2008js}
\bibinfo{author}{\bibfnamefont{J.}~\bibnamefont{Healy}} \bibnamefont{et~al.},
  \bibinfo{journal}{Phys. Rev. Lett.} \textbf{\bibinfo{volume}{102}},
  \bibinfo{pages}{041101} (\bibinfo{year}{2009}), \eprint{0807.3292}.

\bibitem[{\citenamefont{Herrmann
  et~al.}(2007{\natexlab{b}})\citenamefont{Herrmann, Hinder, Shoemaker, and
  Laguna}}]{Herrmann:2007zz}
\bibinfo{author}{\bibfnamefont{F.}~\bibnamefont{Herrmann}},
  \bibinfo{author}{\bibfnamefont{I.}~\bibnamefont{Hinder}},
  \bibinfo{author}{\bibfnamefont{D.}~\bibnamefont{Shoemaker}},
  \bibnamefont{and} \bibinfo{author}{\bibfnamefont{P.}~\bibnamefont{Laguna}},
  \bibinfo{journal}{Class. Quant. Grav.} \textbf{\bibinfo{volume}{24}},
  \bibinfo{pages}{S33} (\bibinfo{year}{2007}{\natexlab{b}}).

\bibitem[{\citenamefont{Herrmann
  et~al.}(2007{\natexlab{c}})\citenamefont{Herrmann, Hinder, Shoemaker, Laguna,
  and Matzner}}]{Herrmann:2007ex}
\bibinfo{author}{\bibfnamefont{F.}~\bibnamefont{Herrmann}},
  \bibinfo{author}{\bibfnamefont{I.}~\bibnamefont{Hinder}},
  \bibinfo{author}{\bibfnamefont{D.~M.} \bibnamefont{Shoemaker}},
  \bibinfo{author}{\bibfnamefont{P.}~\bibnamefont{Laguna}}, \bibnamefont{and}
  \bibinfo{author}{\bibfnamefont{R.~A.} \bibnamefont{Matzner}},
  \bibinfo{journal}{Phys. Rev.} \textbf{\bibinfo{volume}{D76}},
  \bibinfo{pages}{084032} (\bibinfo{year}{2007}{\natexlab{c}}),
  \eprint{0706.2541}.

\bibitem[{\citenamefont{Tichy and Marronetti}(2007)}]{Tichy:2007hk}
\bibinfo{author}{\bibfnamefont{W.}~\bibnamefont{Tichy}} \bibnamefont{and}
  \bibinfo{author}{\bibfnamefont{P.}~\bibnamefont{Marronetti}},
  \bibinfo{journal}{Phys. Rev.} \textbf{\bibinfo{volume}{D76}},
  \bibinfo{pages}{061502} (\bibinfo{year}{2007}), \eprint{gr-qc/0703075}.

\bibitem[{\citenamefont{Koppitz et~al.}(2007)\citenamefont{Koppitz, Pollney,
  Reisswig, Rezzolla, Thornburg et~al.}}]{Koppitz:2007ev}
\bibinfo{author}{\bibfnamefont{M.}~\bibnamefont{Koppitz}},
  \bibinfo{author}{\bibfnamefont{D.}~\bibnamefont{Pollney}},
  \bibinfo{author}{\bibfnamefont{C.}~\bibnamefont{Reisswig}},
  \bibinfo{author}{\bibfnamefont{L.}~\bibnamefont{Rezzolla}},
  \bibinfo{author}{\bibfnamefont{J.}~\bibnamefont{Thornburg}},
  \bibnamefont{et~al.}, \bibinfo{journal}{Phys. Rev. Lett.}
  \textbf{\bibinfo{volume}{99}}, \bibinfo{pages}{041102}
  (\bibinfo{year}{2007}), \eprint{gr-qc/0701163}.

\bibitem[{\citenamefont{Miller and Matzner}(2009)}]{Miller:2008en}
\bibinfo{author}{\bibfnamefont{S.~H.} \bibnamefont{Miller}} \bibnamefont{and}
  \bibinfo{author}{\bibfnamefont{R.~A.} \bibnamefont{Matzner}},
  \bibinfo{journal}{Gen. Rel. Grav.} \textbf{\bibinfo{volume}{41}},
  \bibinfo{pages}{525} (\bibinfo{year}{2009}), \eprint{0807.3028}.

\bibitem[{\citenamefont{Lousto and
  Zlochower}(2011{\natexlab{a}})}]{Lousto:2011kp}
\bibinfo{author}{\bibfnamefont{C.~O.} \bibnamefont{Lousto}} \bibnamefont{and}
  \bibinfo{author}{\bibfnamefont{Y.}~\bibnamefont{Zlochower}},
  \bibinfo{journal}{Phys. Rev. Lett.} \textbf{\bibinfo{volume}{107}},
  \bibinfo{pages}{231102} (\bibinfo{year}{2011}{\natexlab{a}}),
  \eprint{1108.2009}.

\bibitem[{\citenamefont{Zlochower et~al.}(2011)\citenamefont{Zlochower,
  Campanelli, and Lousto}}]{Zlochower:2010sn}
\bibinfo{author}{\bibfnamefont{Y.}~\bibnamefont{Zlochower}},
  \bibinfo{author}{\bibfnamefont{M.}~\bibnamefont{Campanelli}},
  \bibnamefont{and} \bibinfo{author}{\bibfnamefont{C.~O.}
  \bibnamefont{Lousto}}, \bibinfo{journal}{Class. Quant. Grav.}
  \textbf{\bibinfo{volume}{28}}, \bibinfo{pages}{114015}
  (\bibinfo{year}{2011}), \eprint{1011.2210}.

\bibitem[{\citenamefont{Lousto and
  Zlochower}(2011{\natexlab{b}})}]{Lousto:2010xk}
\bibinfo{author}{\bibfnamefont{C.~O.} \bibnamefont{Lousto}} \bibnamefont{and}
  \bibinfo{author}{\bibfnamefont{Y.}~\bibnamefont{Zlochower}},
  \bibinfo{journal}{Phys. Rev.} \textbf{\bibinfo{volume}{D83}},
  \bibinfo{pages}{024003} (\bibinfo{year}{2011}{\natexlab{b}}),
  \eprint{1011.0593}.

\bibitem[{\citenamefont{Lousto et~al.}(2012)\citenamefont{Lousto, Zlochower,
  Dotti, and Volonteri}}]{Lousto:2012su}
\bibinfo{author}{\bibfnamefont{C.~O.} \bibnamefont{Lousto}},
  \bibinfo{author}{\bibfnamefont{Y.}~\bibnamefont{Zlochower}},
  \bibinfo{author}{\bibfnamefont{M.}~\bibnamefont{Dotti}}, \bibnamefont{and}
  \bibinfo{author}{\bibfnamefont{M.}~\bibnamefont{Volonteri}},
  \bibinfo{journal}{Phys. Rev.} \textbf{\bibinfo{volume}{D85}},
  \bibinfo{pages}{084015} (\bibinfo{year}{2012}), \eprint{1201.1923}.

\bibitem[{\citenamefont{Sekiguchi and Shibata}(2011)}]{Sekiguchi:2010ja}
\bibinfo{author}{\bibfnamefont{Y.}~\bibnamefont{Sekiguchi}} \bibnamefont{and}
  \bibinfo{author}{\bibfnamefont{M.}~\bibnamefont{Shibata}},
  \bibinfo{journal}{Astrophys. J.} \textbf{\bibinfo{volume}{737}},
  \bibinfo{pages}{6} (\bibinfo{year}{2011}), \eprint{1009.5303}.

\bibitem[{\citenamefont{Etienne et~al.}(2012)\citenamefont{Etienne, Liu,
  Paschalidis, and Shapiro}}]{Etienne:2011ea}
\bibinfo{author}{\bibfnamefont{Z.~B.} \bibnamefont{Etienne}},
  \bibinfo{author}{\bibfnamefont{Y.~T.} \bibnamefont{Liu}},
  \bibinfo{author}{\bibfnamefont{V.}~\bibnamefont{Paschalidis}},
  \bibnamefont{and} \bibinfo{author}{\bibfnamefont{S.~L.}
  \bibnamefont{Shapiro}}, \bibinfo{journal}{Phys.Rev.}
  \textbf{\bibinfo{volume}{D85}}, \bibinfo{pages}{064029}
  (\bibinfo{year}{2012}), \eprint{1112.0568}.

\bibitem[{\citenamefont{{Etienne} et~al.}(2009)\citenamefont{{Etienne}, {Liu},
  {Shapiro}, and {Baumgarte}}}]{Etienne:2008re}
\bibinfo{author}{\bibfnamefont{Z.~B.} \bibnamefont{{Etienne}}},
  \bibinfo{author}{\bibfnamefont{Y.~T.} \bibnamefont{{Liu}}},
  \bibinfo{author}{\bibfnamefont{S.~L.} \bibnamefont{{Shapiro}}},
  \bibnamefont{and} \bibinfo{author}{\bibfnamefont{T.~W.}
  \bibnamefont{{Baumgarte}}}, \bibinfo{journal}{Phys. Rev. D}
  \textbf{\bibinfo{volume}{79}}, \bibinfo{eid}{044024} (\bibinfo{year}{2009}),
  \eprint{0812.2245}.

\bibitem[{\citenamefont{Etienne et~al.}(2008)\citenamefont{Etienne, Faber, Liu,
  Shapiro, Taniguchi, and Baumgarte}}]{Etienne:2007jg}
\bibinfo{author}{\bibfnamefont{Z.~B.} \bibnamefont{Etienne}},
  \bibinfo{author}{\bibfnamefont{J.~A.} \bibnamefont{Faber}},
  \bibinfo{author}{\bibfnamefont{Y.~T.} \bibnamefont{Liu}},
  \bibinfo{author}{\bibfnamefont{S.~L.} \bibnamefont{Shapiro}},
  \bibinfo{author}{\bibfnamefont{K.}~\bibnamefont{Taniguchi}},
  \bibnamefont{and} \bibinfo{author}{\bibfnamefont{T.~W.}
  \bibnamefont{Baumgarte}}, \bibinfo{journal}{Phys. Rev. D}
  \textbf{\bibinfo{volume}{77}}, \bibinfo{pages}{084002}
  (\bibinfo{year}{2008}), \eprint{arXiv:0712.2460 [astro-ph]}.

\bibitem[{\citenamefont{Rezzolla et~al.}(2011)\citenamefont{Rezzolla,
  Giacomazzo, Baiotti, Granot, Kouveliotou et~al.}}]{Rezzolla:2011da}
\bibinfo{author}{\bibfnamefont{L.}~\bibnamefont{Rezzolla}},
  \bibinfo{author}{\bibfnamefont{B.}~\bibnamefont{Giacomazzo}},
  \bibinfo{author}{\bibfnamefont{L.}~\bibnamefont{Baiotti}},
  \bibinfo{author}{\bibfnamefont{J.}~\bibnamefont{Granot}},
  \bibinfo{author}{\bibfnamefont{C.}~\bibnamefont{Kouveliotou}},
  \bibnamefont{et~al.}, \bibinfo{journal}{Astrophys. J.}
  \textbf{\bibinfo{volume}{732}}, \bibinfo{pages}{L6} (\bibinfo{year}{2011}),
  \eprint{1101.4298}.

\bibitem[{\citenamefont{Hotokezaka et~al.}(2011)\citenamefont{Hotokezaka,
  Kyutoku, Okawa, Shibata, and Kiuchi}}]{Hotokezaka:2011dh}
\bibinfo{author}{\bibfnamefont{K.}~\bibnamefont{Hotokezaka}},
  \bibinfo{author}{\bibfnamefont{K.}~\bibnamefont{Kyutoku}},
  \bibinfo{author}{\bibfnamefont{H.}~\bibnamefont{Okawa}},
  \bibinfo{author}{\bibfnamefont{M.}~\bibnamefont{Shibata}}, \bibnamefont{and}
  \bibinfo{author}{\bibfnamefont{K.}~\bibnamefont{Kiuchi}},
  \bibinfo{journal}{Phys. Rev.} \textbf{\bibinfo{volume}{D83}},
  \bibinfo{pages}{124008} (\bibinfo{year}{2011}), \eprint{1105.4370}.

\bibitem[{\citenamefont{Sekiguchi et~al.}(2011)\citenamefont{Sekiguchi, Kiuchi,
  Kyutoku, and Shibata}}]{Sekiguchi:2011zd}
\bibinfo{author}{\bibfnamefont{Y.}~\bibnamefont{Sekiguchi}},
  \bibinfo{author}{\bibfnamefont{K.}~\bibnamefont{Kiuchi}},
  \bibinfo{author}{\bibfnamefont{K.}~\bibnamefont{Kyutoku}}, \bibnamefont{and}
  \bibinfo{author}{\bibfnamefont{M.}~\bibnamefont{Shibata}},
  \bibinfo{journal}{Phys. Rev. Lett.} \textbf{\bibinfo{volume}{107}},
  \bibinfo{pages}{051102} (\bibinfo{year}{2011}), \eprint{1105.2125}.

\bibitem[{\citenamefont{Foucart et~al.}(2012)\citenamefont{Foucart, Duez,
  Kidder, Scheel, Szilagyi et~al.}}]{Foucart:2011mz}
\bibinfo{author}{\bibfnamefont{F.}~\bibnamefont{Foucart}},
  \bibinfo{author}{\bibfnamefont{M.~D.} \bibnamefont{Duez}},
  \bibinfo{author}{\bibfnamefont{L.~E.} \bibnamefont{Kidder}},
  \bibinfo{author}{\bibfnamefont{M.~A.} \bibnamefont{Scheel}},
  \bibinfo{author}{\bibfnamefont{B.}~\bibnamefont{Szilagyi}},
  \bibnamefont{et~al.}, \bibinfo{journal}{Phys. Rev.}
  \textbf{\bibinfo{volume}{D85}}, \bibinfo{pages}{044015}
  (\bibinfo{year}{2012}), \bibinfo{note}{11 pages, 11 figures - Updated to
  match published version}, \eprint{1111.1677}.

\bibitem[{\citenamefont{Duez et~al.}(2008)\citenamefont{Duez, Foucart, Kidder,
  Pfeiffer, Scheel et~al.}}]{Duez:2008rb}
\bibinfo{author}{\bibfnamefont{M.~D.} \bibnamefont{Duez}},
  \bibinfo{author}{\bibfnamefont{F.}~\bibnamefont{Foucart}},
  \bibinfo{author}{\bibfnamefont{L.~E.} \bibnamefont{Kidder}},
  \bibinfo{author}{\bibfnamefont{H.~P.} \bibnamefont{Pfeiffer}},
  \bibinfo{author}{\bibfnamefont{M.~A.} \bibnamefont{Scheel}},
  \bibnamefont{et~al.}, \bibinfo{journal}{Phys.Rev.}
  \textbf{\bibinfo{volume}{D78}}, \bibinfo{pages}{104015}
  (\bibinfo{year}{2008}), \eprint{0809.0002}.

\bibitem[{\citenamefont{Lousto and Zlochower}(2008)}]{Lousto:2007rj}
\bibinfo{author}{\bibfnamefont{C.~O.} \bibnamefont{Lousto}} \bibnamefont{and}
  \bibinfo{author}{\bibfnamefont{Y.}~\bibnamefont{Zlochower}},
  \bibinfo{journal}{Phys. Rev.} \textbf{\bibinfo{volume}{D77}},
  \bibinfo{pages}{024034} (\bibinfo{year}{2008}), \eprint{0711.1165}.

\bibitem[{\citenamefont{Campanelli et~al.}(2008)\citenamefont{Campanelli,
  Lousto, and Zlochower}}]{Campanelli:2007ea}
\bibinfo{author}{\bibfnamefont{M.}~\bibnamefont{Campanelli}},
  \bibinfo{author}{\bibfnamefont{C.~O.} \bibnamefont{Lousto}},
  \bibnamefont{and}
  \bibinfo{author}{\bibfnamefont{Y.}~\bibnamefont{Zlochower}},
  \bibinfo{journal}{Phys. Rev.} \textbf{\bibinfo{volume}{D77}},
  \bibinfo{pages}{101501(R)} (\bibinfo{year}{2008}), \eprint{0710.0879}.

\bibitem[{\citenamefont{Galaviz and Bruegmann}(2011)}]{Galaviz:2010te}
\bibinfo{author}{\bibfnamefont{P.}~\bibnamefont{Galaviz}} \bibnamefont{and}
  \bibinfo{author}{\bibfnamefont{B.}~\bibnamefont{Bruegmann}},
  \bibinfo{journal}{Phys. Rev.} \textbf{\bibinfo{volume}{D83}},
  \bibinfo{pages}{084013} (\bibinfo{year}{2011}), \eprint{1012.4423}.

\bibitem[{\citenamefont{Campanelli et~al.}(2009)\citenamefont{Campanelli,
  Lousto, and Zlochower}}]{Campanelli:2008dv}
\bibinfo{author}{\bibfnamefont{M.}~\bibnamefont{Campanelli}},
  \bibinfo{author}{\bibfnamefont{C.~O.} \bibnamefont{Lousto}},
  \bibnamefont{and}
  \bibinfo{author}{\bibfnamefont{Y.}~\bibnamefont{Zlochower}},
  \bibinfo{journal}{Phys. Rev.} \textbf{\bibinfo{volume}{D79}},
  \bibinfo{pages}{084012} (\bibinfo{year}{2009}), \eprint{0811.3006}.

\bibitem[{\citenamefont{Owen}(2010)}]{Owen:2010vw}
\bibinfo{author}{\bibfnamefont{R.}~\bibnamefont{Owen}}, \bibinfo{journal}{Phys.
  Rev.} \textbf{\bibinfo{volume}{D81}}, \bibinfo{pages}{124042}
  (\bibinfo{year}{2010}), \eprint{1004.3768}.

\bibitem[{\citenamefont{Campanelli
  et~al.}(2006{\natexlab{b}})\citenamefont{Campanelli, Lousto, and
  Zlochower}}]{Campanelli:2006uy}
\bibinfo{author}{\bibfnamefont{M.}~\bibnamefont{Campanelli}},
  \bibinfo{author}{\bibfnamefont{C.~O.} \bibnamefont{Lousto}},
  \bibnamefont{and}
  \bibinfo{author}{\bibfnamefont{Y.}~\bibnamefont{Zlochower}},
  \bibinfo{journal}{Phys. Rev.} \textbf{\bibinfo{volume}{D74}},
  \bibinfo{pages}{041501(R)} (\bibinfo{year}{2006}{\natexlab{b}}),
  \eprint{gr-qc/0604012}.

\bibitem[{\citenamefont{Ponce et~al.}(2011)\citenamefont{Ponce, Lousto, and
  Zlochower}}]{Ponce:2010fq}
\bibinfo{author}{\bibfnamefont{M.}~\bibnamefont{Ponce}},
  \bibinfo{author}{\bibfnamefont{C.}~\bibnamefont{Lousto}}, \bibnamefont{and}
  \bibinfo{author}{\bibfnamefont{Y.}~\bibnamefont{Zlochower}},
  \bibinfo{journal}{Class. Quant. Grav.} \textbf{\bibinfo{volume}{28}},
  \bibinfo{pages}{145027} (\bibinfo{year}{2011}), \eprint{1008.2761}.

\bibitem[{\citenamefont{Shibata and Yoshino}(2010)}]{Shibata:2010wz}
\bibinfo{author}{\bibfnamefont{M.}~\bibnamefont{Shibata}} \bibnamefont{and}
  \bibinfo{author}{\bibfnamefont{H.}~\bibnamefont{Yoshino}},
  \bibinfo{journal}{Phys. Rev.} \textbf{\bibinfo{volume}{D81}},
  \bibinfo{pages}{104035} (\bibinfo{year}{2010}), \eprint{1004.4970}.

\bibitem[{\citenamefont{Zilhao et~al.}(2010)\citenamefont{Zilhao, Witek,
  Sperhake, Cardoso, Gualtieri et~al.}}]{Zilhao:2010sr}
\bibinfo{author}{\bibfnamefont{M.}~\bibnamefont{Zilhao}},
  \bibinfo{author}{\bibfnamefont{H.}~\bibnamefont{Witek}},
  \bibinfo{author}{\bibfnamefont{U.}~\bibnamefont{Sperhake}},
  \bibinfo{author}{\bibfnamefont{V.}~\bibnamefont{Cardoso}},
  \bibinfo{author}{\bibfnamefont{L.}~\bibnamefont{Gualtieri}},
  \bibnamefont{et~al.}, \bibinfo{journal}{Phys. Rev.}
  \textbf{\bibinfo{volume}{D81}}, \bibinfo{pages}{084052}
  (\bibinfo{year}{2010}), \eprint{1001.2302}.

\bibitem[{\citenamefont{Witek et~al.}(2010)\citenamefont{Witek, Zilhao,
  Gualtieri, Cardoso, Herdeiro et~al.}}]{Witek:2010xi}
\bibinfo{author}{\bibfnamefont{H.}~\bibnamefont{Witek}},
  \bibinfo{author}{\bibfnamefont{M.}~\bibnamefont{Zilhao}},
  \bibinfo{author}{\bibfnamefont{L.}~\bibnamefont{Gualtieri}},
  \bibinfo{author}{\bibfnamefont{V.}~\bibnamefont{Cardoso}},
  \bibinfo{author}{\bibfnamefont{C.}~\bibnamefont{Herdeiro}},
  \bibnamefont{et~al.}, \bibinfo{journal}{Phys. Rev.}
  \textbf{\bibinfo{volume}{D82}}, \bibinfo{pages}{104014}
  (\bibinfo{year}{2010}), \eprint{1006.3081}.

\bibitem[{\citenamefont{Lousto and
  Zlochower}(2011{\natexlab{c}})}]{Lousto:2010ut}
\bibinfo{author}{\bibfnamefont{C.~O.} \bibnamefont{Lousto}} \bibnamefont{and}
  \bibinfo{author}{\bibfnamefont{Y.}~\bibnamefont{Zlochower}},
  \bibinfo{journal}{Phys. Rev. Lett.} \textbf{\bibinfo{volume}{106}},
  \bibinfo{pages}{041101} (\bibinfo{year}{2011}{\natexlab{c}}),
  \eprint{1009.0292}.

\bibitem[{\citenamefont{Sperhake et~al.}(2011)\citenamefont{Sperhake, Cardoso,
  Ott, Schnetter, and Witek}}]{Sperhake:2011ik}
\bibinfo{author}{\bibfnamefont{U.}~\bibnamefont{Sperhake}},
  \bibinfo{author}{\bibfnamefont{V.}~\bibnamefont{Cardoso}},
  \bibinfo{author}{\bibfnamefont{C.~D.} \bibnamefont{Ott}},
  \bibinfo{author}{\bibfnamefont{E.}~\bibnamefont{Schnetter}},
  \bibnamefont{and} \bibinfo{author}{\bibfnamefont{H.}~\bibnamefont{Witek}},
  \bibinfo{journal}{Phys. Rev.} \textbf{\bibinfo{volume}{D84}},
  \bibinfo{pages}{084038} (\bibinfo{year}{2011}), \eprint{1105.5391}.

\bibitem[{\citenamefont{Lovelace et~al.}(2011)\citenamefont{Lovelace, Scheel,
  and Szilagyi}}]{Lovelace:2010ne}
\bibinfo{author}{\bibfnamefont{G.}~\bibnamefont{Lovelace}},
  \bibinfo{author}{\bibfnamefont{M.~A.} \bibnamefont{Scheel}},
  \bibnamefont{and} \bibinfo{author}{\bibfnamefont{B.}~\bibnamefont{Szilagyi}},
  \bibinfo{journal}{Phys. Rev.} \textbf{\bibinfo{volume}{D83}},
  \bibinfo{pages}{024010} (\bibinfo{year}{2011}), \eprint{1010.2777}.

\bibitem[{\citenamefont{Lovelace et~al.}(2012)\citenamefont{Lovelace, Boyle,
  Scheel, and Szilagyi}}]{Lovelace:2011nu}
\bibinfo{author}{\bibfnamefont{G.}~\bibnamefont{Lovelace}},
  \bibinfo{author}{\bibfnamefont{M.}~\bibnamefont{Boyle}},
  \bibinfo{author}{\bibfnamefont{M.~A.} \bibnamefont{Scheel}},
  \bibnamefont{and} \bibinfo{author}{\bibfnamefont{B.}~\bibnamefont{Szilagyi}},
  \bibinfo{journal}{Class. Quant. Grav.} \textbf{\bibinfo{volume}{29}},
  \bibinfo{pages}{045003} (\bibinfo{year}{2012}), \eprint{1110.2229}.

\bibitem[{\citenamefont{Br{\"u}gmann et~al.}(2008)}]{Brugmann:2008zz}
\bibinfo{author}{\bibfnamefont{B.}~\bibnamefont{Br{\"u}gmann}}
  \bibnamefont{et~al.}, \bibinfo{journal}{Phys. Rev.}
  \textbf{\bibinfo{volume}{D77}}, \bibinfo{pages}{024027}
  (\bibinfo{year}{2008}), \eprint{gr-qc/0610128}.

\bibitem[{\citenamefont{Ansorg et~al.}(2004)\citenamefont{Ansorg, Br\"ugmann,
  and Tichy}}]{Ansorg:2004ds}
\bibinfo{author}{\bibfnamefont{M.}~\bibnamefont{Ansorg}},
  \bibinfo{author}{\bibfnamefont{B.}~\bibnamefont{Br\"ugmann}},
  \bibnamefont{and} \bibinfo{author}{\bibfnamefont{W.}~\bibnamefont{Tichy}},
  \bibinfo{journal}{Phys. Rev.} \textbf{\bibinfo{volume}{D70}},
  \bibinfo{pages}{064011} (\bibinfo{year}{2004}), \eprint{gr-qc/0404056}.

\bibitem[{\citenamefont{Brandt and Br{\"u}gmann}(1997)}]{Brandt97b}
\bibinfo{author}{\bibfnamefont{S.}~\bibnamefont{Brandt}} \bibnamefont{and}
  \bibinfo{author}{\bibfnamefont{B.}~\bibnamefont{Br{\"u}gmann}},
  \bibinfo{journal}{Phys. Rev. Lett.} \textbf{\bibinfo{volume}{78}},
  \bibinfo{pages}{3606} (\bibinfo{year}{1997}), \eprint{gr-qc/9703066}.

\bibitem[{\citenamefont{Bowen and York}(1980)}]{Bowen80}
\bibinfo{author}{\bibfnamefont{J.~M.} \bibnamefont{Bowen}} \bibnamefont{and}
  \bibinfo{author}{\bibfnamefont{J.~W.} \bibnamefont{York},
  \bibfnamefont{Jr.}}, \bibinfo{journal}{Phys. Rev.}
  \textbf{\bibinfo{volume}{D21}}, \bibinfo{pages}{2047} (\bibinfo{year}{1980}).

\bibitem[{\citenamefont{Zlochower et~al.}(2005)\citenamefont{Zlochower, Baker,
  Campanelli, and Lousto}}]{Zlochower:2005bj}
\bibinfo{author}{\bibfnamefont{Y.}~\bibnamefont{Zlochower}},
  \bibinfo{author}{\bibfnamefont{J.~G.} \bibnamefont{Baker}},
  \bibinfo{author}{\bibfnamefont{M.}~\bibnamefont{Campanelli}},
  \bibnamefont{and} \bibinfo{author}{\bibfnamefont{C.~O.}
  \bibnamefont{Lousto}}, \bibinfo{journal}{Phys. Rev.}
  \textbf{\bibinfo{volume}{D72}}, \bibinfo{pages}{024021}
  (\bibinfo{year}{2005}), \eprint{gr-qc/0505055}.

\bibitem[{\citenamefont{Marronetti et~al.}(2008)\citenamefont{Marronetti,
  Tichy, Br{\"u}gmann, Gonzalez, and Sperhake}}]{Marronetti:2007wz}
\bibinfo{author}{\bibfnamefont{P.}~\bibnamefont{Marronetti}},
  \bibinfo{author}{\bibfnamefont{W.}~\bibnamefont{Tichy}},
  \bibinfo{author}{\bibfnamefont{B.}~\bibnamefont{Br{\"u}gmann}},
  \bibinfo{author}{\bibfnamefont{J.}~\bibnamefont{Gonzalez}}, \bibnamefont{and}
  \bibinfo{author}{\bibfnamefont{U.}~\bibnamefont{Sperhake}},
  \bibinfo{journal}{Phys. Rev.} \textbf{\bibinfo{volume}{D77}},
  \bibinfo{pages}{064010} (\bibinfo{year}{2008}), \eprint{0709.2160}.

\bibitem[{\citenamefont{Nakamura et~al.}(1987)\citenamefont{Nakamura, Oohara,
  and Kojima}}]{Nakamura87}
\bibinfo{author}{\bibfnamefont{T.}~\bibnamefont{Nakamura}},
  \bibinfo{author}{\bibfnamefont{K.}~\bibnamefont{Oohara}}, \bibnamefont{and}
  \bibinfo{author}{\bibfnamefont{Y.}~\bibnamefont{Kojima}},
  \bibinfo{journal}{Prog. Theor. Phys. Suppl.} \textbf{\bibinfo{volume}{90}},
  \bibinfo{pages}{1} (\bibinfo{year}{1987}).

\bibitem[{\citenamefont{Shibata and Nakamura}(1995)}]{Shibata95}
\bibinfo{author}{\bibfnamefont{M.}~\bibnamefont{Shibata}} \bibnamefont{and}
  \bibinfo{author}{\bibfnamefont{T.}~\bibnamefont{Nakamura}},
  \bibinfo{journal}{Phys. Rev.} \textbf{\bibinfo{volume}{D52}},
  \bibinfo{pages}{5428} (\bibinfo{year}{1995}).

\bibitem[{\citenamefont{Baumgarte and Shapiro}(1999)}]{Baumgarte99}
\bibinfo{author}{\bibfnamefont{T.~W.} \bibnamefont{Baumgarte}}
  \bibnamefont{and} \bibinfo{author}{\bibfnamefont{S.~L.}
  \bibnamefont{Shapiro}}, \bibinfo{journal}{Phys. Rev.}
  \textbf{\bibinfo{volume}{D59}}, \bibinfo{pages}{024007}
  (\bibinfo{year}{1999}), \eprint{gr-qc/9810065}.

\bibitem[{cac()}]{cactus_web}
\bibinfo{note}{Cactus Computational Toolkit home page: {\tt
  http://cactuscode.org}}.

\bibitem[{ein()}]{einsteintoolkit}
\bibinfo{note}{Einstein Toolkit home page: {\tt http://einsteintoolkit.org}}.

\bibitem[{\citenamefont{Loffler et~al.}(2012)\citenamefont{Loffler, Faber,
  Bentivegna, Bode, Diener et~al.}}]{Loffler:2011ay}
\bibinfo{author}{\bibfnamefont{F.}~\bibnamefont{Loffler}},
  \bibinfo{author}{\bibfnamefont{J.}~\bibnamefont{Faber}},
  \bibinfo{author}{\bibfnamefont{E.}~\bibnamefont{Bentivegna}},
  \bibinfo{author}{\bibfnamefont{T.}~\bibnamefont{Bode}},
  \bibinfo{author}{\bibfnamefont{P.}~\bibnamefont{Diener}},
  \bibnamefont{et~al.}, \bibinfo{journal}{Class. Quant. Grav.}
  \textbf{\bibinfo{volume}{29}}, \bibinfo{pages}{115001}
  (\bibinfo{year}{2012}), \eprint{1111.3344}.

\bibitem[{\citenamefont{Schnetter et~al.}(2004)\citenamefont{Schnetter, Hawley,
  and Hawke}}]{Schnetter-etal-03b}
\bibinfo{author}{\bibfnamefont{E.}~\bibnamefont{Schnetter}},
  \bibinfo{author}{\bibfnamefont{S.~H.} \bibnamefont{Hawley}},
  \bibnamefont{and} \bibinfo{author}{\bibfnamefont{I.}~\bibnamefont{Hawke}},
  \bibinfo{journal}{Class. Quant. Grav.} \textbf{\bibinfo{volume}{21}},
  \bibinfo{pages}{1465} (\bibinfo{year}{2004}), \eprint{gr-qc/0310042}.

\bibitem[{\citenamefont{Thornburg}(2004)}]{Thornburg2003:AH-finding}
\bibinfo{author}{\bibfnamefont{J.}~\bibnamefont{Thornburg}},
  \bibinfo{journal}{Class. Quant. Grav.} \textbf{\bibinfo{volume}{21}},
  \bibinfo{pages}{743} (\bibinfo{year}{2004}), \eprint{gr-qc/0306056}.

\bibitem[{\citenamefont{Dreyer et~al.}(2003)\citenamefont{Dreyer, Krishnan,
  Shoemaker, and Schnetter}}]{Dreyer02a}
\bibinfo{author}{\bibfnamefont{O.}~\bibnamefont{Dreyer}},
  \bibinfo{author}{\bibfnamefont{B.}~\bibnamefont{Krishnan}},
  \bibinfo{author}{\bibfnamefont{D.}~\bibnamefont{Shoemaker}},
  \bibnamefont{and}
  \bibinfo{author}{\bibfnamefont{E.}~\bibnamefont{Schnetter}},
  \bibinfo{journal}{Phys. Rev.} \textbf{\bibinfo{volume}{D67}},
  \bibinfo{pages}{024018} (\bibinfo{year}{2003}), \eprint{gr-qc/0206008}.

\bibitem[{\citenamefont{Alcubierre et~al.}(2003)\citenamefont{Alcubierre,
  Br\"ugmann, Diener, Koppitz, Pollney, Seidel, and Takahashi}}]{Alcubierre02a}
\bibinfo{author}{\bibfnamefont{M.}~\bibnamefont{Alcubierre}},
  \bibinfo{author}{\bibfnamefont{B.}~\bibnamefont{Br\"ugmann}},
  \bibinfo{author}{\bibfnamefont{P.}~\bibnamefont{Diener}},
  \bibinfo{author}{\bibfnamefont{M.}~\bibnamefont{Koppitz}},
  \bibinfo{author}{\bibfnamefont{D.}~\bibnamefont{Pollney}},
  \bibinfo{author}{\bibfnamefont{E.}~\bibnamefont{Seidel}}, \bibnamefont{and}
  \bibinfo{author}{\bibfnamefont{R.}~\bibnamefont{Takahashi}},
  \bibinfo{journal}{Phys. Rev.} \textbf{\bibinfo{volume}{D67}},
  \bibinfo{pages}{084023} (\bibinfo{year}{2003}), \eprint{gr-qc/0206072}.

\bibitem[{\citenamefont{van Meter et~al.}(2006)\citenamefont{van Meter, Baker,
  Koppitz, and Choi}}]{vanMeter:2006vi}
\bibinfo{author}{\bibfnamefont{J.~R.} \bibnamefont{van Meter}},
  \bibinfo{author}{\bibfnamefont{J.~G.} \bibnamefont{Baker}},
  \bibinfo{author}{\bibfnamefont{M.}~\bibnamefont{Koppitz}}, \bibnamefont{and}
  \bibinfo{author}{\bibfnamefont{D.-I.} \bibnamefont{Choi}},
  \bibinfo{journal}{Phys. Rev.} \textbf{\bibinfo{volume}{D73}},
  \bibinfo{pages}{124011} (\bibinfo{year}{2006}), \eprint{gr-qc/0605030}.

\bibitem[{\citenamefont{Schnetter}(2010)}]{Schnetter:2010cz}
\bibinfo{author}{\bibfnamefont{E.}~\bibnamefont{Schnetter}},
  \bibinfo{journal}{Class. Quant. Grav.} \textbf{\bibinfo{volume}{27}},
  \bibinfo{pages}{167001} (\bibinfo{year}{2010}), \eprint{1003.0859}.

\bibitem[{\citenamefont{Pollney et~al.}(2011)\citenamefont{Pollney, Reisswig,
  Schnetter, Dorband, and Diener}}]{Pollney:2009yz}
\bibinfo{author}{\bibfnamefont{D.}~\bibnamefont{Pollney}},
  \bibinfo{author}{\bibfnamefont{C.}~\bibnamefont{Reisswig}},
  \bibinfo{author}{\bibfnamefont{E.}~\bibnamefont{Schnetter}},
  \bibinfo{author}{\bibfnamefont{N.}~\bibnamefont{Dorband}}, \bibnamefont{and}
  \bibinfo{author}{\bibfnamefont{P.}~\bibnamefont{Diener}},
  \bibinfo{journal}{Phys. Rev.} \textbf{\bibinfo{volume}{D83}},
  \bibinfo{pages}{044045} (\bibinfo{year}{2011}), \eprint{0910.3803}.

\bibitem[{\citenamefont{Szilagyi et~al.}(2009)\citenamefont{Szilagyi, Lindblom,
  and Scheel}}]{Szilagyi:2009qz}
\bibinfo{author}{\bibfnamefont{B.}~\bibnamefont{Szilagyi}},
  \bibinfo{author}{\bibfnamefont{L.}~\bibnamefont{Lindblom}}, \bibnamefont{and}
  \bibinfo{author}{\bibfnamefont{M.~A.} \bibnamefont{Scheel}},
  \bibinfo{journal}{Phys. Rev.} \textbf{\bibinfo{volume}{D80}},
  \bibinfo{pages}{124010} (\bibinfo{year}{2009}), \eprint{0909.3557}.

\bibitem[{\citenamefont{Babiuc et~al.}(2011)\citenamefont{Babiuc, Szilagyi,
  Winicour, and Zlochower}}]{Babiuc:2010ze}
\bibinfo{author}{\bibfnamefont{M.}~\bibnamefont{Babiuc}},
  \bibinfo{author}{\bibfnamefont{B.}~\bibnamefont{Szilagyi}},
  \bibinfo{author}{\bibfnamefont{J.}~\bibnamefont{Winicour}}, \bibnamefont{and}
  \bibinfo{author}{\bibfnamefont{Y.}~\bibnamefont{Zlochower}},
  \bibinfo{journal}{Phys. Rev.} \textbf{\bibinfo{volume}{D84}},
  \bibinfo{pages}{044057} (\bibinfo{year}{2011}), \eprint{1011.4223}.

\bibitem[{\citenamefont{Reisswig et~al.}(2009)\citenamefont{Reisswig, Bishop,
  Pollney, and Szilagyi}}]{Reisswig:2009us}
\bibinfo{author}{\bibfnamefont{C.}~\bibnamefont{Reisswig}},
  \bibinfo{author}{\bibfnamefont{N.~T.} \bibnamefont{Bishop}},
  \bibinfo{author}{\bibfnamefont{D.}~\bibnamefont{Pollney}}, \bibnamefont{and}
  \bibinfo{author}{\bibfnamefont{B.}~\bibnamefont{Szilagyi}},
  \bibinfo{journal}{Phys. Rev. Lett.} \textbf{\bibinfo{volume}{103}},
  \bibinfo{pages}{221101} (\bibinfo{year}{2009}), \eprint{0907.2637}.

\bibitem[{\citenamefont{Garcia et~al.}(2012)\citenamefont{Garcia, Lovelace,
  Kidder, Boyle, Teukolsky et~al.}}]{Garcia:2012dc}
\bibinfo{author}{\bibfnamefont{B.}~\bibnamefont{Garcia}},
  \bibinfo{author}{\bibfnamefont{G.}~\bibnamefont{Lovelace}},
  \bibinfo{author}{\bibfnamefont{L.~E.} \bibnamefont{Kidder}},
  \bibinfo{author}{\bibfnamefont{M.}~\bibnamefont{Boyle}},
  \bibinfo{author}{\bibfnamefont{S.~A.} \bibnamefont{Teukolsky}},
  \bibnamefont{et~al.}, \bibinfo{journal}{Phys.Rev.}
  \textbf{\bibinfo{volume}{D86}}, \bibinfo{pages}{084054}
  (\bibinfo{year}{2012}), \eprint{1206.2943}.

\bibitem[{\citenamefont{Nakano et~al.}(2011)\citenamefont{Nakano, Zlochower,
  Lousto, and Campanelli}}]{Nakano:2011pb}
\bibinfo{author}{\bibfnamefont{H.}~\bibnamefont{Nakano}},
  \bibinfo{author}{\bibfnamefont{Y.}~\bibnamefont{Zlochower}},
  \bibinfo{author}{\bibfnamefont{C.~O.} \bibnamefont{Lousto}},
  \bibnamefont{and}
  \bibinfo{author}{\bibfnamefont{M.}~\bibnamefont{Campanelli}},
  \bibinfo{journal}{Phys. Rev.} \textbf{\bibinfo{volume}{D84}},
  \bibinfo{pages}{124006} (\bibinfo{year}{2011}), \eprint{1108.4421}.

\bibitem[{\citenamefont{Alic et~al.}(2012)\citenamefont{Alic, Bona-Casas, Bona,
  Rezzolla, and Palenzuela}}]{Alic:2011gg}
\bibinfo{author}{\bibfnamefont{D.}~\bibnamefont{Alic}},
  \bibinfo{author}{\bibfnamefont{C.}~\bibnamefont{Bona-Casas}},
  \bibinfo{author}{\bibfnamefont{C.}~\bibnamefont{Bona}},
  \bibinfo{author}{\bibfnamefont{L.}~\bibnamefont{Rezzolla}}, \bibnamefont{and}
  \bibinfo{author}{\bibfnamefont{C.}~\bibnamefont{Palenzuela}},
  \bibinfo{journal}{Phys.Rev.} \textbf{\bibinfo{volume}{D85}},
  \bibinfo{pages}{064040} (\bibinfo{year}{2012}), \eprint{1106.2254}.

\bibitem[{\citenamefont{Hannam et~al.}(2007)\citenamefont{Hannam, Husa,
  Bruegmann, Gonzalez, and Sperhake}}]{Hannam:2006zt}
\bibinfo{author}{\bibfnamefont{M.}~\bibnamefont{Hannam}},
  \bibinfo{author}{\bibfnamefont{S.}~\bibnamefont{Husa}},
  \bibinfo{author}{\bibfnamefont{B.}~\bibnamefont{Bruegmann}},
  \bibinfo{author}{\bibfnamefont{J.~A.} \bibnamefont{Gonzalez}},
  \bibnamefont{and} \bibinfo{author}{\bibfnamefont{U.}~\bibnamefont{Sperhake}},
  \bibinfo{journal}{Class. Quant. Grav.} \textbf{\bibinfo{volume}{24}},
  \bibinfo{pages}{S15} (\bibinfo{year}{2007}), \eprint{gr-qc/0612001}.

\bibitem[{\citenamefont{Lovelace}(2009)}]{Lovelace:2008hd}
\bibinfo{author}{\bibfnamefont{G.}~\bibnamefont{Lovelace}},
  \bibinfo{journal}{Class. Quant. Grav.} \textbf{\bibinfo{volume}{26}},
  \bibinfo{pages}{114002} (\bibinfo{year}{2009}), \eprint{0812.3132}.

\bibitem[{\citenamefont{Mundim et~al.}(2011)\citenamefont{Mundim, Kelly,
  Zlochower, Nakano, and Campanelli}}]{Mundim:2010hu}
\bibinfo{author}{\bibfnamefont{B.~C.} \bibnamefont{Mundim}},
  \bibinfo{author}{\bibfnamefont{B.~J.} \bibnamefont{Kelly}},
  \bibinfo{author}{\bibfnamefont{Y.}~\bibnamefont{Zlochower}},
  \bibinfo{author}{\bibfnamefont{H.}~\bibnamefont{Nakano}}, \bibnamefont{and}
  \bibinfo{author}{\bibfnamefont{M.}~\bibnamefont{Campanelli}},
  \bibinfo{journal}{Class. Quant. Grav.} \textbf{\bibinfo{volume}{28}},
  \bibinfo{pages}{134003} (\bibinfo{year}{2011}), \eprint{1012.0886}.

\bibitem[{\citenamefont{Kelly et~al.}(2010)\citenamefont{Kelly, Tichy,
  Zlochower, Campanelli, and Whiting}}]{Kelly:2009js}
\bibinfo{author}{\bibfnamefont{B.~J.} \bibnamefont{Kelly}},
  \bibinfo{author}{\bibfnamefont{W.}~\bibnamefont{Tichy}},
  \bibinfo{author}{\bibfnamefont{Y.}~\bibnamefont{Zlochower}},
  \bibinfo{author}{\bibfnamefont{M.}~\bibnamefont{Campanelli}},
  \bibnamefont{and} \bibinfo{author}{\bibfnamefont{B.~F.}
  \bibnamefont{Whiting}}, \bibinfo{journal}{Class. Quant. Grav.}
  \textbf{\bibinfo{volume}{27}}, \bibinfo{pages}{114005}
  (\bibinfo{year}{2010}), \eprint{0912.5311}.

\bibitem[{\citenamefont{Kelly et~al.}(2007)\citenamefont{Kelly, Tichy,
  Campanelli, and Whiting}}]{Kelly:2007uc}
\bibinfo{author}{\bibfnamefont{B.~J.} \bibnamefont{Kelly}},
  \bibinfo{author}{\bibfnamefont{W.}~\bibnamefont{Tichy}},
  \bibinfo{author}{\bibfnamefont{M.}~\bibnamefont{Campanelli}},
  \bibnamefont{and} \bibinfo{author}{\bibfnamefont{B.~F.}
  \bibnamefont{Whiting}}, \bibinfo{journal}{Phys. Rev.}
  \textbf{\bibinfo{volume}{D76}}, \bibinfo{pages}{024008}
  (\bibinfo{year}{2007}), \eprint{0704.0628}.

\bibitem[{\citenamefont{Babiuc et~al.}(2008)}]{Babiuc:2007vr}
\bibinfo{author}{\bibfnamefont{M.~C.} \bibnamefont{Babiuc}}
  \bibnamefont{et~al.}, \bibinfo{journal}{Class. Quant. Grav.}
  \textbf{\bibinfo{volume}{25}}, \bibinfo{pages}{125012}
  (\bibinfo{year}{2008}), \eprint{0709.3559}.

\bibitem[{\citenamefont{Hannam et~al.}(2009)}]{Hannam:2009hh}
\bibinfo{author}{\bibfnamefont{M.}~\bibnamefont{Hannam}} \bibnamefont{et~al.},
  \bibinfo{journal}{Phys. Rev.} \textbf{\bibinfo{volume}{D79}},
  \bibinfo{pages}{084025} (\bibinfo{year}{2009}), \eprint{0901.2437}.

\bibitem[{\citenamefont{Scheel et~al.}(2006)}]{Scheel:2006gg}
\bibinfo{author}{\bibfnamefont{M.~A.} \bibnamefont{Scheel}}
  \bibnamefont{et~al.}, \bibinfo{journal}{Phys. Rev.}
  \textbf{\bibinfo{volume}{D74}}, \bibinfo{pages}{104006}
  (\bibinfo{year}{2006}), \eprint{gr-qc/0607056}.

\bibitem[{\citenamefont{Boyle et~al.}(2007)\citenamefont{Boyle, Lindblom,
  Pfeiffer, Scheel, and Kidder}}]{Boyle:2006ne}
\bibinfo{author}{\bibfnamefont{M.}~\bibnamefont{Boyle}},
  \bibinfo{author}{\bibfnamefont{L.}~\bibnamefont{Lindblom}},
  \bibinfo{author}{\bibfnamefont{H.}~\bibnamefont{Pfeiffer}},
  \bibinfo{author}{\bibfnamefont{M.}~\bibnamefont{Scheel}}, \bibnamefont{and}
  \bibinfo{author}{\bibfnamefont{L.~E.} \bibnamefont{Kidder}},
  \bibinfo{journal}{Phys. Rev.} \textbf{\bibinfo{volume}{D75}},
  \bibinfo{pages}{024006} (\bibinfo{year}{2007}), \eprint{gr-qc/0609047}.

\end{thebibliography}

\end{document}